\documentclass[12pt, letterpaper]{article}
\usepackage[utf8]{inputenc}
\usepackage{amsmath, amssymb}
\usepackage{graphicx}
\usepackage{caption}
\usepackage{subcaption}
\usepackage{booktabs}
\usepackage{threeparttable}
\usepackage{tabularx}
\usepackage{rotating}
\usepackage{algorithm}
\usepackage{algpseudocode}
\usepackage[none]{hyphenat}
\usepackage{microtype}
\usepackage{hyperref}
\usepackage{setspace}
\usepackage{mathrsfs}

\usepackage[utf8]{inputenc}
\usepackage[english]{babel}
\DeclareUnicodeCharacter{0308}{\"{}}

\usepackage{biblatex}
\usepackage[letterpaper, top=1in, bottom=1in, left=1in, right=1in]{geometry}

\title{Introducing the $\sigma$-Cell: Unifying GARCH, Stochastic Fluctuations and Evolving Mechanisms in RNN-based Volatility Forecasting}
\author{German Rodikov \\ SNS  \and Nino Antulov-Fantulin \\ ETH Zürich \& Aisot Technologies AG}

\begin{document}

\maketitle

\begin{abstract}
This paper introduces the $\sigma$-Cell, a novel Recurrent Neural Network (RNN) architecture for financial volatility modeling. Bridging traditional econometric approaches like GARCH with deep learning, the $\sigma$-Cell incorporates stochastic layers and time-varying parameters to capture dynamic volatility patterns. Our model serves as a generative network, approximating the conditional distribution of latent variables. We employ a log-likelihood-based loss function and a specialized activation function to enhance performance. Experimental results demonstrate superior forecasting accuracy compared to traditional GARCH and Stochastic Volatility models, making the next step in integrating domain knowledge with neural networks.
\end{abstract}



\section{Introduction}
Volatility is a key metric in econometrics for understanding financial asset price variability, with significant advancements in the field over the years \cite{bachelier2011louis}. Early stochastic models like Brownian motion were seminal but insufficient for capturing all aspects of asset price fluctuations \cite{cont2001empirical}. Volatility serves as a critical risk indicator in financial markets, especially in the growing derivatives market \cite{Engle1982,Bollerslev1986,poon2003forecasting}. While classical models like ARCH and GARCH have been groundbreaking, they face limitations such as unobservable intrinsic volatility and assumptions that may not fully reflect market dynamics \cite{poon2003forecasting,andersen1998answering,hansen2005forecast}.

In recent years, Neural Networks (NNs) have gained traction in financial econometrics, offering promising results in various applications ranging from bond rating to stock price prediction~\cite{dutta1988bond}.  However, their application in volatility forecasting has been limited, often serving as a supplementary tool to traditional models~\cite{hajizadeh2012hybrid}.

This study bridges the gap between traditional and modern approaches by introducing the $\sigma$-Cell. Our approach integrates the capabilities of Recurrent Neural Networks (RNNs) with the proven methodologies of GARCH and the theoretical foundation of latent stochastic processes. We evaluate the $\sigma$-Cell's performance using synthetic data, the S\&P 500 index, and the cryptocurrency pair of Bitcoin-USD (BTCUSDT), demonstrating its predictive performance.

\section{Preliminaries: Volatility Models and RNN}

The ARCH model, introduced by Engle in 1982, laid the groundwork for identifying time-varying volatility \cite{Engle1982}. Bollerslev extended this with the GARCH model in 1986, incorporating past variances into current estimates \cite{Bollerslev1986}. Engle and Kroner further expanded the model in 1995 to handle multiple financial series, leveraging historical data dependencies \cite{EngleKroner1995}.

We use the GARCH and GJR-GARCH models as deterministic baselines for comparison~\cite{Bollerslev1986, engle1982autoregressive}. The GARCH(1,1) model is especially popular for its simplicity and effectiveness in capturing financial volatility \cite{francq2010garch}, equation \ref{eq:garch11}. The GJR-GARCH model adds complexity by accounting for asymmetric volatility reactions to market returns~\cite{GlostenJagannathanRunkle1993}, equation \ref{eq:GJR_GARCH_11}. These models serve as robust benchmarks for evaluating our proposed approach.

\begin{equation} \label{eq:garch11}
\begin{aligned}
    \sigma_t^2 &= \alpha_0 + \sum_{i=1}^p \alpha_i x_{t-i}^2 + \sum_{j=1}^q \beta_j \sigma_{t-j}^2 \\
    x_t &\sim \mathscr{N}\left(0, \sigma_t^2\right)
\end{aligned}
\end{equation}

\begin{equation} \label{eq:GJR_GARCH_11}
\sigma_t^2 = \alpha_0 + \sum_{i=1}^p \alpha_i x_{t-i}^2 + \gamma x_{t-1}^2 I_{t-1} + \sum_{j=1}^q \beta_j \sigma_{t-j}^2
\end{equation}

Similarly, the TARCH model is another extension \ref{eq:tarch},
where $I_{t-1}$ is defined as in the GJR-GARCH model \cite{Zakoian1994}.

\begin{equation} \label{eq:tarch}
\sigma_t^2 = \alpha_0 + \sum_{i=1}^p \alpha_i x_{t-i}^2 + \gamma |x_{t-1}| I_{t-1} + \sum_{j=1}^q \beta_j \sigma_{t-j}^2
\end{equation}

Another notable variant is Nelson's EGARCH model, which formulates dependencies in log variance $\log(\sigma_t^2)$, equations \ref{eq:egarch}
This model provides a more nuanced understanding of the asymmetric relationships between observations and subsequent volatility shifts \cite{Nelson1991}.

\begin{equation} \label{eq:egarch}
\log(\sigma_t^2) = \alpha_0 + \sum_{i=1}^p \alpha_i \left(|\frac{x_{t-i}}{\sigma_{t-i}}| - \sqrt{2/\pi}\right) + \gamma \frac{x_{t-1}}{\sigma_{t-1}} + \sum_{j=1}^q \beta_j \log(\sigma_{t-j}^2)
\end{equation}

Realized volatility (RV), derived from high-frequency intraday returns, serves as a nonparametric volatility measure \cite{andersen2003modeling}. To capture volatility dynamics, we use Heterogeneous Autoregressive (HAR) models, which incorporate lagged RV components at daily, weekly, and monthly frequencies \cite{corsi2009simple}. The model reflects the varying trading horizons of market participants \cite{muller1993fractals}.

\begin{equation} \label{eq:har}
RV_t = c + \beta_d RV_{t-1}^{(d)} + \beta_w RV_{t-1}^{(w)} + \beta_m RV_{t-1}^{(m)} + \epsilon_t
\end{equation}

In equation \ref{eq:har}, $RV_t^{(d)}$, $RV_t^{(w)}$, and $RV_t^{(m)}$ represent daily, 5-day, and 22-day RV, respectively. This structure allows for the influence of multiple time horizons on volatility forecasts.

SV models propose that unseen latent processes influence current volatility, as exemplified by Heston's model \cite{Heston1993}. While theoretically robust, these models may require assumptions not always empirically supported. MCMC-based frameworks offer enhanced forecasting but can be data-intensive and less effective with multivariate series \cite{KastnerFruhwirthSchnatter2014, WuHernandezLobatoGhahramani2014}

We employ SV Heston model as stochastic baselines \cite{Heston1993}. These models use stochastic differential equations to capture volatility dynamics, equation \ref{eq:heston_1}. Including the SV model serves two purposes: benchmarking our model against established stochastic frameworks and testing its robustness and predictive accuracy. This comparative analysis aims to provide a comprehensive evaluation of various volatility modeling approaches.

\begin{equation}\label{eq:heston_1}
\begin{aligned}
    d S_t &= \mu S_t dt + \sqrt{v_t} S_t dW_t^1, \\
    d v_t &= \kappa(\theta - v_t) dt + \sigma \sqrt{v_t} dW_t^2
\end{aligned}
\end{equation}

\subsection{Hybrid approaches}

Deep learning techniques, initially successful in image and speech recognition, are increasingly applied to volatility modeling \cite{LeCunBengioHinton2015, KrizhevskySutskeverHinton2012}. RNNs and their variants like LSTM, BRNN, and GRU are particularly suited for this task due to their ability to model sequential data. Recent innovations aim to add flexibility and randomness to these networks, enhancing their forecasting power \cite{HochreiterSchmidhuber1997, SchusterPaliwal1997, Cho2014}.

Hybrid approaches in financial volatility modeling are gaining traction, combining deterministic and stochastic models for more nuanced time series analysis. One such approach integrates GARCH models with Long Short-Term Memory (LSTM) networks to capture complex volatility patterns \cite{HochreiterSchmidhuber1997,hu2020hybrid}. Specifically, GARCH models are first applied to forecast volatility, such as in copper returns, and these forecasts are then used as input features for LSTM networks, combining the strengths of both methods \cite{hu2020hybrid}.

The GARCH-MIDAS model integrates high-frequency financial data with low-frequency macroeconomic indicators to capture their impact on volatility \cite{engle2013stock}. Another hybrid approach combines ARCH-type and SV models, leveraging both autoregressive and stochastic volatility features for more accurate predictions \cite{shephard2005stochastic}. Additionally, a GARCH-Markov Switching model is used to capture both autoregressive patterns and regime-switching behavior in financial time series data \cite{hamilton2020time}.

Another approach combines the Neural Stochastic Volatility Model (NSVM) and multi-layer perceptrons (MLPs). The Neural Stochastic Volatility Model (NSVM) is combined with multi-layer perceptrons (MLPs) to compute means and variances of distributions. In this probabilistic model, observable variables depend on previous observables and latent variables, while latent variables only depend on their past values. The distributions are modeled as normal distributions with specified mean and variance parameters \cite{luo2018neural}. The Neural Stochastic Volatility Model (NSVM) is integrated with multi-layer perceptrons (MLPs) to compute distribution parameters. This probabilistic approach models observable and latent variables based on their past values, using normal distributions with specified mean and variance \cite{luo2018neural}.

In financial volatility forecasting, hybrid models have gained prominence for their ability to merge various modeling strengths. These can be categorized into statistical-statistical, machine-learning - machine-learning, statistical-machine-learning, and ensemble approaches. This paper introduces a novel hybrid model, the $\sigma$-Cell RNN, which combines the stochastic principles of GARCH with the time-series capabilities of RNNs, offering a versatile and adaptive approach for volatility forecasting.

\subsection{Recurrent Neural Networks}

RNNs cyclically process past and current inputs, mathematically captured by equations for hidden and output states \cite{zhang2021dive}, equations \ref{eq:general_rnn1}, \ref{eq:general_rnn2}.

\begin{equation}
\mathbf{h}_t=\phi_h\left(\mathbf{X}_t \mathbf{W}_{x h}+\mathbf{h}_{t-1} \mathbf{W}_{h h}+\mathbf{b}_h\right) 
\label{eq:general_rnn1}
\end{equation}

\begin{equation}
\mathbf{O}_t=\phi_o\left(\mathbf{h}_t \mathbf{W}_{h o}+\mathbf{b}_o\right)
\label{eq:general_rnn2}
\end{equation}

Training RNNs involves Backpropagation Through Time (BPTT) and a loss function $\mathcal{L}(\mathbf{O}, \mathbf{Y})$, defined in \ref{eq:general_rnn_loss}.

\begin{equation}
\mathcal{L}(\mathbf{O}, \mathbf{Y})=\sum_{t=1}^T \ell_t\left(\mathbf{O}_t, \mathbf{Y}_t\right)
\label{eq:general_rnn_loss}
\end{equation}

We employ a probabilistic loss using MLE. Neural networks are deterministic but can be extended to handle uncertainty through latent variables and variational inference \cite{kingma2013auto, rezende2014stochastic}. Other approaches involve combining observable and hidden variables \cite{chung2015recurrent} or leveraging the network's Markovian properties \cite{fraccaro2016sequential}.

\section{$\sigma$-Cell RNNs Volatility Models}

In this section, we introduce the estimation of conditional volatility in a time series using a modified GARCH process integrated with RNN dynamics. Given a time series $X=$ $\left\{x_1, x_2, \ldots, x_T\right\}$, our primary aim is to ascertain the conditional volatility, $\sigma_t$, at each discrete time $t$. The volatility, $\sigma_t^2$, is characterized by the subsequent relation, equation \ref{eq:x_series}. 

\begin{equation}
\sigma_t^2=F\left(x_t\right)_1 \cdot \sigma_{t-1}^2+F\left(x_t\right)_2 \cdot \epsilon_t^2 + b
\label{eq:x_series}
\end{equation}

In equation \ref{eq:x_series}, $F: \mathbb{R}^d \rightarrow \mathbb{R}^2$ denotes a function mapping the value $x_t$ onto a vector in $\mathbb{R}^2$. Notably, unlike traditional GARCH$(1,1)$ where parameters remain constant, the entities of this vector, denoted by $F\left(x_t\right)_1$ and $F\left(x_t\right)_2$, act as dynamic parameters, varying based on the data point $x_t$. And $b$ is a constant term, providing an additional degree of flexibility to the model. The residual error $\epsilon_t$ at an instance $t$ is computed as shown in equation \ref{eq:genera_rnn_garch_cell}, where $G: \mathbb{R}^d \rightarrow \mathbb{R}$ signifies a function that associates the observed value $x_t$ with its residual \ref{eq:genera_rnn_garch_cell_resid}.


\begin{equation}
\epsilon_t=x_t-G\left(x_t\right)
\label{eq:genera_rnn_garch_cell}
\end{equation}

\begin{equation}
G(x_t) \sim N(0,\sigma_t^2)
\label{eq:genera_rnn_garch_cell_resid}
\end{equation}

Equation \ref{eq:x_series} defines the process in which our methodology operates. It delineates a GARCH-like mechanism tailored for the precise estimation of conditional volatility inherent in a provided time series. Central to this approach is the explicit estimation of volatility, $\sigma_t$, for every point $t$ within the time series $X$, and dynamic volatility modeling wherein the model determines $\sigma_t^2$ by integrating the prevailing time series value, $x_t$, with the preceding volatility, $\sigma_{t-1}^2$, and the corresponding error term, $\epsilon_t^2$. The function $F$ is a pivotal element, offering a dynamic mapping from the current observation, $x_t$, to a bivariate vector, bestowing the model with real-time modulation of the GARCH parameters and thus affording enhanced adaptability. The error quantification term $\epsilon_t$ captures the difference between the actual and the anticipated values at time $t$, serving as a crucial metric to gauge the precision of the model.

For simplicity, in our model, the time series is assumed to be univariate with dimensionality $d = 1$, but in the general form, it could be multivariate. This dimensionality allows for capturing intricate patterns from multiple time series variables, enriching the model's capability in volatility estimation.

Our methodology provides a method to forecast volatility in multivariate time series data, by fusing the principles of the GARCH process with increased parameter flexibility through function $F$. Based on this, subsequent sections delve into how this model can be seamlessly merged with RNNs, thus potentially benefiting GARCH and RNN architectures to enhance volatility predictions.

\subsection{$\sigma$-Cell: Nonlinear GARCH-based RNN Cell}

This section introduces the nonlinear GARCH-based RNN Cell, abbreviated as $\sigma$-Cell. This innovative approach combines the predictive power of the GARCH model, known in econometrics for its volatility forecasting capabilities, with the ability of recurrent neural networks to process and learn from sequential data.

The motivation for the $\sigma$-Cell comes from the need for better volatility management in sequences, especially in financial data. RNNs, capable of processing information across long sequences, are combined with GARCH principles to handle this challenge more effectively.

Another innovation of the $\sigma$-Cell is its introduction of nonlinearity into the traditional GARCH model using an activation function. This approach allows the model to detect more complex patterns in sequential data, significantly improving its effectiveness compared to linear models \cite{franses1996forecasting}. The $\sigma$-Cell calculates conditional volatility at each time point using a nonlinear transformation, represented by $\phi$. This transformation adjusts the weighted mix of squared past volatility and squared input, as expressed in equation \ref{sigma_cell}:


\begin{equation}
\tilde{\sigma}_{t}^2 = \phi(\tilde{\sigma}_{t-1}^2 W_s + x_{t-1}^2 W_r + b_h)
\label{sigma_cell}
\end{equation}
\begin{equation}
\sigma_{t}^2=\phi_o(\tilde{\sigma}_{t}^2 W_o + b_o)
\label{sigma_cell2}
\end{equation}

In equation \ref{sigma_cell}, $\sigma_t^2$ denotes the hidden state of the RNN at time $t$. Weight matrices $W_s$ and $W_r$ correspond to the previous volatility and input, respectively.

In the RNN described by equation \ref{sigma_cell}, the weight matrices $W_s$ and $W_r$ each have parameters equal to the product of the input size and hidden size. In this specific case, our input is scalar, and we have chosen a small hidden size of 10, making the number of parameters for both $W_s$ and $W_r$ relatively small. The bias vector $b_h$ has a number of parameters equal to the hidden size. In equation \ref{sigma_cell2}, the weight matrix $W_o$ has parameters corresponding to the product of the hidden size and output size, with our output consisting of just one value. The bias vector $b_o$ has parameters equal to the output size. Summing up the parameters from all these elements, the total number of parameters in this system comes to 41.

The optimization of the $\sigma$-Cell RNN model is carried out in two phases, with each phase refining one set of weights to enhance the model's stability. This novel approach augments the traditional GARCH model with nonlinearity, better capturing complex volatility fluctuations. By integrating the $\sigma$-Cell, we expect to achieve a more effective model for forecasting volatile patterns, benefiting from the flexibility and adaptability of deep learning architectures.

The selection of the optimal activation function, denoted as $\phi$ and $\phi_o$, is pivotal to neural network efficacy. In this study, the Adjuated-Softplus activation function serves as $\phi$, introducing nonlinearity in hidden layers. Conversely, the ReLU activation function is adopted for $\phi_o$ to impart efficient nonlinearity to the output. Prior research underscores the pronounced influence of activation functions on network performance
\cite{karlik2011performance,ramachandran2017searching}. Thus, careful consideration should be given when selecting an activation function from the various options available \cite{he2015delving}.






\subsection{$\sigma$-Cell-N: Integrating Stochastich Layer}

Building on the $\sigma$-Cell RNN volatility model introduced in the previous section, we present the enhanced $\sigma$-Cell-N model. This iteration integrates a stochastic layer into the $\sigma$-Cell's RNN dynamics, adding depth and complexity to its predictive capabilities.
The stochastic layer introduces a stochastic component to the residuals. Specifically, for each time instance $t$, the residual is given by \ref{eq:N_layer_sigma_cell}.

\begin{equation}
\tilde{x}_{t-1} = x_{t-1} - N(0, \sigma_{t-1})
\label{eq:N_layer_sigma_cell}
\end{equation}


In equation \ref{eq:N_layer_sigma_cell}, $N$ represents the Gaussian distribution, and $\sigma_{t-1}$ is the volatility from the previous time point. The stochastic layer in the \ref{eq:N_layer_sigma_cell} model is formulated to directly incorporate the volatility from the previous time point as a variance measure, thereby coupling the past volatility's influence with the current observation in a manner reminiscent of GARCH model dynamics. This design choice ensures a continuous and coherent propagation of uncertainty through the series, allowing for more nuanced volatility predictions. While the notation draws inspiration from RNN structures, it essentially mirrors the error term in traditional GARCH models, as depicted in equation \ref{eq:x_series}.

The variance dynamics over time are represented by equations \ref{eq:sigma_cell_N}, \ref{eq:sigma_cell_N2}.


\begin{equation}
\tilde{\sigma}_{t}^2 = \phi(\tilde{\sigma}_{t-1}^2 W_s + \tilde{x}_{t-1}^2 W_r + b_h)
\label{eq:sigma_cell_N}
\end{equation}

\begin{equation}
\sigma_{t}^2=\phi_o(\tilde{\sigma}_{t}^2 W_o + b_o)
\label{eq:sigma_cell_N2}
\end{equation}

Equations \ref{eq:sigma_cell_N}, \ref{eq:sigma_cell_N2}, $\sigma_{t-1}^2 W_s$ capture the influence of past variance on current variance, echoing GARCH models legacy effects. The $\tilde{x}_{t-1}^2 W_r$ term assesses the squared residual effects on current variance, paralleling GARCH models' disturbance influence. 

Weights $W_s, W_r$, and $W_o$ are learned parameters potentially derived from neural network structures. Functions $\phi$ and $\phi_o$ are the Adjusted Softplus and ReLU, respectively, adding nonlinearity to the model, while $b_h$ and $b_o$ are bias terms.

The $\sigma$-Cell-N offers a more comprehensive perspective on volatility dynamics by combining past variance, disturbances, and a stochastic layer. This model exemplifies the blend of GARCH principles with neural architecture-derived weights and nonlinear activations, providing a richer understanding of time series volatility.



\subsection{$\sigma$-Cell-RL: Integrating Residuals RNN Layer}

The enhanced model introduces a novel mechanism for calculating residuals, diverging from the purely stochastic approach in the $\sigma$-Cell-N configuration. Instead of solely relying on stochastic deviations, this version uses the discrepancies between empirical data and predictions provided by an added RNN layer $G$, as shown in equations \ref{eq:RL_resid_layer}, \ref{eq:RL_resid_layer_output}, where we use the hyperbolic tangent (tanh) activation function $\varphi$ \cite{lecun1989backpropagation}.

\begin{equation}
h_t = \varphi\left(x_{t-1} W_{x h}+h_{t-1} W_{h h}+b_h\right)
\label{eq:RL_resid_layer}
\end{equation}

\begin{equation}
G(h_t) = \varphi \left(h_t W_{h o}+b_o\right)
\label{eq:RL_resid_layer_output}
\end{equation}

\begin{equation}
\tilde{x}_{t-1} = x_{t-1} - G(h_t)
\label{eq:RL_layer_sigma_cell}
\end{equation}

Here, $\tilde{x}_{t-1}$ represents the residuals at the $t^{\text{th}}$ time instance. The residual computation takes into account the actual value $x_{t-1}$ as input for $G$. $G$ is computed using the prior hidden state $h_{t-1}$ and the current input $x_{t-1}$, reflecting the behavior of recurrent cells. In essence, $h_{t-1}$ captures historical context, which, combined with $x_{t-1}$, aids in current forecasting. For variance, the modeling remains fundamentally consistent, as shown in equation \ref{eq:sigma_cell_RL}.


\begin{equation}
\tilde{\sigma}_{t}^2 = \phi(\tilde{\sigma}_{t-1}^2 W_s + \tilde{x}_{t-1}^2 W_r + b_h)
\label{eq:sigma_cell_RL}
\end{equation}
\begin{equation}
\sigma_{t}^2=\phi_o(\tilde{\sigma}_{t}^2 W_o + b_o)
\label{eq:sigma_cell_RL2}
\end{equation}

The critical distinction is that the residuals, $\tilde{x}_t$, now stem from the RNN layer's predictions. This tie-in of the RNN layer inherently adjusts the variance equation based on the predictive capabilities of the RNN cell.

The $\sigma$-Cell-RL model integrates an RNN component to craft residuals. Instead of using stochastic elements for unpredictability, it harnesses the RNN's forecasting deviations. This marriage of time series modeling with neural networks enhances adaptability, potentially elevating the model's capability to detect complex data patterns in sequences.

\subsection{$\sigma$-Cell-NTV: Integrating Time-Varying Approach}

Building upon previous discussions on fixed parameter weights $W_r$ and $W_s$, this section delves into a time-varying approach for these parameters.
First, the input vector $\mathbf{x}_{t-1}$ undergoes a transformation \ref{eq:sigma_cell-NTV_time_var_layer}.

\begin{equation}
w_{t-1}=\tilde{\varphi}\left(\mathbf{W} \mathbf{x}_{t-1}+\mathbf{b}\right)
\label{eq:sigma_cell-NTV_time_var_layer}
\end{equation}

Here, the linear combination of $\mathbf{x}_{t-1}$ with weight matrix $\mathbf{W}$ and bias vector $\mathbf{b}$ is passed through a nonlinear function, $\tilde{\varphi}$, resulting in the vector $w_{t-1}$.

Further, $w_{t-1}$ is split into two components \ref{eq:sigma_cell-NTV_Ws}, \ref{eq:sigma_cell-NTV_Wr}.

\begin{equation}
W_{s, t}=\pi_1\left(w_{t-1}\right)
\label{eq:sigma_cell-NTV_Ws}
\end{equation}

\begin{equation}
W_{r, t}=\pi_2\left(w_{t-1}\right)
\label{eq:sigma_cell-NTV_Wr}
\end{equation}

Using $\pi_1$ and $\pi_2$, the first and second halves of $w_{t-1}$ are extracted, respectively. For a $w_{t-1}$ with $2n$ elements \ref{eq:sigma_cell-NTV_P1}, \ref{eq:sigma_cell-NTV_P2}.

\begin{equation}
\pi_1\left(w_{t-1}\right)=w_{t-1}[1: n]
\label{eq:sigma_cell-NTV_P1}
\end{equation}

\begin{equation}
\pi_2\left(w_{t-1}\right)=w_{t-1}[n+1: 2n]
\label{eq:sigma_cell-NTV_P2}
\end{equation}

The residual computation employs stochasticity similar to \ref{eq:N_layer_sigma_cell}, we provide it in equation \ref{eq:sigma_cell-NTV_residulas}.

\begin{equation}
\tilde{x}_{t-1} = x_{t-1} - N(0, \sigma_{t-1})
\label{eq:sigma_cell-NTV_residulas}
\end{equation}

Lastly, the variance evolution, similar to $\sigma$-Cell dynamics, is described in equation \ref{eq:sigma_cell_NTV}


\begin{equation}
\tilde{\sigma}_{t}^2 = \phi(\tilde{\sigma}_{t-1}^2 W_{s, t} + \tilde{x}_{t-1}^2 W_{r, t} + b_h)
\label{eq:sigma_cell_NTV}
\end{equation}

\begin{equation}
\sigma_{t}^2=\phi_o(\tilde{\sigma}_{t}^2 W_o + b_o)
\label{eq:sigma_cell_NTV2}
\end{equation}

In \ref{eq:sigma_cell_NTV}, the variance $\sigma_t^2$ hinges on past variance, $\sigma_{t-1}^2$, modulated by the time-varying parameter $W_{s, t}$ and the squared residuals $\tilde{x}t^2$ modulated by $W{r, t}$. 

Conclusively, by integrating time-varying parameters $W_{s,t}$ and $W_{r,t}$, we achieve a fusion of traditional time series techniques with deep learning approaches, with refining variance modeling by time-varying parameters.

\subsection{$\sigma$-Cell-RLTV: Integrating Time-Varying Approach}

The proposed model marries the dynamic attributes of the time-varying method with the recurrent, residual features of the $\sigma$-Cell-RL. For each time step $t$ we describe it \ref{eq:sigma_cell-RLTV1}-\ref{eq:sigma_cell-RLTV5}.

\begin{equation}
w_t=\tilde{\varphi}(\mathbf{W} \mathbf{x}_{t-1}+\mathbf{b})
\label{eq:sigma_cell-RLTV1}
\end{equation}

\begin{equation}
W_{s, t}=\pi_1(w_t)
\label{eq:sigma_cell-RLTV2}
\end{equation}

\begin{equation}
W_{r, t}=\pi_2(w_t)
\label{eq:sigma_cell-RLTV3}
\end{equation}


\begin{equation}
h_t = \varphi\left(x_{t-1} W_{x h}+h_{t-1} W_{h h}+b_h\right)
\label{eq:sigma_cell-RLTV4}
\end{equation}

\begin{equation}
G(h_t) = \varphi \left(h_t W_{h o}+b_o\right)
\label{eq:sigma_cell-RLTV4_2}
\end{equation}

\begin{equation}
\tilde{x}_{t-1} = x_{t-1} - G(h_t)
\label{eq:sigma_cell-RLTV4_3}
\end{equation}

\begin{equation}
\tilde{\sigma}_{t}^2 = \phi(\tilde{\sigma}_{t-1}^2 W_{s, t} + \tilde{x}_{t-1}^2 W_{r, t} + b_h)
\label{eq:sigma_cell-RLTV5}
\end{equation}

\begin{equation}
\sigma_{t}^2=\phi_o(\tilde{\sigma}_{t}^2 W_o + b_o)
\label{eq:sigma_cell-RLTV6}
\end{equation}

Here, $\sigma_{t-1}^2$ represents the historical variance, modulated by the recurrent, time-varying weights $W_{s, t}$ and $W_{r, t}$.
Parameters - $\mathbf{W}$ and $\mathbf{b}$ denote weights and biases. And $\tilde{\varphi}, \pi_1, \pi_2, f$, and $\phi$ are the operational functions in the network. However, $h_{t-1}$, $\tilde{\sigma}_{t-1}$ are the hidden states from the prior time step. Analogous with $\sigma$-RL, we implement $\tilde{x}_{t-1}$ in equations \ref{eq:sigma_cell-RLTV4}, \ref{eq:sigma_cell-RLTV4_2}, \ref{eq:sigma_cell-RLTV4_3}.

In this approach, we improved memory retention and heightened resistance to specific noise disturbances. However, the model comes with an increased computational burden and a potential to overfit, particularly in sparser data sets.

\subsection{Log-likelihood}
Our objective is to identify optimal functions $F$ and $G$ using maximum likelihood estimation, which seeks to maximize the likelihood of observed data given a model. We focus on minimizing the negative log-likelihood, expressed as the loss function $L$ in equation \ref{eq:MLE_loss}.

\begin{equation}
L=\sum_t\left[\log \left(\sigma_t^2\right)+\frac{\left(x_t-G\left(x_t\right)\right)^2}{\sigma_t^2}\right]
\label{eq:MLE_loss}
\end{equation}

The loss $L$ comprises two terms: the log-variance $\log \left(\sigma_t^2\right)$ and the squared error between observed and predicted values, scaled by the inverse variance. Minimizing $L$ effectively adjusts the model's predictions closer to observed values \ref{eq:argmin_}.

\begin{equation}
\underset{F, G}{\text{argmin}}~ L(F, G)
\label{eq:argmin_}
\end{equation}

This minimization is typically achieved through gradient-based methods like Gradient Descent. The negative log-likelihood is used to convert the maximization problem into a minimization problem, aligning it with standard optimization algorithms.

\subsection{Activation Function}

Activation functions like ReLU and Softplus are crucial in neural networks for capturing nonlinear patterns, especially in complex systems like financial markets. These markets are influenced by interconnected variables that often interact nonlinearly, making the ability to model these relationships vital for accurate predictions.

ReLU has become a default choice due to its simplicity and computational efficiency, mitigating the vanishing gradient problem in deep networks \cite{nair2010rectified,glorot2011deep,ramachandran2017searching}. ReLU outputs the input if positive and zero otherwise.

Softplus is a smooth approximation to ReLU, without a sharp transition at zero and avoiding absolute zero activation \cite{glorot2011deep}.

We introduce a modified Softplus activation function, defined in equation \ref{eq:adj_softplus}. Unlike the standard Softplus, our version outputs zero for negative inputs and is scaled so that Softplus(1) equals 1. This adaptation is helpful in scenarios requiring smooth, non-negative, and normalized output values.

\begin{equation}
    \operatorname{Adjusted} \operatorname{Softplus}(x)=\max \left(0,\left(\frac{1}{\beta} \log \left(1+e^{\beta x}\right)-\frac{\log (2)}{\beta}\right) \cdot \frac{1}{\frac{1}{\beta} \log \left(1+e^{\beta}\right)-\frac{\log (2)}{\beta}}\right)
    \label{eq:adj_softplus}
\end{equation}

\subsection{Traning}

We initialize the $\sigma$-Cell RNN weights with the Xavier Uniform distribution and zero biases \cite{glorot2010understanding, sutskever2013importance}. Gradient clipping is applied to prevent exploding gradients, with a maximum norm of 1.0 \cite{pascanu2013difficulty}. Utilizing the Adam optimizer, known for its efficiency in training deep learning models \cite{kingma2014adam}, we adapt learning rates based on historical gradients. The RNN weights are updated each epoch using the Log-likelihood loss equation \ref{eq:MLE_loss}, iterating until convergence. This approach effectively captures the complex volatility dynamics in financial time series.

\section{Experimental Approach: Synthetic and Real Data}

Synthetic data sets, where the true generative process is known, are valuable for evaluating statistical and machine-learning models \cite{gretton2009covariate,moreno2012unifying}. With real-world data, the underlying data-generating process is usually unknown. Synthetic data provides a means of assessing model performance against a ground truth \cite{quinonero2008dataset}. In time series analysis and econometrics, synthetic data enables testing volatility predictions when the true volatility path is observable \cite{francq2019garch}. This helps determine model accuracy in volatility estimation prior to deployment on real financial data.

Synthetic data generation requires specifying a data model that captures key properties of the real data. For financial data, this may include stylistic properties like autocorrelation, heteroskedasticity, jumps, and fat tails \cite{cont2001empirical}. Controlled experiments can then evaluate model performance under different controlled generative processes \cite{athey2015machine}. Cross-validation on real data is still needed, but synthetic data provides an additional diagnostic.

Real-world financial data sets also have advantages complementary to synthetic data. They capture the nuances of actual market conditions \cite{fama1998market}. Economic events, investor behavior, and market microstructure are naturally embedded \cite{o1998market}. Models developed and tested solely on synthetic data may fail to generalize to real data. Testing on real data sets from different time periods and markets is essential \cite{pagan1996econometrics}.

In practice, a combination of synthetic and real data is ideal for developing and evaluating financial models \cite{francq2019garch,francq2015risk}. Synthetic data allows diagnosing accuracy and tuning models. Real data then evaluates real-world performance across different markets and time periods. The dual use of synthetic and real data leverages the strengths of each in developing robust and generalizable models.

\subsection{Estimation and Forecasting Estimation}

Our study utilized multiple models to evaluate their forecasting performance. To assess the accuracy of these models, we employed multiple approaches. The $R^2$ of Mincer-Zarnowitz forecasting regressions \cite{mincer1969evaluation}.

Mean Absolute Error (MAE) metric evaluates the average magnitude of errors between predicted and observed values without considering their direction, equation \ref{eq:mae}, where $\sigma_t$ is the observed value and $\hat{\sigma}_t$ is the predicted value at observation $t$, and $T$ is the total number of observations. In contrast to Root Mean Squared Error (RMSE), MAE treats all errors equally. 

\begin{equation}
\text{MAE} = \frac{1}{T} \sum_{i=1}^n |\sigma_t - \hat{\sigma}_t|
\label{eq:mae}
\end{equation}

Root Mean Squared Error metric assesses the average magnitude of errors between predicted and observed values, equation \ref{eq:rmse}, where $\sigma_t$ is the observed value and $\hat{\sigma}_t$ is the predicted value at observation $t$, and $T$ is the total number of observations. It is particularly sensitive to outliers since it gives more weight to larger errors than smaller ones. A model with a lower RMSE value is considered better fitting.

\begin{equation}
\text{RMSE} = \sqrt{\frac{1}{T} \sum_{i=1}^n (\sigma_t - \hat{\sigma}_t)^2}
\label{eq:rmse}
\end{equation}

The heteroskedasticity adjusted root mean square error (HRMSE) \cite{bollerslev1996periodic}, which is calculated in equation \ref{eq:hrmse}, where  $\sigma_t$ is the variance at time $t$ and $\widehat{\sigma}_t$ is the corresponding forecast. HRMSE is a modified version of the RMSE that takes into account the presence of heteroscedasticity in the data.

\begin{equation}
    \mathrm{HRMSE}=\sqrt{\frac{1}{T} \sum_{t=1}^T\left(\frac{\mathrm{\sigma}_t-\widehat{\mathrm{\sigma}}_t}{\mathrm{\sigma}_t}\right)^2}
    \label{eq:hrmse}
\end{equation}

However, as the HRMSE is not considered a robust loss function \cite{patton2011volatility}, we also employed the QLIKE loss function, defined in equation \ref{eq:qlike}.

\begin{equation}
    \mathrm{QLIKE}=\frac{1}{T} \sum_{t=1}^T\left(\log \mathrm{\sigma}_t+\frac{\widehat{\mathrm{\sigma}}_t}{\mathrm{\sigma}_t}\right)
    \label{eq:qlike}
\end{equation}

The QLIKE loss function measures how well a model predicts a set of observations, considering both the mean and variance of the predicted values. It is particularly useful for evaluating volatility models where the focus is on forecasting the variance of returns, which is robust in the context \cite{patton2011volatility}. 


The Negative Log-Likelihood (NLL) metric quantifies the fit of a model to observed data by calculating the negative logarithm of the likelihood of the observed data given the model. Lower NLL values indicate a better fit of the model to the observed data. In this study, NLL is defined as shown in equation \ref{eq:NLL}, where $r_t$ is the observed return at time $t, \hat{\sigma}_t$ is the model's predicted volatility at time $t$, and $P\left(r_t \mid \hat{\sigma}_t\right)$ represents the likelihood of observing return $r_t$ given the predicted volatility $\hat{\sigma}_t$.

\begin{equation}
\text{NLL} = - \sum_{t=1}^T \log P(r_t | \hat{\sigma}_t)
\label{eq:NLL}
\end{equation}

We employ the Diebold-Mariano (DM) test to assess the performance of volatility forecasting models using Mean Squared Error (MSE) and Mean Absolute Deviation (MAD) as loss functions \cite{diebold1995paring}. The test is conducted at a 5\% confidence level to identify significant model differences. Additionally, the Model Confidence Set (MCS) test is used for a more comprehensive comparison of multiple models \cite{hansen2011model}. This test employs bootstrapping with 10,000 samples to identify the best-performing models at a 5\% confidence level.

\subsection{Synthetic Data Generation}

Our synthetic data generation is inspired by the cyclical and often fluctuating nature of financial market volatility. We consider a sequence of 2000 data points where volatility ($\sigma_{i}$) at each point $i$ is generated using equation \ref{eq:synth_sigma}.

\begin{equation}
\sigma_i=1+A \sin \left(\frac{\pi i}{B}\right)
\label{eq:synth_sigma}
\end{equation}

Here, the parameters $A$ and $B$ govern the Amplitude and frequency of the sine wave, respectively. These are set to $A = 0.7$ and $B = 50$ in our specific synthetic data generation process. The $\sin$ function imbues our model with cyclical fluctuations in the volatility, embodying the frequently changing volatility regimes often observed in financial markets.

The synthetic return ($r_{i}$) at each point $i$ is then created using the equation:

\begin{equation}
r_i=\sigma_i \cdot \epsilon_i
\label{eq:synth_r}
\end{equation}

In \ref{eq:synth_r} $\epsilon_{i}$ is a random number drawn from a standard normal distribution ($\epsilon_{i} \sim N(0,1)$). This ensures that our synthetic returns ($r_{i}$) are directly influenced by our generated volatility ($\sigma_{i}$).

We split this synthetic data set into a training set (the first half of the data) and a test set (the second half) to evaluate the model's performance on unseen data.

This synthetic data provides a robust testing ground for our model, enabling us to compare the predicted volatility values with the true known volatility. This synthetic data set, with known underlying dynamics, is a valuable benchmark for evaluating the performance of the volatility modeling. It is essential to note that the synthetic data has been structured to reflect some stylized facts of financial returns, which will aid in better understanding the model's efficacy in real-world scenarios.




In the following sections, we will expand our experimental analysis to real financial data, bringing additional complexity and testing the model's performance in capturing more intricate, real-world dynamics.

\subsection{Real Data}
In this study, we investigate the proposed models for estimating and predicting realized volatility in diverse market structures. We focus on two specific asset types: an index, represented by the S\&P 500, and a cryptocurrency, represented by the Bitcoin-USD (BTCUSDT) pair. We analyze daily data to examine the effects and dependencies associated with this granularity \ref{tab:real_data_len}.

For the S\&P 500 data, we use 1-minute price observations from March 10, 2007, to March 1, 2022, resulting in 3,800 days of realized volatility observations. RV and returns are calculated based on the last daily closing price. Our experimental data set consists of intraday returns and corresponding RV. We partition the data set into training, validation, and test subsets. The validation and test subsets each contain 252 points, equivalent to one trading year, with the remaining data allocated to the training subset.

For the cryptocurrency data, we study the Bitcoin-USD pair. We use 1-minute price data from January 1, 2013, to April 20, 2020. This data set provides extensive price points, from which we calculate 2,667 RV observations. As with the S\&P 500 data set, we divide the cryptocurrency data into training, validation, and test subsets, with the validation and test subsets each containing 252 points.

\begin{table}[ht]
\centering
\caption{Data set Description for S\&P 500 and BTCUSDT}
\label{tab:real_data_len}
\begin{tabular}{lclllr}
\toprule
Asset    & Time Frame & From    & To      & RV Points \\
\midrule
S\&P 500 & 1 minute         & 10.03.07 & 01.03.22 & 3,800    \\
BTCUSDT  & 1 minute         & 01.01.13 & 20.04.20 & 2,667    \\
\bottomrule
\end{tabular}
\begin{tablenotes}
\item[\textit{Note:}]The table presents an overview of the data used in the analysis, including the asset, the time frame of the data, the date range of the data, and the number of realized volatility (RV) points. The last 252 data points are used for out-of-sample testing, and the preceding 252 data points are used for validation.
\end{tablenotes}
\end{table}

The mean, median, standard deviation, skewness, and kurtosis were calculated for each asset's returns, giving us valuable insights into the underlying data distributions \ref{tab:stat_SnP_BTC}.

\begin{table}[ht]
\centering
\caption{Statistical Summary of Intraday Returns for S\&P 500 and BTCUSDT.}
\label{tab:stat_SnP_BTC}
\begin{tabular}{llllll}
\toprule
Asset    & Mean    & Median & STD      & Skewness & Kurtosis \\
\midrule
S\&P 500 & 0.00029 & 0.0007 & 0.0128   & -0.1886  & 13.9915  \\
BTCUSDT  & 0.00349 & 0.0020 & 0.0466   & -0.0726  & 14.3508  \\
\bottomrule
\end{tabular}
\begin{tablenotes}
\item[\textit{Note:}] The table presents key statistical metrics of the intraday returns for the S\&P 500 index and the Bitcoin-USD (BTCUSDT) trading pair. The metrics include mean, median, standard deviation (STD), skewness, and kurtosis. A positive skewness indicates a right-side heavier tail of the probability density function, while a negative skewness indicates a left-side heavier tail. Kurtosis measures the "tailedness" of the probability distribution of returns. Higher kurtosis indicates a heavier tail, signifying a higher probability of extreme outcomes. For the analysis, we use the entire data set without dividing it into validation and test sets.
\end{tablenotes}
\end{table}




By focusing on these two contrasting asset types, we aim to gain a comprehensive understanding of how various $\sigma$-Cell can estimate and predict realized volatility across different market structures accurately.

\section{Results}

In this study, we sought to gauge the predictive efficacy of the proposed $\sigma$-Cell models. Our evaluation encompassed various experiments using synthetic and real-world financial data sets. The proposed models' performance was contrasted against traditional GARCH family models and the Stochastic Volatility (SV) model in the synthetic data scenario. However, the comparison was made against GARCH, SV models, as well as the Heterogeneous Autoregressive (HAR) model proposed by Corsi (2009) in the actual data context \cite{corsi2009simple}.




\subsection{Synthetic data set}

The comparative analysis delineated in Table \ref{tab:synth_valid_metrics} elucidates the performance of various $\sigma$-Cell model variants vis-à-vis established models across multiple evaluation metrics, namely RMSE, MAE, NLL, $\delta$ Mean, and $\delta$ Amplitude. Notably, the $\sigma$-Cell-RLTV variant demonstrates superior performance in RMSE and MAE metrics, outclassing all other models under consideration. Conversely, the TARCH model lags behind, registering the least favorable scores in these categories.

In terms of the NLL metric, the $\sigma$-Cell-NTV variant emerges as the most efficient model, while the TARCH model once again exhibits suboptimal performance. Furthermore, the $\sigma$-Cell-NTV variant also leads in the $\delta$ Mean and $\delta$ Amplitude metrics, underscoring its robustness across multiple dimensions of evaluation.

Turning our attention to Table \ref{tab:synth_test_metrics}, which assesses out-of-sample performance on synthetic test data, the $\sigma$-Cell-RLTV variant continues to distinguish itself. It excels not only in RMSE and MAE but also registers the lowest NLL among its $\sigma$-Cell counterparts. 

When juxtaposed with other volatility models such as GARCH(1,1), EGARCH, TARCH, GJR-GARCH, and SV, the $\sigma$-Cell-RLTV variant maintains its competitive edge in RMSE and MAE metrics.

Overall, the $\sigma$-Cell-RLTV model shows superior performance in RMSE, MAE, and NLL metrics compared to other models. Its performance in the $\delta$ Mean and $\delta$ Amplitude metrics also suggests its consistency and accuracy in predicting volatility changes. These results highlight the potential effectiveness of the $\sigma$-Cell-RLTV model in predicting realized volatility on synthetic test data.

\begin{table}
\centering 
\caption{Comparative Performance Metrics of Volatility Models on In-Sample Synthetic Data}
\label{tab:synth_valid_metrics} 
\begin{threeparttable}
\begin{tabular}{lrrrrr}
\toprule
 Model             &   RMSE &    MAE &    NLL &   $\delta$ Mean  &  $\delta$ Amplitude \\
\midrule
$\sigma$-Cell        & 0.3207 & 0.2362 & 0.9703 & -0.0810 &                -1.4876 \\
$\sigma$-Cell-N      & 0.3039 & 0.2292 & 0.9651 & -0.0316 &                -1.2188 \\
$\sigma$-Cell-NTV    & 0.2741 & 0.2223 & 0.9565 &  0.0318 &                -0.4003 \\
$\sigma$-Cell-RL     & 0.3217 & 0.2401 & 0.9707 & -0.0496 &                -1.6247 \\
$\sigma$-Cell-RLTV   & 0.2614 & 0.2043 & 0.9531 & -0.0265 &                -0.7845 \\
 GARCH(1,1)          & 0.3058 & 0.2295 & 1.1977 & -0.0537 &                -1.4702 \\
 EGARCH              & 0.3712 & 0.2376 & 1.2255 & -0.0594 &                -4.8949 \\
 TARCH               & 0.3844 & 0.2471 & 1.2182 & -0.0361 &                -5.1248 \\
 GJR-GARCH           & 0.3160 & 0.2383 & 1.1890 & -0.0304 &                -1.6385 \\
 SV                  & 0.3246 & 0.2762 & 0.9716 & -0.1170 &                -0.7082 \\
\bottomrule
\end{tabular}
\begin{tablenotes}
\item[\textit{Note:}] The table presents the performance metrics for five variants of the $\sigma$-Cell model and five other volatility models using synthetic in-sample data for validation. The evaluation metrics include RMSE, MAE, NLL, $\delta$ Mean, and $\delta$ Amplitude. The $\sigma$-Cell-RLTV variant is notable for its performance in the RMSE and MAE metrics, while the $\sigma$-Cell-NTV variant stands out in the NLL, $\delta$ Mean, and $\delta$ Amplitude metrics. Values highlighted in bold indicate the best performance for each metric.
\end{tablenotes}
\end{threeparttable}
\end{table}

\begin{table}
\centering 
\caption{Comparative Performance Metrics of Volatility Models on Out-of-Sample Synthetic Data}
\label{tab:synth_test_metrics} 
\begin{threeparttable}
\begin{tabular}{lrrrrr}
\toprule
 Model              &   RMSE &  MAE &   NLL &     $\delta$ Mean  &  $\delta$ Amplitude \\
\midrule
$\sigma$-Cell       & 0.3269 & 0.2584 & 0.9724 & -0.0426 & -1.3954 \\
$\sigma$-Cell-N     & 0.3271 & 0.2510 & 0.9724 &  0.0075 & -1.0481 \\
$\sigma$-Cell-NTV   & 0.3176 & 0.2479 & 0.9694 &  0.0027 & -1.4280  \\
$\sigma$-Cell-RL    & 0.3382 & 0.2633 & 0.9761 & -0.0231 & -1.5759 \\
$\sigma$-Cell-RLTV  & 0.2873 & 0.2307 & 0.9602 & -0.0110 & -0.9959 \\
 GARCH(1,1)         & 0.3214 & 0.2643 & 1.1571 & -0.0343 & -1.0934 \\
 EGARCH             & 0.3230 & 0.2617 & 1.1609 & -0.0341 & -1.2120  \\
 TARCH              & 0.3602 & 0.2688 & 1.1834 & -0.0460 & -3.5309 \\
 GJR-GARCH          & 0.3244 & 0.2650 & 1.1644 & -0.0354 & -1.3509 \\
 SV                 & 0.4070 & 0.3568 & 1.0565 & -0.0087 &  0.4855 \\
\bottomrule
\end{tabular}
\begin{tablenotes}
\item[\textit{Note:}] The table presents the performance metrics for various volatility models on out-of-sample synthetic data. The metrics include Root Mean Squared Error (RMSE), Mean Absolute Error (MAE), Negative Log-Likelihood (NLL), the difference in mean ($\delta$ Mean), and difference in Amplitude ($\delta$ Amplitude). The models being compared include five variants of the $\sigma$-Cell model and five other well-known volatility models. The $\sigma$-Cell-RLTV variant stands out with a comparatively low RMSE, MAE, and NLL, indicating superior accuracy and consistency among the $\sigma$-Cell variants. Values highlighted in bold indicate the best performance for each metric.
\end{tablenotes}
\end{threeparttable}
\end{table}

\newpage
\subsubsection{Exploring Time-Varying Parameters of $\sigma$-Cell-NTV and $\sigma$-Cell-RLTV}

The $\sigma$-Cell-NTV and $\sigma$-Cell-RLTV models represent a fusion of time series modeling techniques with the power of deep learning. At their core, these models leverage recurrent neural networks to model the time-dependent structure in financial data, coupled with the introduction of time-varying parameters to capture dynamic shifts in the underlying financial processes.

The $\sigma$-Cell-NTV model introduces time-varying weights $W_{s,t}$ and $W_{r,t}$ to model the time-varying nature of financial time series. This is achieved through the transformation of the input vector $\mathbf{x}_t$ via a nonlinear function, followed by the separation of the transformed vector into two components. These components are used to modulate the past variance and the squared residuals in the variance evolution equation. This enables the model to adapt to changing market conditions and capture complex temporal relationships in the data.

On the other hand, the $\sigma$-Cell-RLTV model builds upon the $\sigma$-Cell-RL by incorporating recurrent, time-varying weights into the variance evolution equation. This provides the model with an improved memory retention capability and heightened resistance to specific noise disturbances. By integrating the recurrent nature of the $\sigma$-Cell-RL and the dynamic attributes of the time-varying method, the $\sigma$-Cell-RLTV model offers a more robust and nuanced understanding of financial time series data.

\begin{figure}[htp]
\centering
\begin{subfigure}{.25\textwidth}
  \centering
  \caption{}
  \label{fig:}
  \includegraphics[width=\linewidth]{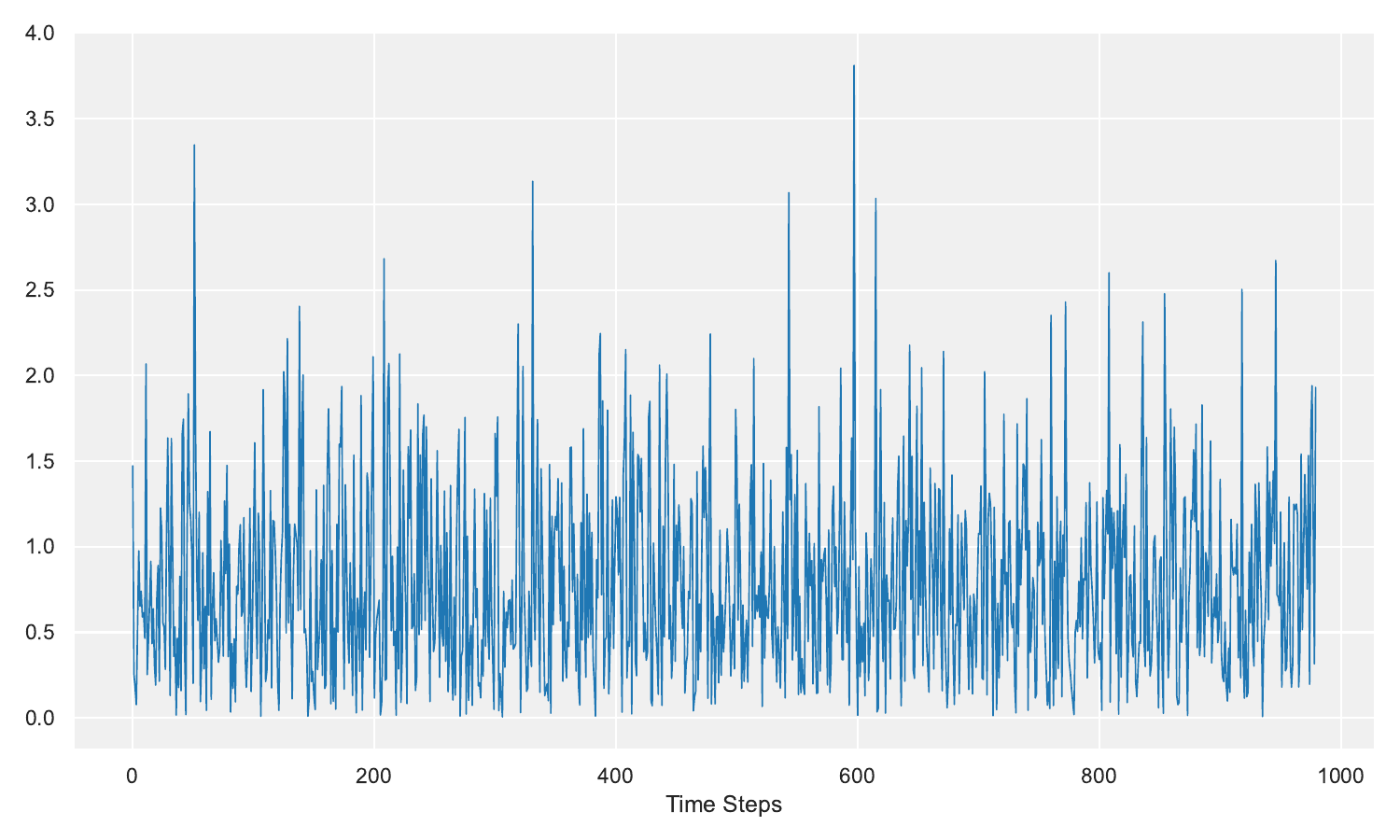}
\end{subfigure}%
\begin{subfigure}{.25\textwidth}
  \centering
  \caption{}
  \label{fig:}
  \includegraphics[width=\linewidth]{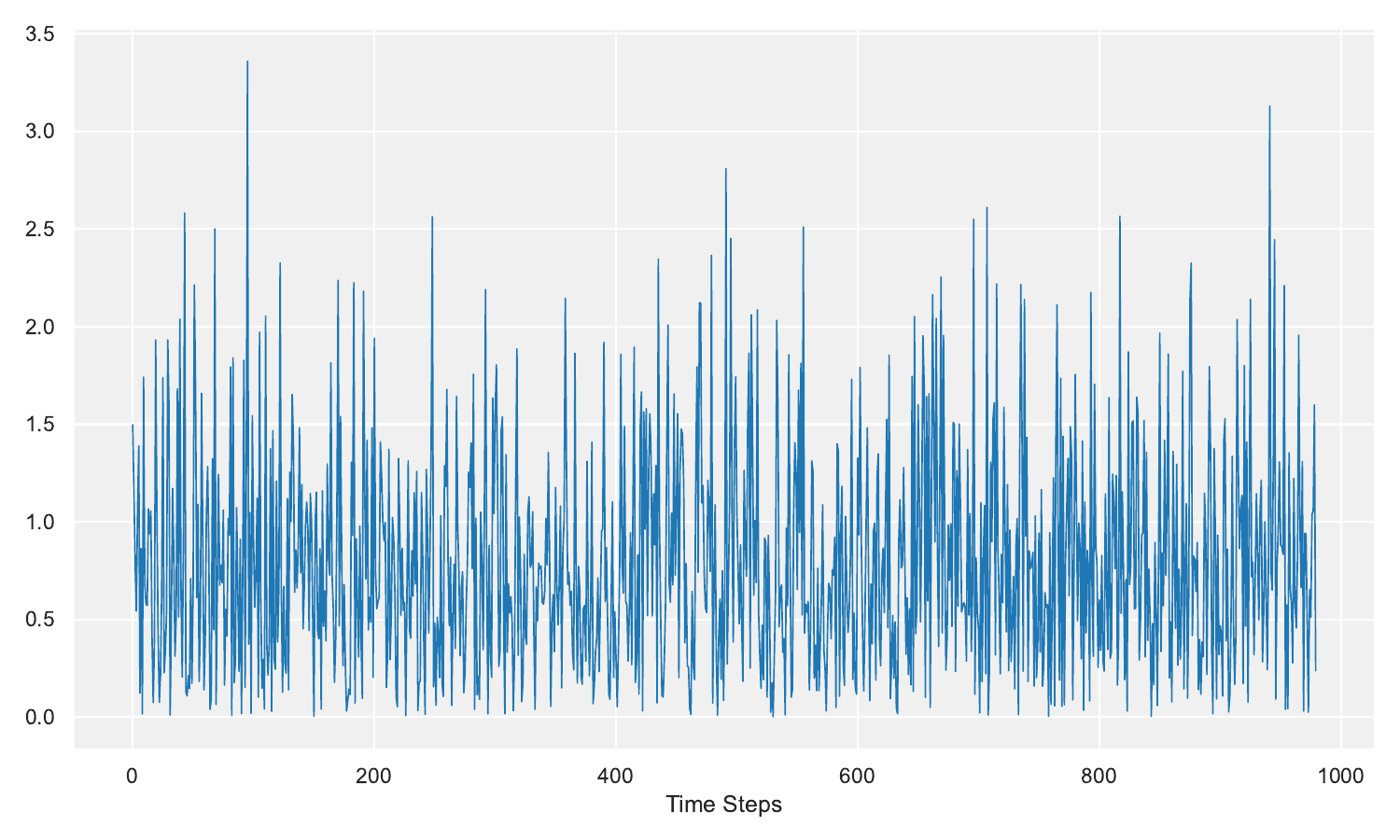}
\end{subfigure}

\begin{subfigure}{.25\textwidth}
  \centering
  \caption{}
  \label{fig:}
  \includegraphics[width=\linewidth]{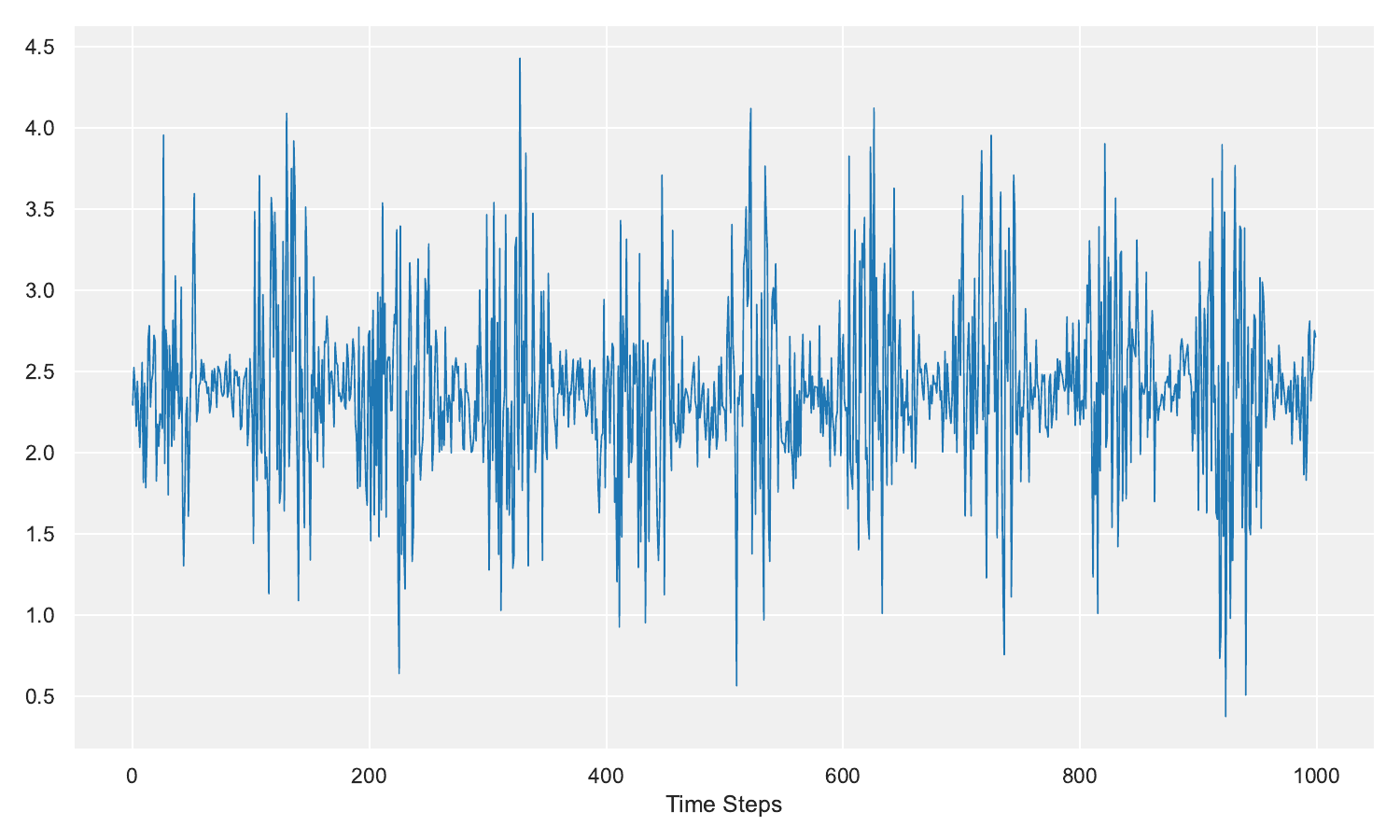}
\end{subfigure}%
\begin{subfigure}{.25\textwidth}
  \centering
  \caption{}
  \label{fig:}
  \includegraphics[width=\linewidth]{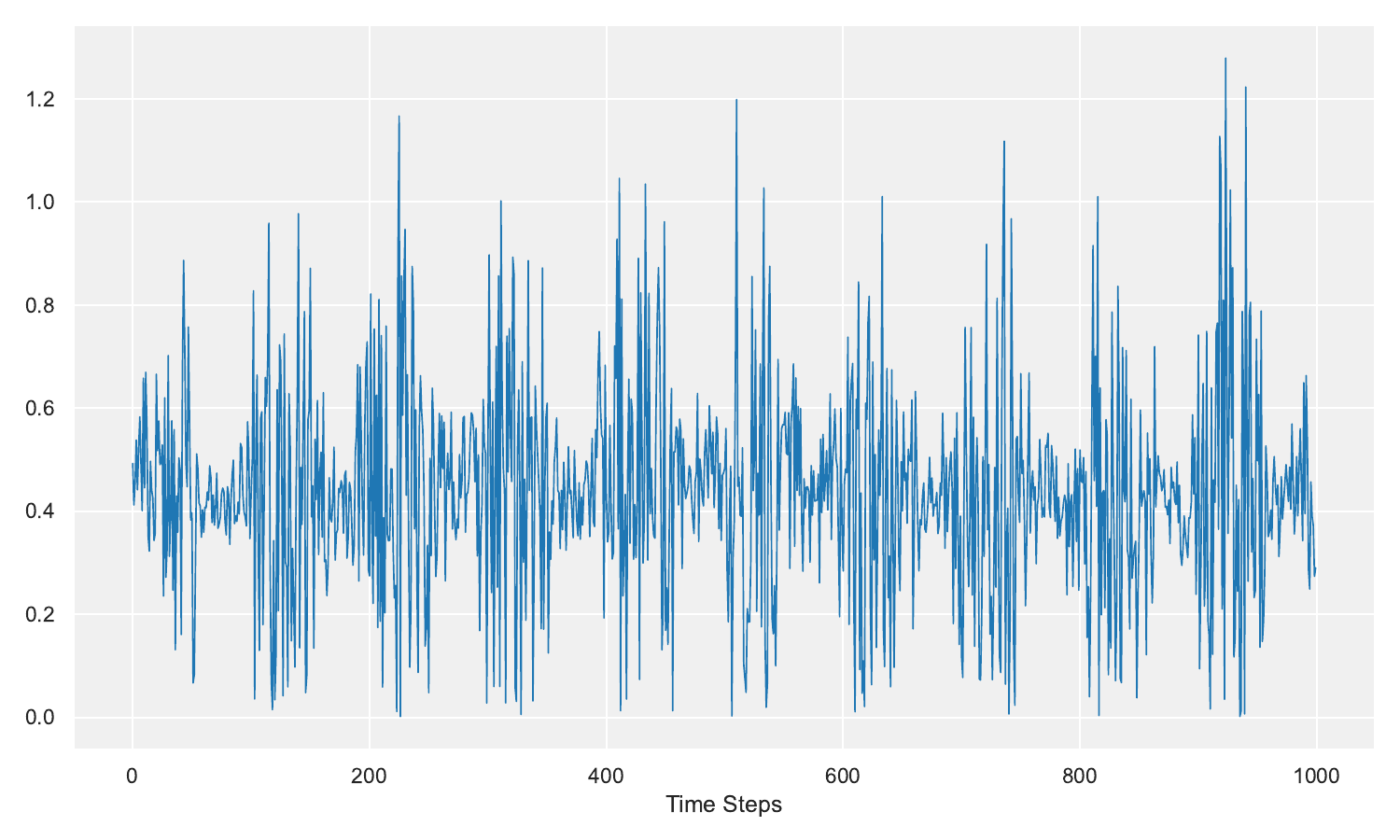}
\end{subfigure}
\caption{The following plot illustrates the evolution of $|W_r|$ and $|W_s|$ in the $\sigma$-Cell-NTV model during training. The plots illustrate the progression of norms at two different training epochs, highlighting the emergence of a structured pattern in $|W_r|$ and $|W_s|$ as training progresses. (a)  $|W_r|$ at epoch 1: Distribution of the  $|W_r|$ during the initial stages of training is mostly noise.
(b)  $|W_s|$ at epoch 1: Distribution of the  $|W_s|$ at the start of training is mostly noise.
(c)  $|W_r|$ at epoch 100: After 100 epochs, a distinct pattern is visible in the distribution of $|W_r|$.
(d)  $|W_s|$ at epoch 100: The distribution of $|W_s|$ after 100 epochs, revealing the emergence of a clear structure.}
\label{fig:Norm_sigma-Cell-NTV}
\end{figure}

\begin{figure}[htp]
\centering
\begin{subfigure}{.25\textwidth}
  \centering
  \caption{}
  \label{fig:dell_dist}
  \includegraphics[width=\linewidth]{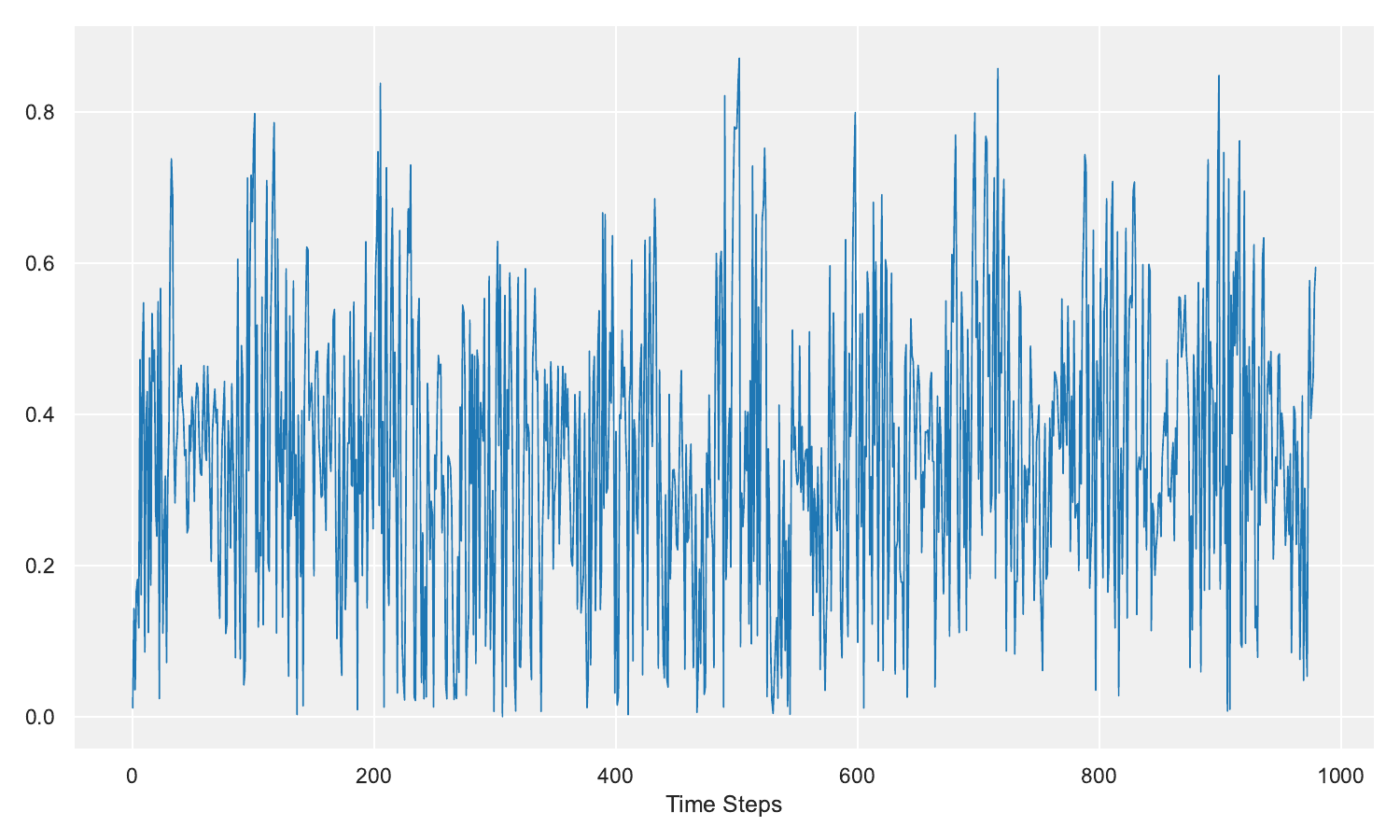}
\end{subfigure}%
\begin{subfigure}{.25\textwidth}
  \centering
  \caption{}
  \label{fig:dell_boxplot}
  \includegraphics[width=\linewidth]{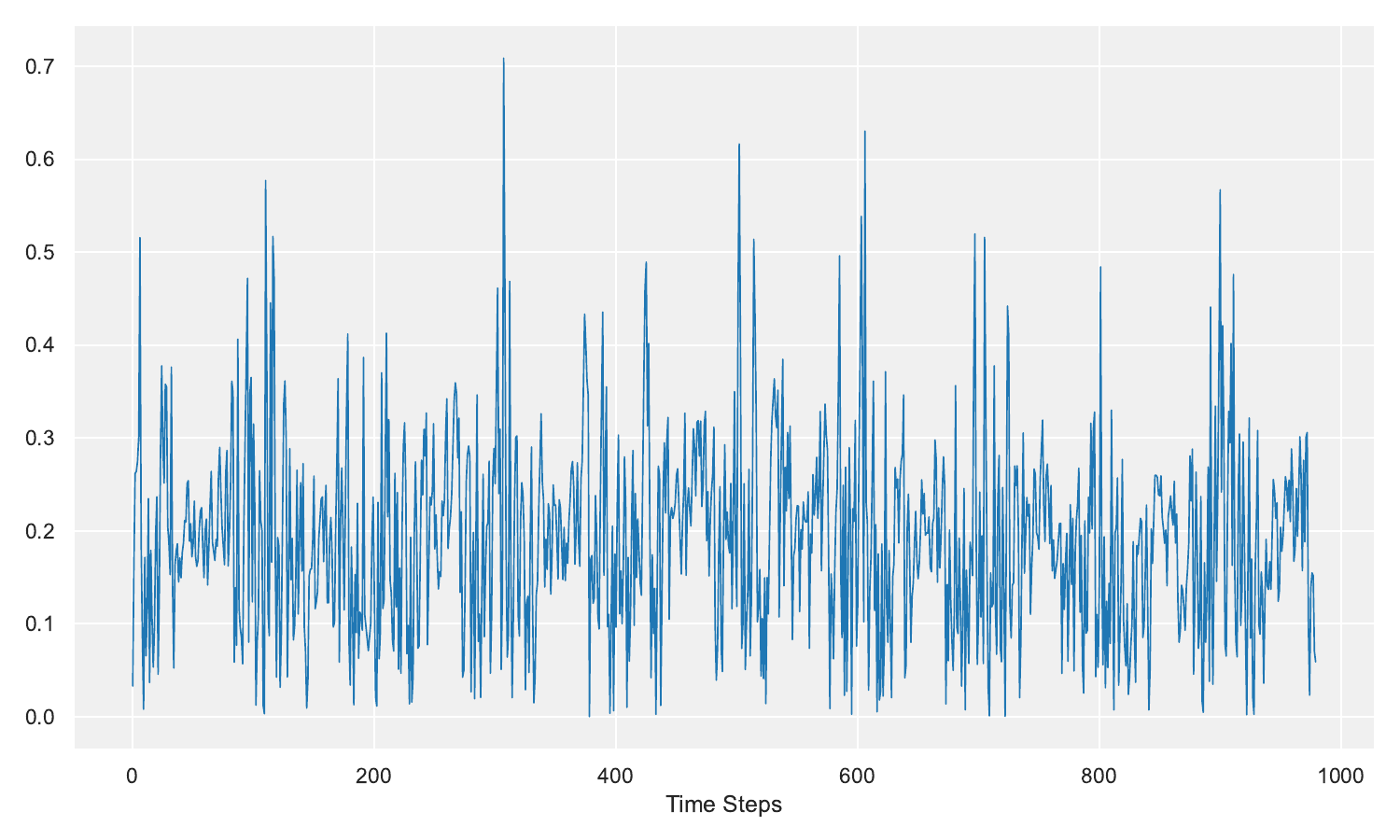}
\end{subfigure}

\begin{subfigure}{.25\textwidth}
  \centering
  \caption{}
  \label{fig:bitcoin_dist}
  \includegraphics[width=\linewidth]{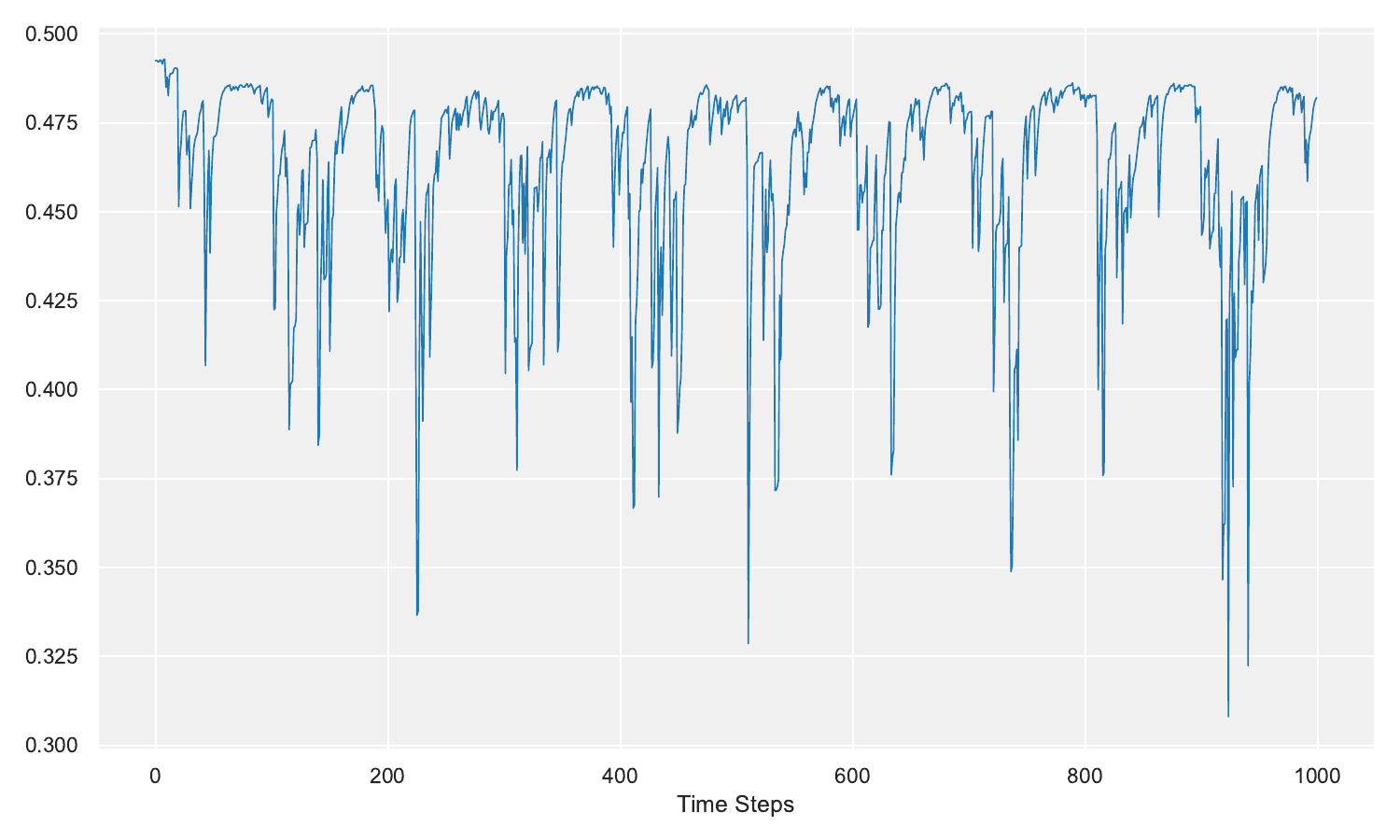}
\end{subfigure}%
\begin{subfigure}{.25\textwidth}
  \centering
  \caption{}
  \label{fig:bitcoin_boxplot}
  \includegraphics[width=\linewidth]{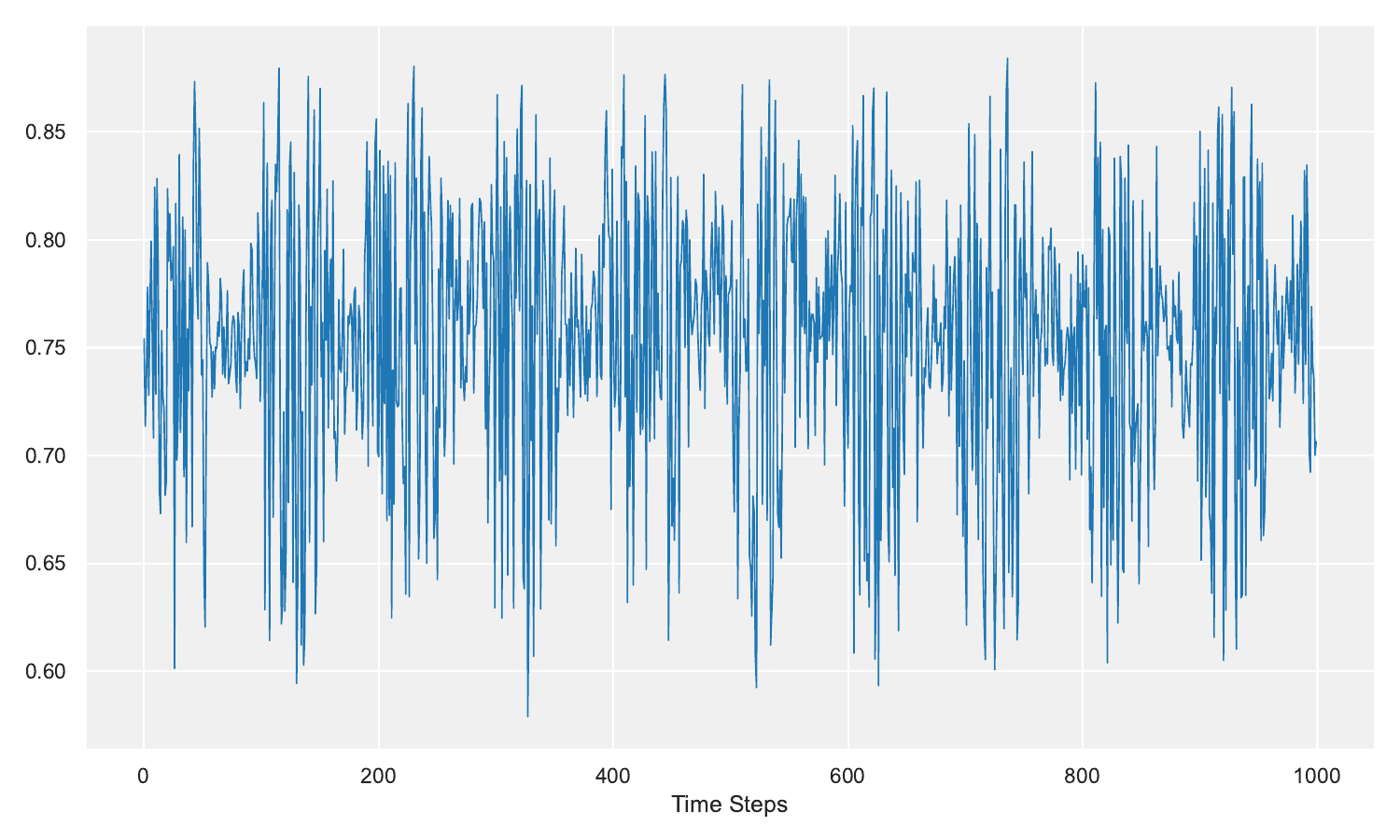}
\end{subfigure}
\caption{The following plot illustrates the evolution of $|W_r|$ and $|W_s|$ in the $\sigma$-Cell-RLTV model during training. The plots illustrate the progression of norms at two different training epochs, highlighting the emergence of a structured pattern in $|W_r|$ and $|W_s|$ as training progresses. 
(a)  $|W_r|$ at epoch 1: the  $|W_r|$ during the initial stages of training. The pattern is largely unstructured at this point.
(b)  $|W_s|$ at epoch 1: the  $|W_s|$ at the start of training, showing a lack of clear structure.
(c)  $|W_r|$ at epoch 100: After 100 epochs, a distinct inverse pattern of variance is visible in the $|W_r|$.
(d)  $|W_s|$ at epoch 100: After 100 epochs $|W_s|$ revealing the emergence of a clear structure.}
\label{fig:Norm_sigma-Cell-RLTV}
\end{figure}

Both models demonstrate a clear progression in the evolution of the norms $|W_r|$ and $|W_s|$ during training \ref{fig:Norm_sigma-Cell-NTV}, \ref{fig:Norm_sigma-Cell-RLTV} Initially, these norms exhibit an unstructured pattern, but as training progresses, a clear structure emerges. The observed structure in the norms reflects the model's ability to capture intricate relationships in the data and provides valuable insights into the underlying financial processes.

While these models offer a promising approach for modeling financial time series data, they come with increased computational complexity and a potential risk of overfitting, especially in sparser data sets. Thus, careful consideration should be given to model selection and hyperparameter tuning to strike a balance between model complexity and predictive performance.

In summary, the $\sigma$-Cell-NTV and $\sigma$-Cell-RLTV models offer a novel approach to modeling financial time series data by combining traditional time series techniques with the flexibility and adaptability of deep learning. These models show promise in capturing complex temporal relationships and provide valuable insights into the dynamics of financial processes.

\newpage
\subsection{Real Data}

Table \ref{tab:SnP_valid_metrics} shows the evaluation of the in-sample performance metrics for a diverse array of volatility forecasting models specifically trained on the S\&P 500 index. Within the ambit of $\sigma$-Cell models, the $\sigma$-Cell-NTV variant merits particular attention for its $R^2$, which signifies a high degree of predictive accuracy during the in-sample period. This model also manifests superior point forecast accuracy, as evidenced by its comparatively low RMSE. The $\sigma$-Cell-RLTV variant is not far behind, also demonstrating robust predictive capabilities as indicated by its $R^2$.

In juxtaposition with other models, the HAR model is a formidable contender, boasting in-sample solid performance. Its relatively low MAE and RMSE metrics corroborate its point forecast accuracy, while its elevated $R^2$ underscores its predictive prowess.

Conversely, the SV model languishes at the lower end of the performance spectrum, marred by elevated MAE and RMSE values, which suggest suboptimal point forecast accuracy. Its $R^2$ further attests to its diminished predictive efficacy relative to the other models under consideration.

Occupying a middle ground, the GARCH variants and EGARCH models exhibit moderate performance metrics. Their $R^2$ values and RMSE metrics place them in an intermediary position, falling short of the high-performing $\sigma$-Cell and HAR models yet surpassing the underperforming SV model.

In summation, the $\sigma$-Cell-NTV and $\sigma$-Cell-RLTV models distinguish themselves with superior in-sample performance metrics, closely following the traditional HAR model.

\begin{table}
\centering 
\caption{In-Sample Performance Metrics for S\&P 500 Volatility Forecasting Models}
\label{tab:SnP_valid_metrics} 
\begin{threeparttable}
\begin{tabular}{lrrrrr}
\toprule
 Model             &    MAE $10^3$ &    RMSE $10^3$ &   HRMSE &    QLIKE &   $R^2$ \\
\midrule
$\sigma$-Cell      & 4.5433 &  7.1825 &  2.5177 & -3.35507 &  0.683718 \\
$\sigma$-Cell-N    & 4.2755 &  7.3268 &  3.9454 & -3.37082 &  0.703984 \\
$\sigma$-Cell-NTV  & 4.4954 &  6.7988 &  2.9629 & -3.37653 &  0.741065 \\
$\sigma$-Cell-RL   & 4.6405 &  7.4594 &  3.7466 & -3.33337 &  0.669921 \\
$\sigma$-Cell-RLTV & 4.2681 &  6.7396 &  4.6134 & -3.36883 &  0.72224  \\
 GARCH(1,1)        & 5.3125 &  8.2736 &  3.349  & -3.36648 &  0.660096 \\
 EGARCH            & 5.2580  &  8.3533 &  3.3518 & -3.36965 &  0.653453 \\
 TARCH             & 5.1771 &  8.4349 &  3.232  & -3.36978 &  0.669526 \\
 GJR-GARCH         & 5.1116 &  8.193  &  3.2048 & -3.36603 &  0.667675 \\
 HAR               & 4.1901 &  7.3104 &  3.7523 & -3.3793  &  0.675116 \\
 SV                & 8.3227 & 15.8116 & 13.3486 & -3.30144 &  0.317564 \\
\bottomrule
\end{tabular}
\begin{tablenotes}
\item[\textit{Note:}] The table presents the in-sample performance of various volatility forecasting models applied to S\&P 500 index. The performance metrics include Mean Absolute Error (MAE), Root Mean Squared Error (RMSE), Heteroscedasticity-Adjusted RMSE (HRMSE), Quasi-Likelihood (QLIKE), and the coefficient of determination ($R^2$). Metrics MAE, RMSE are evaluated at a scale of $10^3$. 
\end{tablenotes}
\end{threeparttable}
\end{table}

Table \ref{tab:SnP_test_metrics} compares the out-of-sample performance of various volatility forecasting models on the S\&P 500 index. It particularly focuses on how the $\sigma$-Cell models measure up against other models.

Among the $\sigma$-Cell models, the $\sigma$-Cell-RLTV model performs the best. It has the second-highest $R^2$ value among all models, indicating strong predictive accuracy. It is lower RMSE, compared to other $\sigma$-Cell models, also suggests better point forecast accuracy.

The HAR model also performs well, especially considering it uses realized volatility (RV) data as input, which gives it more information. It has the highest $R^2$ and the lowest RMSE, confirming its strong predictive performance for the S\&P 500 index's volatility.

In contrast, the SV model performs poorly. Its MAE and RMSE are much higher than those of the other models, indicating that it may struggle to make accurate predictions for the S\&P 500 index's volatility. This could be due to large errors in the model's forecasts.

In summary, the $\sigma$-Cell-RLTV model shows the most promise among the $\sigma$-Cell models for forecasting S\&P 500 volatility.





In summary, among the $\sigma$-Cell models, the $\sigma$-Cell-RLTV model appears to be the most promising for forecasting the volatility of the S\&P 500 index. However, the traditional HAR model also stands out as a strong performer, highlighting the need for further investigation into the comparative advantages of these models in the context of S\&P 500 volatility forecasting.

\begin{table}
\centering 
\caption{Out-of-Sample Performance Metrics for S\&P 500 Volatility Forecasting Models}
\label{tab:SnP_test_metrics} 
\begin{threeparttable}
\begin{tabular}{lrrrrr}
\toprule
 Model             &     MAE $10^3$ &    RMSE $10^3$ &   HRMSE &    QLIKE &   $R^2$ \\
\midrule
$\sigma$-Cell      &  2.9022 &  4.3506 &  2.1655 & -3.79978 &  0.292862 \\
$\sigma$-Cell-N    &  2.7759 &  4.2079 &  2.4303 & -3.85495 &  0.25181  \\
$\sigma$-Cell-NTV  &  2.8566 &  4.2148 &  2.2502 & -3.8158  &  0.327333 \\
$\sigma$-Cell-RL   &  2.8681 &  4.3061 &  2.4395 & -3.83052 &  0.257238 \\
$\sigma$-Cell-RLTV &  2.4940  &  3.6792 &  2.1279 & -3.86195 &  0.464835 \\
 GARCH(1,1)        &  2.6258 &  3.9908 &  2.6406 & -3.868   &  0.319658 \\
 EGARCH            &  2.8062 &  4.123  &  2.671  & -3.86277 &  0.274502 \\
 TARCH             &  2.5811 &  3.961  &  2.5103 & -3.86102 &  0.33776  \\
 GJR-GARCH         &  2.4939 &  3.8503 &  2.6177 & -3.87638 &  0.372308 \\
 HAR               &  2.3316 &  3.3896 &  2.6567 & -3.88498 &  0.516026 \\
 SV                & 68.2419 & 82.3501 & 24.6661 & -2.64681 &  0.262184 \\
\bottomrule
\end{tabular}
\begin{tablenotes}
\item[\textit{Note:}] The table presents the out-of-Sample performance of various volatility forecasting models applied to S\&P 500 index. The performance metrics include Mean Absolute Error (MAE), Root Mean Squared Error (RMSE), Heteroscedasticity-Adjusted RMSE (HRMSE), Quasi-Likelihood (QLIKE), and the coefficient of determination ($R^2$). Metrics MAE, RMSE are evaluated at a scale of $10^3$.  
\end{tablenotes}
\end{threeparttable}
\end{table}

From Table \ref{tab:BTCUSDT_valid_metrics}, we observe the in-sample performance metrics for BTCUSDT volatility forecasting models. The $\sigma$-Cell-RLTV model stands out among the $\sigma$-Cell models with a reasonable $R^2$ value indicating good predictive accuracy. The $\sigma$-Cell-RLTV model's performance appears to be competitive, highlighting the potential of this variant for forecasting the volatility of the BTCUSDT trading pair.

Among the traditional models, the HAR model performs as expected well with the highest $R^2$, making it the best-performing model in this comparison. Moreover, it has the lowest MAE among all models, further indicating the robust forecasting ability of the HAR model in capturing the complex volatility dynamics of the BTCUSDT trading pair.

\begin{table}
\centering 
\caption{In-Sample Performance Metrics for BTCUSDT Volatility Forecasting Models}
\label{tab:BTCUSDT_valid_metrics} 
\begin{threeparttable}
\begin{tabular}{lrrrrr}
\toprule
 Model             &     MAE $10^3$ &    RMSE $10^3$ &   HRMSE &    QLIKE &   $R^2$ \\
\midrule
$\sigma$-Cell      & 11.4741 & 17.2758 &  2.1228 & -2.3708  &  0.406287 \\
$\sigma$-Cell-N    &  9.5845 & 14.8304 &  2.0743 & -2.40586 &  0.383233 \\
$\sigma$-Cell-NTV  &  9.8954 & 14.5676 &  2.2275 & -2.4149  &  0.40949  \\
$\sigma$-Cell-RL   & 11.8278 & 17.3736 &  1.9941 & -2.30458 &  0.323242 \\
$\sigma$-Cell-RLTV &  9.8975 & 13.9467 &  2.2753 & -2.41649 &  0.490842 \\
 GARCH(1,1)        & 10.645  & 15.2788 &  2.2737 & -2.40314 &  0.339935 \\
 EGARCH            & 10.4241 & 15.4131 &  2.249  & -2.40367 &  0.317588 \\
 TARCH             &  9.815  & 17.0394 &  2.2396 & -2.40786 &  0.251025 \\
 GJR-GARCH         & 10.4658 & 16.0143 &  2.2568 & -2.39506 &  0.278871 \\
 HAR               &  8.3432 & 13.046  &  2.1962 & -2.43033 &  0.519423 \\
 SV                & 16.2048 & 20.2572 &  2.632  & -2.35452 &  0.120267 \\
\bottomrule
\end{tabular}
\begin{tablenotes}
\item[\textit{Note:}] The table presents the in-sample performance of various volatility forecasting models applied to the BTCUSDT trading pair. The performance metrics include Mean Absolute Error (MAE), Root Mean Squared Error (RMSE), Heteroscedasticity-Adjusted RMSE (HRMSE), Quasi-Likelihood (QLIKE), and the coefficient of determination ($R^2$). Metrics MAE, RMSE are evaluated at a scale of $10^3$.  
\end{tablenotes}
\end{threeparttable}
\end{table}

Table \ref{tab:BTCUSDT_test_metrics} shows the out-of-sample test results for the BTCUSD trading pair using various volatility forecasting models. Among the $\sigma$-Cell models, the $\sigma$-Cell-NTV variant stands out with a good $R^2$ value, showcasing good predictive accuracy in forecasting the BTCUSD trading pair volatility. Further, the model also has a relatively low MAE and RMSE, emphasizing its robust forecasting performance. This demonstrates the effectiveness of the $\sigma$-Cell-NTV model in capturing the volatility dynamics of the BTCUSD trading pair.

Regarding traditional models, the GJR-GARCH model performs exceptionally well in terms of $R^2$ with the highest value among all models in the table. However, this model has a relatively low MAE and RMSE but is not comparable to $\sigma$-Cell-NTV.

Overall, these results emphasize the potential of both $\sigma$-Cell and traditional models in predicting the volatility of the BTCUSD trading pair, with the $\sigma$-Cell-NTV and GJR-GARCH models showcasing particularly strong performance.

\begin{table}
\centering 
\caption{Out-of-Sample Performance Metrics for BTCUSDT Volatility Forecasting Models}
\label{tab:BTCUSDT_test_metrics} 
\begin{threeparttable}
\begin{tabular}{lrrrrr}
\toprule
 Model             &     MAE $10^3$ &   RMSE $10^3$ &   HRMSE &   QLIKE &   $R^2$ \\
\midrule
$\sigma$-Cell      & 11.5603 & 18.5519 &  2.0116 & -2.3075 &  0.523151 \\
$\sigma$-Cell-N    &  8.6526 & 15.0106 &  2.0952 & -2.3861 &  0.54771  \\
$\sigma$-Cell-NTV  &  8.694  & 15.1973 &  2.1058 & -2.3841 &  0.552158 \\
$\sigma$-Cell-RL   & 10.2195 & 16.7332 &  2.0523 & -2.3489 &  0.481362 \\
$\sigma$-Cell-RLTV &  8.9647 & 15.3183 &  2.1562 & -2.3821 &  0.512133 \\
 GARCH(1,1)        & 11.7684 & 20.9741 &  2.2144 & -2.3579 &  0.164693 \\
 EGARCH            & 11.9173 & 22.4158 &  2.2097 & -2.3569 &  0.231588 \\
 TARCH             & 11.1243 & 23.8724 &  2.212  & -2.3619 &  0.194875 \\
 GJR-GARCH         & 10.316  & 19.6479 &  2.1261 & -2.3712 &  0.561019 \\
 HAR               &  8.7206 & 16.1625 &  2.1609 & -2.3901 &  0.462834 \\
 SV                & 42.9903 & 47.1292 &  3.5722 & -2.1328 &  0.342456 \\
\bottomrule
\end{tabular}
\begin{tablenotes}
\item[\textit{Note:}] The table presents the out-of-sample performance of various volatility forecasting models applied to the BTCUSDT trading pair. The performance metrics include Mean Absolute Error (MAE), Root Mean Squared Error (RMSE), Heteroscedasticity-Adjusted RMSE (HRMSE), Quasi-Likelihood (QLIKE), and the coefficient of determination ($R^2$). Metrics MAE, RMSE are evaluated at a scale of $10^3$. 
\end{tablenotes}
\end{threeparttable}
\end{table}

Table \ref{tab:SnP_DM_vs_sigmRLTV} presents the results of the Diebold-Mariano (DM) test comparing the performance of various volatility forecasting models with the $\sigma$-Cell-RLTV model, which serves as the base model for the S\&P 500 index. The metrics used for performance evaluation are MSE and MAD losses, and the associated p-values are also reported, signifying the statistical significance of the performance differences.

The $\sigma$-Cell, $\sigma$-Cell-N, $\sigma$-Cell-NTV, and $\sigma$-Cell-RL models all show low p-values in both MSE and MAD metrics, indicating their performance is statistically different from the base model, with the $\sigma$-Cell model in particular showing extremely low p-values.  In contrast, the GARCH(1,1), TARCH, and GJR-GARCH models have relatively high p-values in both metrics, suggesting that their performance is not significantly different from the $\sigma$-Cell-RLTV model.  The EGARCH model exhibits a low p-value in the MSE metric but a relatively high p-value in the MAD metric, indicating that its performance is significantly different in terms of MSE but not in terms of MAD. The HAR model has the lowest MSE and MAD among the models, but its p-values suggest that its performance is not significantly different from the base model.

The SV model has extremely low p-values in both metrics and substantially higher loss values, indicating its significantly inferior performance compared to the base model. In summary, the $\sigma$-Cell-RLTV model demonstrates comparable performance to the GARCH(1,1), TARCH, GJR-GARCH, and HAR models in forecasting S\&P 500 volatility, while the SV model, $\sigma$-Cell, $\sigma$-Cell-N, $\sigma$-Cell-NTV, and $\sigma$-Cell-RL models show significantly different performance.

\begin{table}
\centering 
\caption{S\&P 500 Volatility Forecasting: Diebold-Mariano Test with $\sigma$-Cell-RLTV as the Base Model}
\label{tab:SnP_DM_vs_sigmRLTV} 
\begin{threeparttable}
\begin{tabular}{lrrrr}
\toprule
               Model &  MSE Loss &  MSE p-value &  MAD Loss &  MAD p-value \\
\midrule
      $\sigma$-Cell  &  0.018927 & 1.840e-05 &  2.902154 & 3.608e-13 \\
    $\sigma$-Cell-N  &  0.017707 & 4.069e-03 &  2.775868 & 1.204e-02 \\
  $\sigma$-Cell-NTV  &  0.017764 & 1.588e-05 &  2.856639 & 1.700e-10 \\
   $\sigma$-Cell-RL  &  0.018542 & 9.517e-03 &  2.868097 & 1.937e-02 \\
$\sigma$-Cell-RLTV   &  0.013536 &         -   &  2.494049 &          -  \\
       GARCH(1,1)    &  0.015927 & 9.404e-02 &  2.625849 & 2.729e-01 \\
           EGARCH    &  0.016999 & 1.675e-02 &  2.806154 & 1.275e-02 \\
            TARCH    &  0.015689 & 2.259e-01 &  2.581056 & 5.045e-01 \\
           GJR-GARCH &  0.014825 & 3.834e-01 &  2.493900 & 9.989e-01 \\
                 HAR &  0.011489 & 2.247e-01 &  2.331576 & 2.589e-01 \\
                  SV &  6.781546 & 2.434e-34 & 68.241853 & 1.336e-63 \\
\bottomrule
\end{tabular}
\begin{tablenotes}
\item[\textit{Note:}] The table presents the results of the Diebold-Mariano (DM) test comparing the Mean Squared Error (MSE) and Mean Absolute Deviation (MAD) losses of various volatility forecasting models against the $\sigma$-Cell-RLTV model for the S\&P 500 index. The table reports the loss values scaled by $10^3$ and the associated p-values. P-values below 0.05 indicate a statistically significant difference in performance from the $\sigma$-Cell-RLTV model.
\end{tablenotes}
\end{threeparttable}
\end{table}

Table \ref{tab:SnP_DM_vs_HAR} presents the results of the DM test comparing the performance of various volatility forecasting models with the HAR model, which serves as the base model for the S\&P 500 index.

Most of the models in this table have low p-values in both the MSE and MAD metrics, indicating that their performance is statistically different from the HAR model. Specifically, the $\sigma$-Cell, $\sigma$-Cell-N, $\sigma$-Cell-NTV, $\sigma$-Cell-RL, GARCH(1,1), and EGARCH models all have p-values below 0.05 in both metrics. The extremely low p-values associated with the SV model in both metrics, along with substantially higher loss values, highlight its significantly inferior performance compared to the HAR model.

The $\sigma$-Cell-RLTV model, in contrast, shows p-values greater than 0.05 in both metrics, suggesting that its performance is not significantly different from the HAR model. The TARCH and GJR-GARCH models have p-values below 0.05 in the MSE metric but greater than 0.05 in the MAD metric, indicating that their performance differs from the HAR model in terms of MSE but not in terms of MAD.

In summary, the $\sigma$-Cell-RLTV model shows comparable performance to the HAR model in forecasting S\&P 500 volatility. The TARCH and GJR-GARCH models have mixed performance depending on the metric. In contrast, the $\sigma$-Cell, $\sigma$-Cell-N, $\sigma$-Cell-NTV, $\sigma$-Cell-RL, GARCH(1,1), EGARCH, and SV models exhibit statistically different performance from the HAR model in forecasting S\&P 500 volatility.

\begin{table}
\centering 
\caption{S\&P 500 Volatility Forecasting: Diebold-Mariano Test with HAR as the Base Model}
\label{tab:SnP_DM_vs_HAR} 
\begin{threeparttable}
\begin{tabular}{lrrrr}
\toprule
               Model &  MSE Loss &  MSE p-value &  MAD Loss &  MAD p-value \\
\midrule
      $\sigma$-Cell  &  0.018927 & 4.610e-04 &  2.902154 & 5.639e-04 \\
    $\sigma$-Cell-N  &  0.017707 & 1.192e-03 &  2.775868 & 3.550e-03 \\
  $\sigma$-Cell-NTV  &  0.017764 & 1.693e-03 &  2.856639 & 1.175e-03 \\
   $\sigma$-Cell-RL  &  0.018542 & 1.203e-04 &  2.868097 & 1.916e-03 \\
$\sigma$-Cell-RLTV   &  0.013536 & 2.247e-01 &  2.494049 & 2.589e-01 \\
       GARCH(1,1)    &  0.015927 & 3.345e-03 &  2.625849 & 1.337e-02 \\
           EGARCH    &  0.016999 & 2.934e-04 &  2.806154 & 6.765e-05 \\
            TARCH    &  0.015689 & 4.127e-02 &  2.581056 & 6.787e-02 \\
           GJR-GARCH &  0.014825 & 1.968e-02 &  2.493900 & 1.927e-01 \\
                 HAR &  0.011489 &          - &  2.331576 &          - \\
                  SV &  6.781546 & 2.444e-34 & 68.241853 & 3.198e-63 \\
\bottomrule
\end{tabular}
\begin{tablenotes}
\item[\textit{Note:}] The table presents the results of the Diebold-Mariano (DM) test comparing the Mean Squared Error (MSE) and Mean Absolute Deviation (MAD) losses of various volatility forecasting models against the HAR model for the S\&P 500 index. The table reports the loss values scaled by $10^3$ and the associated p-values. P-values below 0.05 indicate a statistically significant difference in performance from the HAR model.
\end{tablenotes}
\end{threeparttable}
\end{table}

Table \ref{tab:BTCUSD_DM_vs_sigmRLTV} provides the results of a DM test comparing the performance of models to the $\sigma$-Cell-RLTV model, which serves as the base model for the BTCUSDT cryptocurrency.

The $\sigma$-Cell, $\sigma$-Cell-RL, GARCH(1,1), EGARCH, GJR-GARCH, and SV models all have p-values below 0.05 in the MAD metric indicating a statistically significant difference in performance from the $\sigma$-Cell-RLTV model. In particular, the SV model exhibits substantially higher loss values and very low p-values in both metrics, highlighting its significantly inferior performance relative to the $\sigma$-Cell-RLTV model.

On the other hand, the $\sigma$-Cell-N and $\sigma$-Cell-NTV models show p-values greater than 0.05 in both metrics, suggesting that their performance is not significantly different from the $\sigma$-Cell-RLTV model. The TARCH model exhibits p-values above 0.05 in both metrics, indicating comparable performance to the $\sigma$-Cell-RLTV model as well.

The HAR model also shows p-values greater than 0.05 in both MSE and MAD metrics, indicating that its performance is not statistically different from the $\sigma$-Cell-RLTV model.

The $\sigma$-Cell and $\sigma$-Cell-RL models exhibit mixed performance with p-values below 0.05 in the MAD metric but above 0.05 in the MSE metric, suggesting significant differences in performance from the $\sigma$-Cell-RLTV model in terms of MAD but not MSE. Similarly, the GARCH(1,1), EGARCH, and GJR-GARCH models have p-values below 0.05 in the MAD metric but above 0.05 in the MSE metric, indicating their performance differs from the $\sigma$-Cell-RLTV model in terms of MAD but not MSE.

In summary, the $\sigma$-Cell-N, $\sigma$-Cell-NTV, TARCH, and HAR models show comparable performance to the $\sigma$-Cell-RLTV model in forecasting BTCUSDT volatility, whereas the $\sigma$-Cell, $\sigma$-Cell-RL, GARCH(1,1), EGARCH, GJR-GARCH, and SV models exhibit statistically different performance in terms of MAD.

\begin{table}
\centering 
\caption{BTCUSDT Volatility Forecasting: Diebold-Mariano Test with $\sigma$-Cell-RLTV as the Base Model}
\label{tab:BTCUSD_DM_vs_sigmRLTV} 
\begin{threeparttable}
\begin{tabular}{lrrrr}
\toprule
             Model &  MSE Loss &  MSE p-value &  MAD Loss $10^3$ &  MAD p-value \\
\midrule
      $\sigma$-Cell   &  0.344173 & 4.188e-02 & 11.560288 & 7.333e-05 \\
    $\sigma$-Cell-N   &  0.225319 & 7.433e-01 &  8.652624 & 4.743e-01 \\
  $\sigma$-Cell-NTV   &  0.230959 & 9.051e-01 &  8.693979 & 4.971e-01 \\
   $\sigma$-Cell-RL   &  0.279999 & 8.147e-02 & 10.219531 & 5.952e-03 \\
$\sigma$-Cell-RLTV    &  0.234649 &          - &  8.964701 &          - \\
       GARCH(1,1)     &  0.439914 & 9.324e-02 & 11.768381 & 1.272e-03 \\
           EGARCH     &  0.502470 & 6.094e-02 & 11.917297 & 4.814e-03 \\
            TARCH     &  0.569893 & 1.244e-01 & 11.124331 & 8.147e-02 \\
           GJR-GARCH &  0.386041 & 6.668e-02 & 10.315967 & 8.031e-02 \\
                 HAR &  0.261226 & 6.955e-01 &  8.720598 & 7.188e-01 \\
                  SV &  2.221157 & 2.886e-27 & 42.990330 & 1.941e-59 \\
\bottomrule
\end{tabular}
\begin{tablenotes}
\item[\textit{Note:}] The table presents the results of the Diebold-Mariano (DM) test comparing the Mean Squared Error (MSE) and Mean Absolute Deviation (MAD) losses of various volatility forecasting models against the $\sigma$-Cell-RLTV model for BTCUSDT cryptocurrency. The table reports the loss values scaled by $10^3$ and the associated p-values. P-values below 0.05 indicate a statistically significant difference in performance from the $\sigma$-Cell-RLTV model.
\end{tablenotes}
\end{threeparttable}
\end{table}

Table \ref{tab:BTCUSD_DM_vs_HAR} presents the results of a DM test comparing the forecasting performance of various models to that of the HAR model, which serves as the base model for the BTCUSDT cryptocurrency.

The $\sigma$-Cell-N, $\sigma$-Cell-NTV, and $\sigma$-Cell-RLTV models exhibit strong performance in both MSE and MAD metrics, with high p-values indicating their performance is not significantly different from the HAR model. The SV model performs the worst with extremely low p-values, indicating its performance is significantly worse than the HAR model. GARCH(1,1) and EGARCH models have low p-values in both MSE and MAD loss, suggesting their performance is significantly different from the HAR model.

The $\sigma$-Cell-RL model shows mixed performance, with a p-value below 0.05 in the MAD metric but above 0.05 in the MSE metric. This suggests that its performance is significantly different from the HAR model in terms of MAD but not MSE.

In summary, the $\sigma$-Cell-N, $\sigma$-Cell-NTV, and $\sigma$-Cell-RLTV models show comparable performance to the HAR model in forecasting BTCUSDT volatility. On the other hand, the GARCH(1,1), EGARCH, and SV models exhibit statistically different performance from the HAR model in both MSE and MAD metrics. The $\sigma$-Cell-RL model shows a nuanced performance, differing from the HAR model in terms of MAD but not MSE.

\begin{table}
\centering 
\caption{BTCUSDT Volatility Forecasting: Diebold-Mariano Test with HAR as the Base Model}
\label{tab:BTCUSD_DM_vs_HAR} 
\begin{threeparttable}
\begin{tabular}{lrrrr}
\toprule
               Model &  MSE Loss $10^3$ &  MSE p-value &  MAD Loss $10^3$ &  MAD p-value \\
\midrule
      $\sigma$-Cell   &  0.344173 & 2.747e-01 & 11.560288 & 1.865e-04 \\
    $\sigma$-Cell-N   &  0.225319 & 5.879e-01 &  8.652624 & 9.194e-01 \\
  $\sigma$-Cell-NTV   &  0.230959 & 6.520e-01 &  8.693979 & 9.676e-01 \\
   $\sigma$-Cell-RL   &  0.279999 & 7.886e-01 & 10.219531 & 4.722e-02 \\
$\sigma$-Cell-RLTV    &  0.234649 & 6.955e-01 &  8.964701 & 7.188e-01 \\
       GARCH(1,1)     &  0.439914 & 4.965e-03 & 11.768381 & 3.452e-07 \\
           EGARCH     &  0.502470 & 6.950e-03 & 11.917297 & 2.073e-05 \\
            TARCH     &  0.569893 & 8.848e-02 & 11.124331 & 1.264e-02 \\
           GJR-GARCH &  0.386041 & 2.145e-01 & 10.315967 & 7.111e-02 \\
                 HAR &  0.261226 &          - &  8.720598 &          - \\
                  SV &  2.221157 & 6.309e-25 & 42.990330 & 1.001e-58 \\
\bottomrule
\end{tabular}
\begin{tablenotes}
\item[\textit{Note:}] The table presents the results of the Diebold-Mariano (DM) test comparing the Mean Squared Error (MSE) and Mean Absolute Deviation (MAD) losses of various volatility forecasting models against the HAR model for BTCUSDT cryptocurrency. The table reports the loss values scaled by $10^3$ and the associated p-values. P-values below 0.05 indicate a statistically significant difference in performance from the HAR model.
\end{tablenotes}
\end{threeparttable}
\end{table}

Table \ref{tab:MCS_10000} presents the results of a Model Confidence Set (MCS) procedure conducted on various volatility forecasting models applied to two different financial markets, the S\&P 500 index and the BTCUSDT cryptocurrency. The MCS method is employed to identify which models perform significantly better or worse than others in terms of MSE performance on the test data, using 10,000 bootstrap resamples. Three metrics are reported for each model, MSE (scaled by $10^3$ ), MCS $p$-values, and a designation for the set of models that perform at or above the $90 \%$ and $75 \%$ confidence levels, denoted as $\hat{\mathcal{M}}_{90,75 \%}^*$.

For the S\&P 500 index, the $\sigma$-Cell-RLTV model performs exceptionally well, achieving the lowest MSE among the $\sigma$-Cell variants at 0.0135 and a high MCS p-value of 0.791. This high $p$-value suggests that its performance is statistically indistinguishable from the best performing model, the HAR model, which has the lowest overall MSE of 0.0114 and a high $\mathrm{p}$-value of 0.839. All $\sigma$-Cell variants, as well as the GARCH(1,1), TARCH, and GJR-GARCH models, also perform well, with relatively low MSE values and high $\mathrm{p}$-values, indicating their inclusion in the $\hat{\mathcal{M}}_{75 \%}^*$ set. The EGARCH model, however, falls into the $\hat{\mathcal{M}}_{90 \%}^*$ set due to its $p$-value of 0.111. The SV model performs the worst, with a high MSE of 6.7815 and a p-value of 0.000, suggesting it is not well-suited for forecasting S\&P 500 volatility.

Turning to the BTCUSDT data, several $\sigma$-Cell variants exhibit the lowest MSE values, with the $\sigma$-Cell-N model performing best at an MSE of 0.2253. The $\sigma$-Cell-RLTV model stands out with a high MCS p-value of 0.945, closely followed by the $\sigma$-Cell-NTV and HAR models. These models, along with other $\sigma$-Cell variants, fall into the $\hat{\mathcal{M}}_{75 \%}^*$ set, indicating their strong performance in forecasting BTCUSDT volatility. In contrast, the SV model performs the worst, with an MSE of 2.2211 and a p-value of 0.000, indicating poor predictive accuracy.

In summary, the $\sigma$-Cell-RLTV and HAR models are top performers for both the S\&P 500 index and BTCUSDT data, showing superior forecasting abilities. Other $\sigma$-Cell variants and GARCH-type models also perform well across both data sets. However, the SV model consistently shows the least suitability among the models tested for both markets.

\begin{table}
\centering
\caption{MCS with 10,000 bootstraps test sample}
\label{tab:MCS_10000}
\begin{threeparttable}
\begin{tabular}{lrrlrrl}
\toprule
      Model             &  \multicolumn{3}{c}{S\&P 500} & \multicolumn{3}{c}{BTCUSDT} \\
                        &  MSE $10^3$ &  P-value & $\hat{\mathcal{M}}_{90,75 \%}^*$ &  MSE $10^3$ &  P-value & $\hat{\mathcal{M}}_{90,75 \%}^*$ \\
\midrule
     $\sigma$-Cell      &  0.0189 &    0.117 &  *  &  0.3441 &    0.135 &  * \\
   $\sigma$-Cell-N      &  0.0177 &    0.533 & **  &  0.2253 &    0.617 & ** \\
 $\sigma$-Cell-NTV      &  0.0177 &    0.325 & **  &  0.2309 &    0.769 & ** \\
  $\sigma$-Cell-RL      &  0.0185 &    0.639 & **  &  0.2799 &    0.525 & ** \\
$\sigma$-Cell-RLTV      &  0.0135 &    0.791 & **  &  0.2346 &    0.945 & ** \\
       GARCH(1,1)       &  0.0159 &    0.444 & **  &  0.4399 &    0.027 &    \\
           EGARCH       &  0.0169 &    0.111 &  *  &  0.5024 &    0.039 &    \\
            TARCH       &  0.0156 &    0.627 & **  &  0.5698 &    0.000 &    \\
        GJR-GARCH       &  0.0148 &    0.660 & **  &  0.3860 &    0.037 &    \\
              HAR       &  0.0114 &    0.839 & **  &  0.2612 &    0.836 & ** \\
               SV       &  6.7815 &    0.000 &     &  2.2211 &    0.000 &    \\
\bottomrule
\end{tabular}
\begin{tablenotes}
\item[\textit{Note:}] The table presents the average loss over the test sample and the MCS \( p \)-values. The realized volatility forecasts in \( \hat{\mathcal{M}}_{90 \%}^* \) and \( \hat{\mathcal{M}}_{75 \%}^* \) are indicated by one and two asterisks, respectively. Values highlighted in bold indicate superior performance for the given Loss metric. In cases where multiple models exhibit closely matched performance, the top few models are highlighted to emphasize their comparative effectiveness.
\end{tablenotes}
\end{threeparttable}
\end{table}

\section{Conclusion}
In conclusion, our exploration of integrating a well-established econometric volatility model with RNNs has provided new models for volatility prediction. Several designs of new $\sigma$-Cell were provided, inspired by leveraging the GARCH process, time-varying recurrent parameter, and an inductive bias, thereby enhancing the model's ability to capture intricate temporal dynamics inherent in financial time series.

We employed a distinctive loss function grounded in a log-likelihood-based methodology, which optimizes the training process; we developed a particular version of the activation function Adjuated-Softplus to improve the training process further. We evaluated and compared the forecast performance of the proposed models with a well-established model in the field. The proposed $\sigma$-Cell-RLTV and $\sigma$-Cell-NTV models outperform traditional methods in out-of-sample predictive tasks, demonstrating the potential for significant advancements in econometric modeling techniques with deep learning. 

The promising results obtained from our study pave the way for further explorations in integrating traditional econometric models and advanced neural network architectures. Such amalgamations can provide more precise and reliable predictions, crucial in various financial applications such as risk management, portfolio optimization, and algorithmic trading. Therefore, all innovations presented in this paper substantially enhance the capabilities of neural network-based volatility modeling.

\newpage

\appendix
\renewcommand{\thesection}{\arabic{section}} 

\section{Appendix 1:  Pairwise Comparisons using a linear regression framework}
Tables \ref{tab:SnP_reg_a1}, \ref{tab:SnP_reg_a1}, \ref{tab:BTCUSDT_reg_a1} and \ref{tab:BTCUSDT_reg_a2} present the results of pairwise comparisons between different forecasting models for the S\&P 500 and BTCUSDT respectively, using a linear regression framework. These comparisons aim to determine which models provide valuable information for forecasting. The framework is based on a linear regression model of the form $y_{t+1}=\alpha_0+\alpha_1 \hat{y}_{i, t+1}+\alpha_2 \hat{y}_{j, t+1}+u_t$, where $\hat{y}_{i, t+1}$ represents the forecast from the model in the $i-th$ row, $\hat{y}_{j, t+1}$ represents the forecast from the model in the $j-th$ column, and $y_{t+1}$ is the actual value at time $t+1$. Tables \ref{tab:SnP_reg_a1}, \ref{tab:SnP_reg_a1}, \ref{tab:BTCUSDT_reg_a1} and \ref{tab:BTCUSDT_reg_a2} display the estimated coefficients, $\alpha_1$ and $\alpha_2$, and their respective p-values for each pairwise comparison.

The significance levels are indicated with asterisks: * denotes significance at the 1\% level, ** at the 5\% level, and *** at the 10\% level. These significance levels serve as evidence for or against the null hypothesis that the respective coefficient equals zero. A significant a1 coefficient suggests that the model in the row provides valuable information for forecasting, while a significant a2 coefficient suggests that the model in the column provides valuable information.

Overall, this analysis offers insights into the relative performance of different forecasting models for the S\&P 500 and BTCUSDT.

The data presented in Tables \ref{tab:SnP_reg_a1} and \ref{tab:SnP_reg_a2} offer several insights into the comparative performance of the various forecasting models for the S\&P 500 index.
\begin{itemize}
\item The $\sigma$-Cell models, including $\sigma$-Cell, $\sigma$-Cell-N, and $\sigma$-Cell-NTV, seem to have high significance with each other. This suggests that these models contain valuable information for forecasting the S\&P 500 index.
\item The $\sigma$-Cell-NTV model has high significance (indicated by ***) with the $\sigma$-Cell-RLTV model, showing strong evidence against the null hypothesis that the coefficient equals zero. This indicates that the $\sigma$-Cell-NTV model and the $\sigma$-Cell-RLTV model might have a relationship.
\item The GARCH(1,1) model has high significance with the $\sigma$-Cell-RL and EGARCH models, which implies that it is highly informative for forecasting the S\&P 500 index.
\item Interestingly, the TARCH model seems to have significant coefficients with the GJR-GARCH model. This suggests that there might be a connection between these models when forecasting the S\&P 500 index.
\item The HAR model also exhibits significant coefficients with the SV model, indicating that these models might share valuable information for forecasting the S\&P 500 index.
\end{itemize}

In summary, among the $\sigma$-Cell type models, $\sigma$-Cell-N and $\sigma$-Cell-RL seem to have strong performance overall, with significant coefficients against most of the other models. However, the performance of other $\sigma$-Cell type models is mixed and varies depending on the specific models being compared. The GARCH(1,1) and EGARCH models also seem to be significant in forecasting the S\&P 500 index. The relationship between the TARCH and GJR-GARCH models, as well as the HAR and SV models, suggests that there may be shared information among these models that can be utilized for better forecasting of the S\&P 500 index.

\begin{sidewaystable}

\begin{subtable}{.5\textwidth}
\caption{S\&p 500 Pairwise Comparison of Forecasting Model Performance, coefficient a1} 
\label{tab:SnP_reg_a1} 
\begin{threeparttable}
\tiny
\begin{tabular}{llllllllllll}
\toprule
{} & $\sigma$-Cell & $\sigma$-Cell-N & $\sigma$-Cell-NTV & $\sigma$-Cell-RL & $\sigma$-Cell-RLTV & GARCH(1,1) &    EGARCH &     TARCH & GJR-GARCH &       HAR &        SV \\
\midrule
$\sigma$-Cell      &     - &       0.518*** &         -0.827** &        0.507*** &             0.492 &   0.349*** &  0.315*** &     0.068 &     0.099 &  0.253*** &  0.440*** \\
$\sigma$-Cell-N    &      0.308** &       - &            0.201 &        0.578*** &           0.338** &   0.402*** &  0.382*** &    0.189* &    0.209* &  0.317*** &  0.512*** \\
$\sigma$-Cell-NTV  &     1.458*** &       0.584*** &         - &        0.522*** &          1.369*** &   0.427*** &  0.380*** &     0.104 &     0.151 &  0.259*** &  0.447*** \\
$\sigma$-Cell-RL   &     0.436*** &       0.485*** &         0.409*** &        - &          0.438*** &   0.464*** &  0.455*** &  0.264*** &  0.271*** &  0.506*** &  0.496*** \\
$\sigma$-Cell-RLTV &        0.226 &       0.493*** &         -0.731** &        0.496*** &          - &   0.326*** &  0.296*** &     0.042 &     0.070 &  0.248*** &  0.441*** \\
GARCH(1,1)        &     0.801*** &       0.924*** &         0.621*** &        0.972*** &          0.827*** &   - &     0.083 &   -0.442* &    -0.379 &  0.347*** &  0.874*** \\
EGARCH            &     0.955*** &       1.053*** &         0.793*** &        1.087*** &          0.979*** &    1.324** &  - &    -0.283 &    -0.270 &  0.392*** &  0.955*** \\
TARCH             &     1.067*** &       1.011*** &         1.018*** &        0.953*** &          1.092*** &   1.425*** &  1.308*** &  - &  0.894*** &  0.587*** &  0.820*** \\
GJR-GARCH         &     1.070*** &       1.026*** &         0.993*** &        0.968*** &          1.101*** &   1.455*** &  1.364*** &     0.280 &  - &  0.599*** &  0.847*** \\
HAR               &     0.914*** &       0.937*** &         0.885*** &        0.959*** &          0.918*** &   0.930*** &  0.916*** &  0.733*** &  0.778*** &  - &  0.792*** \\
SV                &     0.580*** &       0.607*** &         0.558*** &        0.627*** &          0.585*** &   0.584*** &  0.570*** &  0.496*** &  0.526*** &  0.327*** &  - \\
\bottomrule
\end{tabular}
\vspace{1cm}
\end{threeparttable}
\end{subtable}

\begin{subtable}{.5\textwidth}

\caption{S\&p 500 Pairwise Comparison of Forecasting Model Performance, coefficient a2} 
\label{tab:SnP_reg_a2} 
\begin{threeparttable}
\tiny
\begin{tabular}{llllllllllll}
\toprule
{} & $\sigma$-Cell & $\sigma$-Cell-N & $\sigma$-Cell-NTV & $\sigma$-Cell-RL & $\sigma$-Cell-RLTV & GARCH(1,1) &    EGARCH &     TARCH & GJR-GARCH &       HAR &        SV \\
\midrule
$\sigma$-Cell      &        - &        0.308** &         1.458*** &        0.436*** &             0.226 &   0.801*** &  0.955*** &  1.067*** &  1.070*** &  0.914*** &  0.580*** \\
$\sigma$-Cell-N    &     0.518*** &          - &         0.584*** &        0.485*** &          0.493*** &   0.924*** &  1.053*** &  1.011*** &  1.026*** &  0.937*** &  0.607*** \\
$\sigma$-Cell-NTV  &     -0.827** &          0.201 &            - &        0.409*** &          -0.731** &   0.621*** &  0.793*** &  1.018*** &  0.993*** &  0.885*** &  0.558*** \\
$\sigma$-Cell-RL   &     0.507*** &       0.578*** &         0.522*** &           - &          0.496*** &   0.972*** &  1.087*** &  0.953*** &  0.968*** &  0.959*** &  0.627*** \\
$\sigma$-Cell-RLTV &        0.492 &        0.338** &         1.369*** &        0.438*** &             - &   0.827*** &  0.979*** &  1.092*** &  1.101*** &  0.918*** &  0.585*** \\
GARCH(1,1)        &     0.349*** &       0.402*** &         0.427*** &        0.464*** &          0.326*** &      - &   1.324** &  1.425*** &  1.455*** &  0.930*** &  0.584*** \\
EGARCH            &     0.315*** &       0.382*** &         0.380*** &        0.455*** &          0.296*** &      0.083 &     - &  1.308*** &  1.364*** &  0.916*** &  0.570*** \\
TARCH             &        0.068 &         0.189* &            0.104 &        0.264*** &             0.042 &    -0.442* &    -0.283 &     - &     0.280 &  0.733*** &  0.496*** \\
GJR-GARCH         &        0.099 &         0.209* &            0.151 &        0.271*** &             0.070 &     -0.379 &    -0.270 &  0.894*** &     - &  0.778*** &  0.526*** \\
HAR               &     0.253*** &       0.317*** &         0.259*** &        0.506*** &          0.248*** &   0.347*** &  0.392*** &  0.587*** &  0.599*** &     - &  0.327*** \\
SV                &     0.440*** &       0.512*** &         0.447*** &        0.496*** &          0.441*** &   0.874*** &  0.955*** &  0.820*** &  0.847*** &  0.792*** &     - \\
\bottomrule
\end{tabular}
\begin{tablenotes}
\item[\textit{Note:}]  The table (a) and (b) present the results of pairwise comparisons between different forecasting models for S\&p 500 using the framework of a linear regression model. The linear regression model is specified as $y_{t+1}=\alpha_0+\alpha_1 \hat{y}_{i, t+1}+\alpha_2 \hat{y}_{j, t+1}+u_t$, where $\hat{y}_{i, t+1}$ represents the forecast from the model in the i-th row, $\hat{y}_{j, t+1}$ represents the forecast from the model in the j-th column, and $y_{t+1}$ is the true value at time $t+1$. The table displays the estimated coefficients, $\alpha_1$ and $\alpha_2$, and their respective p-values for each pairwise comparison. The forecast evaluation period covers the last 252 trading days.

In Table (a), a1 coefficients and their significance levels are presented in the first part of the table. Each cell in this part of the table shows the a1 coefficient for the corresponding model pairing, along with asterisks indicating the significance level. Similarly, Table (b) shows the a2 coefficients and their significance levels.

Significance levels are indicated with asterisks: '*' denotes significance at the 1\% level, '**' at the 5\% level, and '***' at the 10\% level. These significance levels serve as evidence for or against the null hypothesis that the respective coefficient equals zero. A significant a1 coefficient suggests that the model in the row provides valuable information for forecasting, while a significant a2 coefficient suggests that the model in the column provides valuable information.
\end{tablenotes}
\end{threeparttable}
\end{subtable}
\end{sidewaystable}

The data presented in Tables \ref{tab:BTCUSDT_reg_a1} 
 and \ref{tab:BTCUSDT_reg_a2} offer several insights into the comparative performance of the various forecasting models for BTCUSDT.
\begin{itemize}
    \item The $\sigma$ models, such as $\sigma$-Cell, $\sigma$-Cell-N, and $\sigma$-Cell-NTV, seem to be highly significant with each other, indicating that they contain valuable information for forecasting the BTCUSDT pair.
    \item The $\sigma$-Cell model has high significance (indicated by ***) with GARCH(1,1), EGARCH, and TARCH models, showing strong evidence against the null hypothesis that the coefficient equals zero.
    \item The GARCH(1,1) model has high significance with HAR and SV models, which also suggests that it is highly informative for forecasting BTCUSDT.

\end{itemize}
In summary, among the $\sigma$-Cell type models, $\sigma$-Cell-N and $\sigma$-Cell-RL seem to have strong performance overall, with significant coefficients against most of the other models. However, the performance of other $\sigma$-Cell type models is mixed and varies depending on the specific models being compared.

\begin{sidewaystable}

\begin{subtable}{.5\textwidth}
\caption{BTCUSDT Pairwise Comparison of Forecasting Model Performance, coefficient a1} 
\label{tab:BTCUSDT_reg_a1} 
\begin{threeparttable}
\tiny
\begin{tabular}{llllllllllll}
\toprule
{} & $\sigma$-Cell & $\sigma$-Cell-N & $\sigma$-Cell-NTV & $\sigma$-Cell-RL & $\sigma$-Cell-RLTV & GARCH(1,1) &    EGARCH &     TARCH & GJR-GARCH &       HAR &        SV \\
\midrule
$\sigma$-Cell      &    - &          0.172 &            0.156 &        0.450*** &          0.361*** &   0.613*** &  0.630*** &  0.594*** &  0.267*** &  0.405*** &  0.478*** \\
$\sigma$-Cell-N    &     0.618*** &      - &         0.411*** &        0.679*** &          0.614*** &   0.819*** &  0.888*** &  0.827*** &  0.423*** &  0.602*** &  0.680*** \\
$\sigma$-Cell-NTV  &     0.606*** &       0.437*** &        - &        0.700*** &          0.719*** &   0.763*** &  0.787*** &  0.764*** &  0.417*** &  0.574*** &  0.648*** \\
$\sigma$-Cell-RL   &       0.206* &         0.170* &            0.107 &       - &           0.270** &   0.682*** &  0.677*** &  0.685*** &  0.256*** &  0.451*** &  0.578*** \\
$\sigma$-Cell-RLTV &      0.415** &         0.297* &            0.107 &        0.663*** &         - &   0.908*** &  0.880*** &  0.887*** &  0.466*** &  0.646*** &  0.780*** \\
GARCH(1,1)        &       -0.049 &          0.061 &            0.105 &         0.158** &           0.179** &  - &    0.214* &  0.350*** &     0.083 &     0.008 &  0.497*** \\
EGARCH            &       -0.058 &         -0.058 &            0.019 &           0.090 &           0.134** &   0.387*** & - &  0.341*** &   -0.107* &    -0.091 &  0.371*** \\
TARCH             &        0.006 &          0.023 &            0.059 &           0.083 &          0.142*** &   0.301*** &  0.199*** & - &    -0.032 &    -0.031 &  0.353*** \\
GJR-GARCH         &     0.357*** &       0.322*** &         0.316*** &        0.410*** &          0.363*** &   0.538*** &  0.615*** &  0.574*** & - &  0.510*** &  0.455*** \\
HAR               &     0.466*** &       0.426*** &         0.428*** &        0.588*** &          0.569*** &   1.044*** &  1.154*** &  1.082*** &     0.113 & - &  0.819*** \\
SV                &     0.326*** &       0.295*** &         0.312*** &        0.348*** &          0.332*** &   0.518*** &  0.481*** &  0.512*** &  0.308*** &  0.357*** & - \\
\bottomrule
\end{tabular}
\vspace{1cm}
\end{threeparttable}
\end{subtable}

\begin{subtable}{.5\textwidth}

\caption{BTCUSDT Pairwise Comparison of Forecasting Model Performance, coefficient a2} 
\label{tab:BTCUSDT_reg_a2} 
\begin{threeparttable}
\tiny
\begin{tabular}{llllllllllll}
\toprule
{} & $\sigma$-Cell & $\sigma$-Cell-N & $\sigma$-Cell-NTV & $\sigma$-Cell-RL & $\sigma$-Cell-RLTV & GARCH(1,1) &    EGARCH &     TARCH & GJR-GARCH &       HAR &        SV \\
\midrule
$\sigma$-Cell      &        - &       0.618*** &         0.606*** &          0.206* &           0.415** &     -0.049 &    -0.058 &     0.006 &  0.357*** &  0.466*** &  0.326*** \\
$\sigma$-Cell-N    &        0.172 &          - &         0.437*** &          0.170* &            0.297* &      0.061 &    -0.058 &     0.023 &  0.322*** &  0.426*** &  0.295*** \\
$\sigma$-Cell-NTV  &        0.156 &       0.411*** &            - &           0.107 &             0.107 &      0.105 &     0.019 &     0.059 &  0.316*** &  0.428*** &  0.312*** \\
$\sigma$-Cell-RL   &     0.450*** &       0.679*** &         0.700*** &           - &          0.663*** &    0.158** &     0.090 &     0.083 &  0.410*** &  0.588*** &  0.348*** \\
$\sigma$-Cell-RLTV &     0.361*** &       0.614*** &         0.719*** &         0.270** &             - &    0.179** &   0.134** &  0.142*** &  0.363*** &  0.569*** &  0.332*** \\
GARCH(1,1)        &     0.613*** &       0.819*** &         0.763*** &        0.682*** &          0.908*** &      - &  0.387*** &  0.301*** &  0.538*** &  1.044*** &  0.518*** \\
EGARCH            &     0.630*** &       0.888*** &         0.787*** &        0.677*** &          0.880*** &     0.214* &     - &  0.199*** &  0.615*** &  1.154*** &  0.481*** \\
TARCH             &     0.594*** &       0.827*** &         0.764*** &        0.685*** &          0.887*** &   0.350*** &  0.341*** &     - &  0.574*** &  1.082*** &  0.512*** \\
GJR-GARCH         &     0.267*** &       0.423*** &         0.417*** &        0.256*** &          0.466*** &      0.083 &   -0.107* &    -0.032 &     - &     0.113 &  0.308*** \\
HAR               &     0.405*** &       0.602*** &         0.574*** &        0.451*** &          0.646*** &      0.008 &    -0.091 &    -0.031 &  0.510*** &     - &  0.357*** \\
SV                &     0.478*** &       0.680*** &         0.648*** &        0.578*** &          0.780*** &   0.497*** &  0.371*** &  0.353*** &  0.455*** &  0.819*** &     - \\
\bottomrule
\end{tabular}
\begin{tablenotes}
\item[\textit{Note:}] Tables (a) and (b) present the results of pairwise comparisons between different forecasting models for BTCUSDT using the framework of a linear regression model. The linear regression model is specified as $y_{t+1}=\alpha_0+\alpha_1 \hat{y}_{i, t+1}+\alpha_2 \hat{y}_{j, t+1}+u_t$, where $\hat{y}_{i, t+1}$ represents the forecast from the model in the i-th row, $\hat{y}_{j, t+1}$ represents the forecast from the model in the j-th column, and $y_{t+1}$ is the true value at time $t+1$. The table displays the estimated coefficients, $\alpha_1$ and $\alpha_2$, and their respective p-values for each pairwise comparison. The forecast evaluation period covers the last 252 trading days.

In Table (a), a1 coefficients and their significance levels are presented in the first part of the table. Each cell in this part of the table shows the a1 coefficient for the corresponding model pairing, along with asterisks indicating the significance level. Similarly, Table (b) shows the a2 coefficients and their significance levels.

Significance levels are indicated with asterisks: '*' denotes significance at the 1\% level, '**' at the 5\% level, and '***' at the 10\% level. These significance levels serve as evidence for or against the null hypothesis that the respective coefficient equals zero. A significant a1 coefficient suggests that the model in the row provides valuable information for forecasting, while a significant a2 coefficient suggests that the model in the column provides valuable information.
\end{tablenotes}
\end{threeparttable}
\end{subtable}
\end{sidewaystable}

\section{Appendix 2: Algorithm for Volatility Prediction with $\sigma$-Cell-RLTV}

\begin{algorithm}
\caption{$\sigma$-Cell-RLTV Algorithm for Volatility Prediction}
\begin{algorithmic}[1]
\Require Sequence of inputs: $\mathbf{x} \in \mathbb{R}^n$
\Ensure Predicted volatility for the next input: $\sigma_{t}$

\State Initialize $\sigma_0$

\For{each time step $t=1, 2, \ldots, n$}
    \State Compute parameter vector $w_t$ using Eq. \ref{eq:sigma_cell-RLTV1}: \\
    \hspace{2em} $w_t \leftarrow \tilde{\varphi}(\mathbf{W} \mathbf{x}_{t-1}+\mathbf{b})$
    \Statex 
    \State Compute component $W_{s, t}$ using Eq. \ref{eq:sigma_cell-RLTV2}: \\
    \hspace{2em} $W_{s, t} \leftarrow \pi_1(w_t)$
    \Statex 
    \State Compute component $W_{r, t}$ using Eq. \ref{eq:sigma_cell-RLTV3}: \\
    \hspace{2em} $W_{r, t} \leftarrow \pi_2(w_t)$
    \Statex 
    \State Compute residual $\tilde{x}_{t-1}$ using Eq. \ref{eq:sigma_cell-RLTV4}: \\
    \hspace{2em} $\tilde{x}_{t-1} \leftarrow x_t - f(h_{t-1}, x_{t-1})$
    \Statex 
    \State Compute estimated volatility $\tilde{\sigma}_{t}^2$ using Eq. \ref{eq:sigma_cell-RLTV5}: \\
    \hspace{2em} $\tilde{\sigma}_{t}^2 \leftarrow \phi(\tilde{\sigma}_{t-1}^2 W_{s, t} + \tilde{x}_{t-1}^2 W_{r, t} + b_h)$
    \Statex 
    \State Predict $\sigma_{t}^2$ using Eq. \ref{eq:sigma_cell-RLTV6}: \\
    \hspace{2em} $\sigma_{t}^2 \leftarrow \phi_o(\tilde{\sigma}_{t}^2 W_o + b_o)$
    \Statex 
\EndFor
\Statex 
\State \Return Predicted volatility $\sigma_{t} \leftarrow \sqrt{\sigma_{t}^2}$
\end{algorithmic}
\end{algorithm}

Algorithm 1 presents the $\sigma$-Cell-RLTV approach, which is designed to forecast the volatility of financial returns. The sequence of returns denoted as $x$ and with dimension $n$, serves as the primary input for this algorithm.

The parameter vector $w_t$ is obtained by passing the input vector $x_{t-1}$ through the function $\tilde{\varphi}$, which involves a linear transformation using the weight matrix $\mathbf{W}$ and the bias vector $\mathbf{b}$. Subsequently, the components $W{s, t}$ and $W_{r, t}$ are derived from $w_t$ using the functions $\pi_1$ and $\pi_2$, respectively.

The sequence is passed through the layer described in Equation \ref{eq:sigma_cell-RLTV1}, which generates $\tilde{x}_{t-1}$. Then, for each time step $t$, ranging from 1 to $n$, $\tilde{x}_t$ is calculated as the difference between the input and the layer $f\left(h_{t-1}, x_{t-1}\right)$, modulated by a function of the RNN's previous hidden state $h_{t-1}$ and the actual input at that time.

Next, $\tilde{\sigma}_t^2$ is computed as a function of $W{s, t}$, $\tilde{\sigma}_{t-1}^2$, $W{r, t}$, and $\tilde{x}_t^2$. Finally, $\sigma_t^2$ is calculated using Equation \ref{eq:sigma_cell-RLTV6}, the function $\phi_o$ ensures that the estimated volatility remains positive. This process generates the estimated volatility for each return in the sequence.

\section{Appendix 3}

\begin{figure}[htp]
\centering
\includegraphics[width=\linewidth]{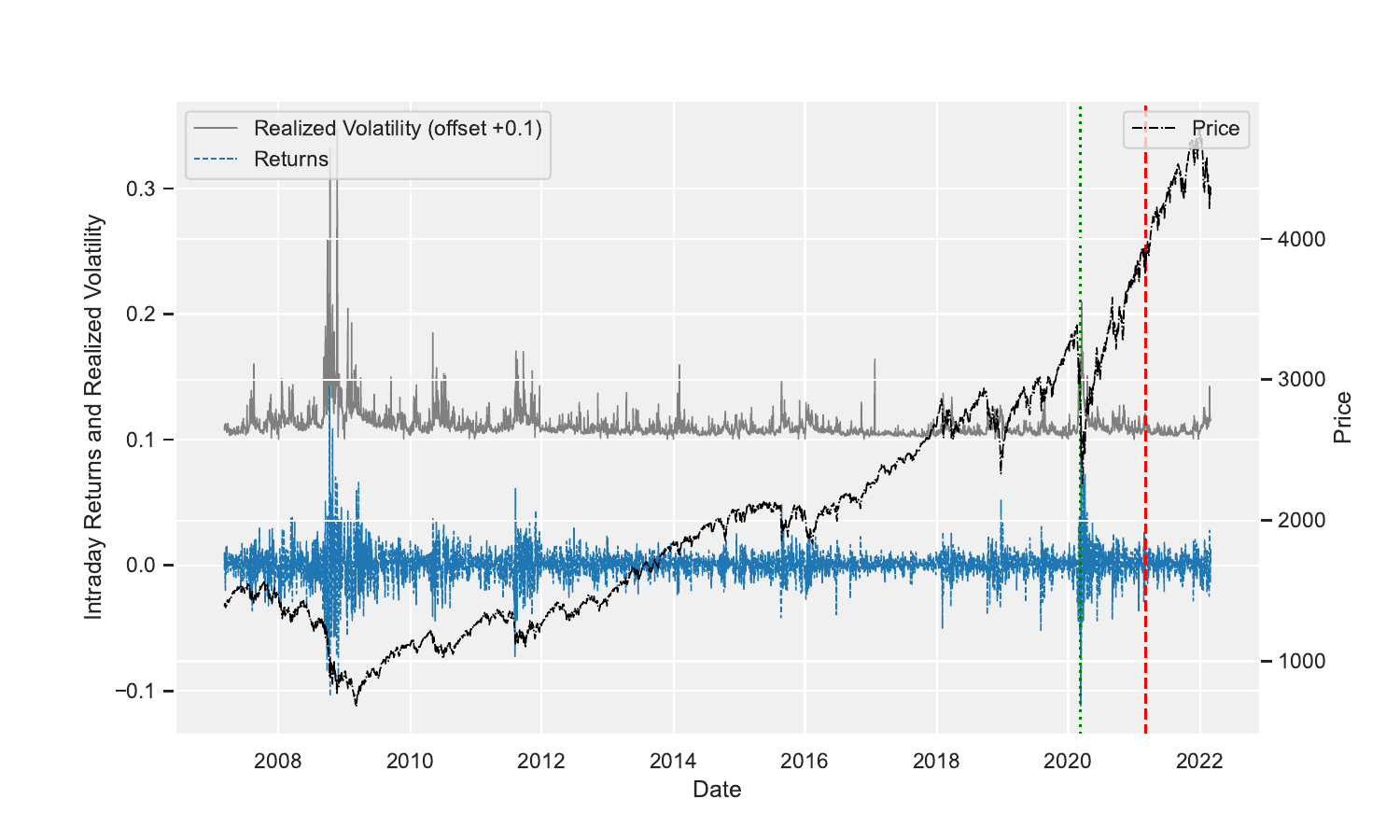}
\caption{The following plot illustrates Realized Volatility, Returns, and Price of S\&P 500 Index over Time. This plot displays the realized volatility (RV, offset +0.1), intraday returns, and price of the S\&P 500 index from March 10, 2007, to March 1, 2022. The gray solid line represents the realized volatility (offset by +0.1), the blue dashed line shows intraday returns, and the black dash-dot line displays the price. The RV and returns are calculated based on daily data and are presented on the primary y-axis, while the price is plotted on a secondary y-axis. The vertical red dashed, and green dotted lines mark the start of the test and validation sets, respectively, each containing 252 points. The remaining data is used as the training set.}
\label{fig:snp_data_split}
\end{figure}

\begin{figure}[htp]
\centering
\begin{subfigure}{.45\textwidth}
  \centering
  \caption{}
  \label{fig:SnP_ACF_returns}
  \includegraphics[width=\linewidth]{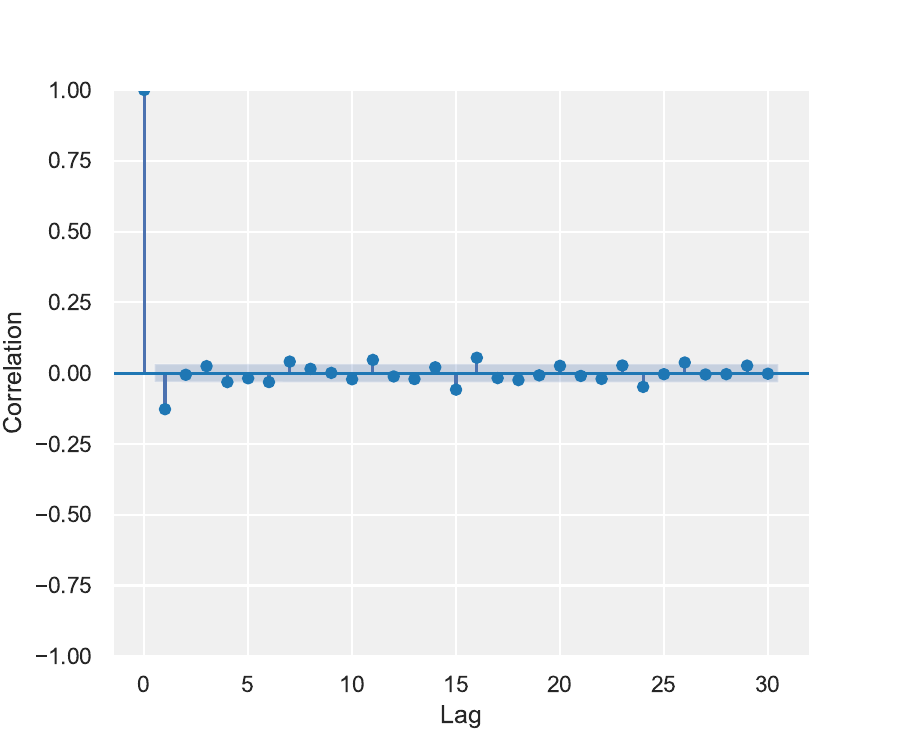}
\end{subfigure}%
\begin{subfigure}{.45\textwidth}
  \centering
  \caption{}
  \label{fig:SnP_PACF_returns}
  \includegraphics[width=\linewidth]{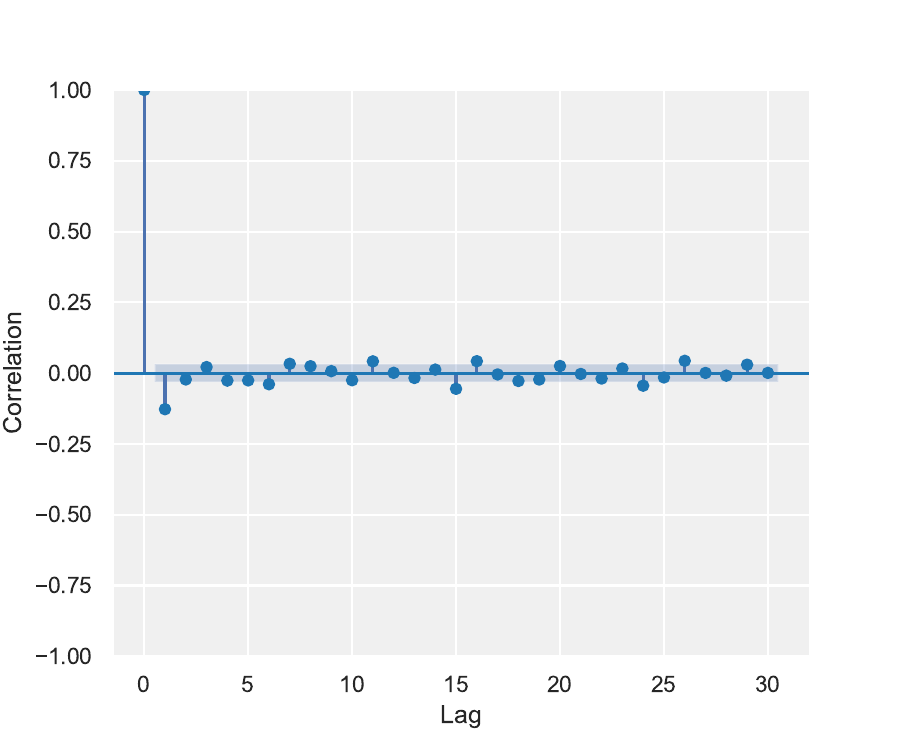}
\end{subfigure}

\begin{subfigure}{.45\textwidth}
  \centering
  \caption{}
  \label{fig:SnP_ACF_volatility}
  \includegraphics[width=\linewidth]{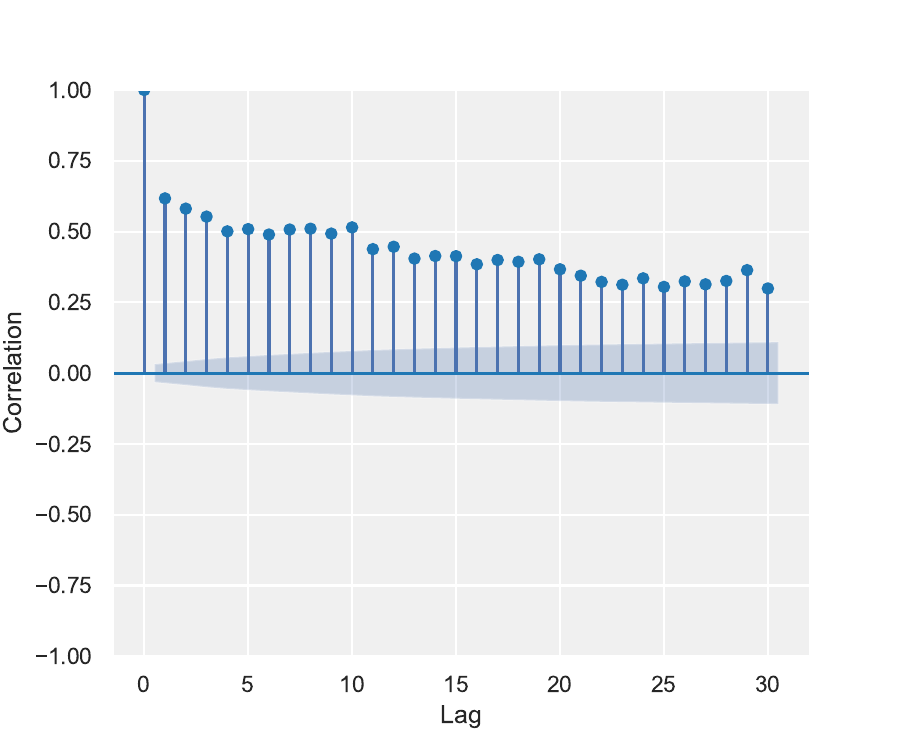}
\end{subfigure}%
\begin{subfigure}{.45\textwidth}
  \centering
  \caption{}
  \label{fig:SnP_PACF_volatility}
  \includegraphics[width=\linewidth]{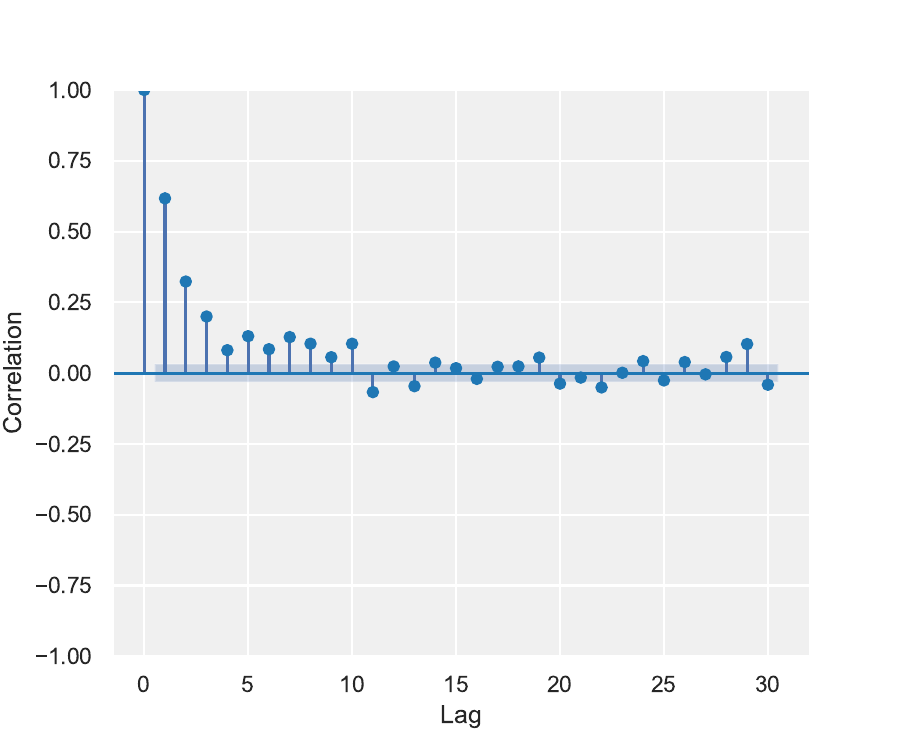}
\end{subfigure}
\caption{The following plot illustrates autocorrelation and partial autocorrelation plots for the returns and volatility of the S\&P 500 Index. The ACF plots show the extent of a linear relationship between lagged values and current values, whereas the PACF plots display the correlation between a variable and its lagged value that is not explained by all shorter lags.
a) autocorrelation plots for the returns
b) partial autocorrelation plots for the returns
c) autocorrelation plots for the Realized Volatility
d) partial autocorrelation plots for the Realized Volatility}
\label{fig:snp_acf_pcf}
\end{figure}

\begin{figure}[htp]
\centering
\includegraphics[width=\linewidth]{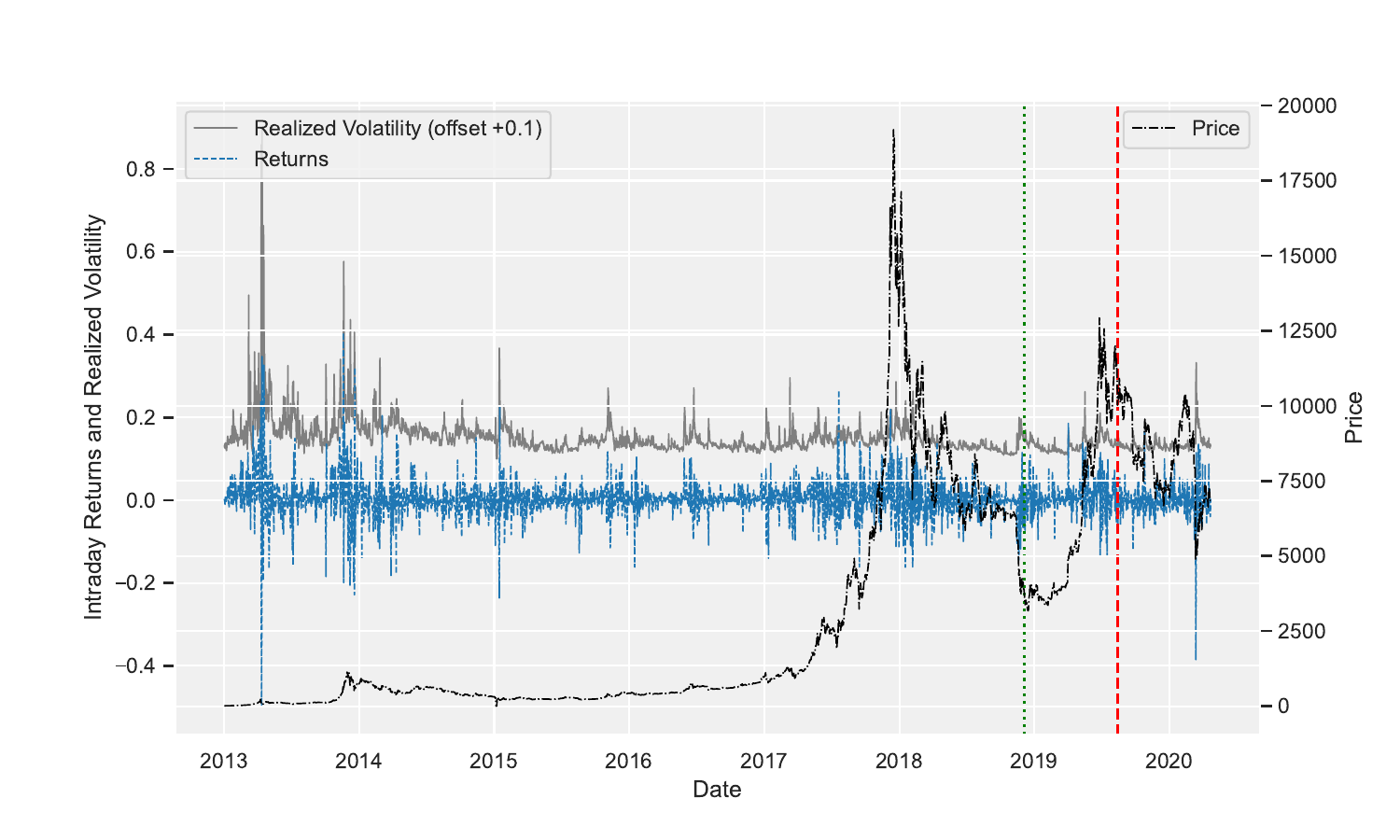}
\caption{The following plot illustrates Realized Volatility, Returns, and Price of Bitcoin-USD (BTCUSDT) Pair over Time. The plot shows the realized volatility (RV, offset +0.1), intraday returns, and price of the BTCUSDT pair from January 1, 2013, to April 20, 2020.  The gray solid line represents the realized volatility (offset by +0.1), the blue dashed line shows intraday returns, and the black dash-dot line displays the price. The RV and returns are calculated based on daily data and are shown on the primary y-axis, while the price is plotted on a secondary y-axis. The vertical red dashed, and green dotted lines mark the start of the test and validation sets, respectively, each containing 252 points. The remaining data is used as the training set.}
\label{fig:snp_data_split}
\end{figure}

\begin{figure}[htp]
\centering
\begin{subfigure}{.45\textwidth}
  \centering
  \caption{}
  \label{fig:BTCUSDT_ACF_returns}
  \includegraphics[width=\linewidth]{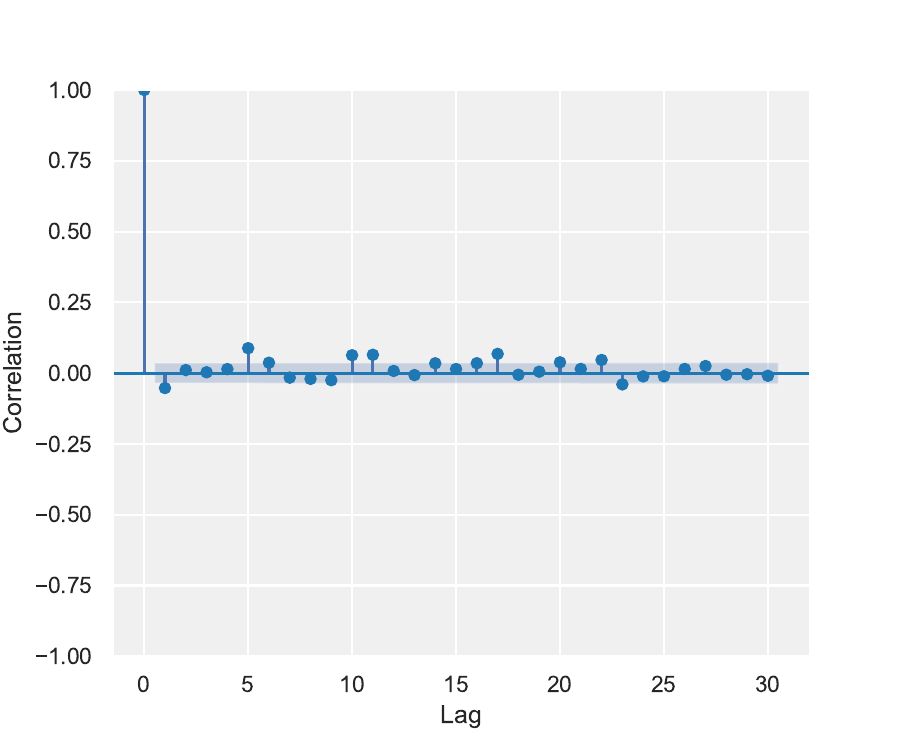}
\end{subfigure}%
\begin{subfigure}{.45\textwidth}
  \centering
  \caption{}
  \label{fig:BTCUSDT_PACF_returns}
  \includegraphics[width=\linewidth]{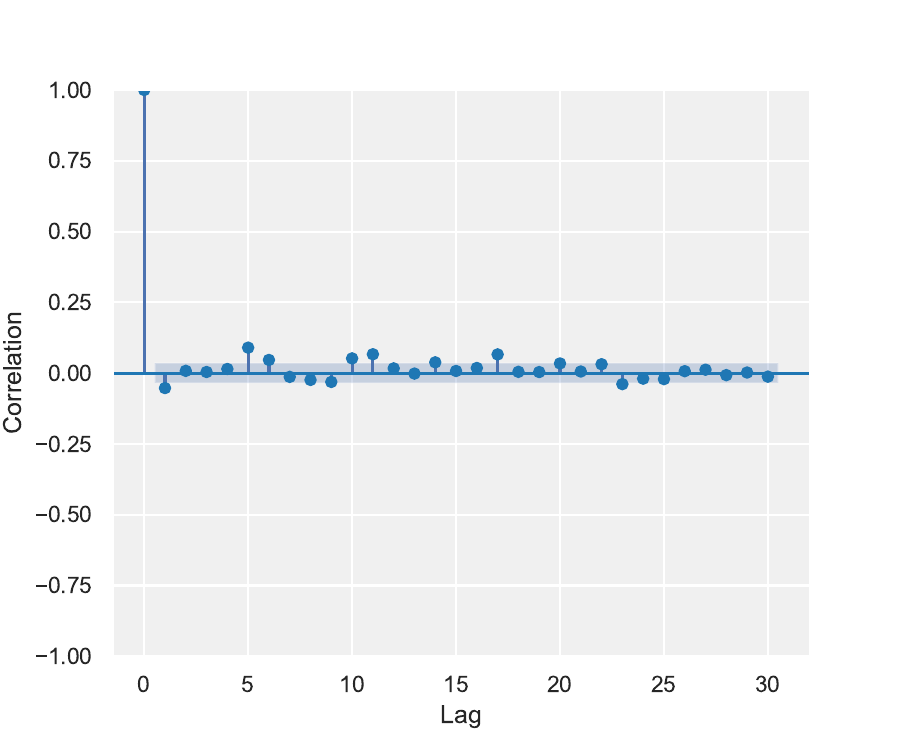}
\end{subfigure}

\begin{subfigure}{.45\textwidth}
  \centering
  \caption{}
  \label{fig:BTCUSDT_ACF_volatility}
  \includegraphics[width=\linewidth]{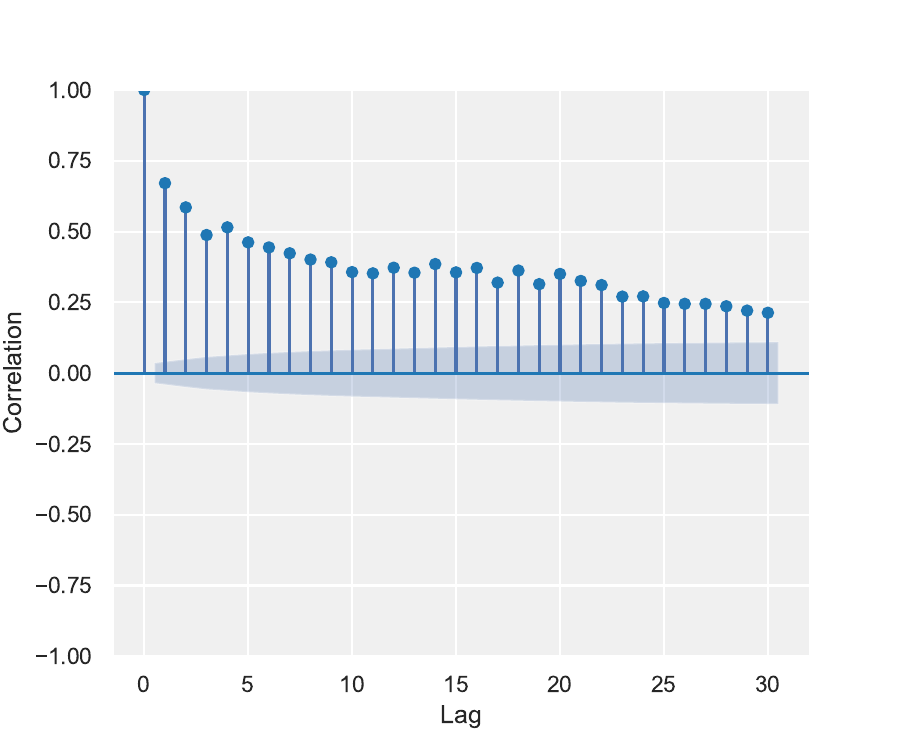}
\end{subfigure}%
\begin{subfigure}{.45\textwidth}
  \centering
  \caption{}
  \label{fig:BTCUSDT_PACF_volatility}
  \includegraphics[width=\linewidth]{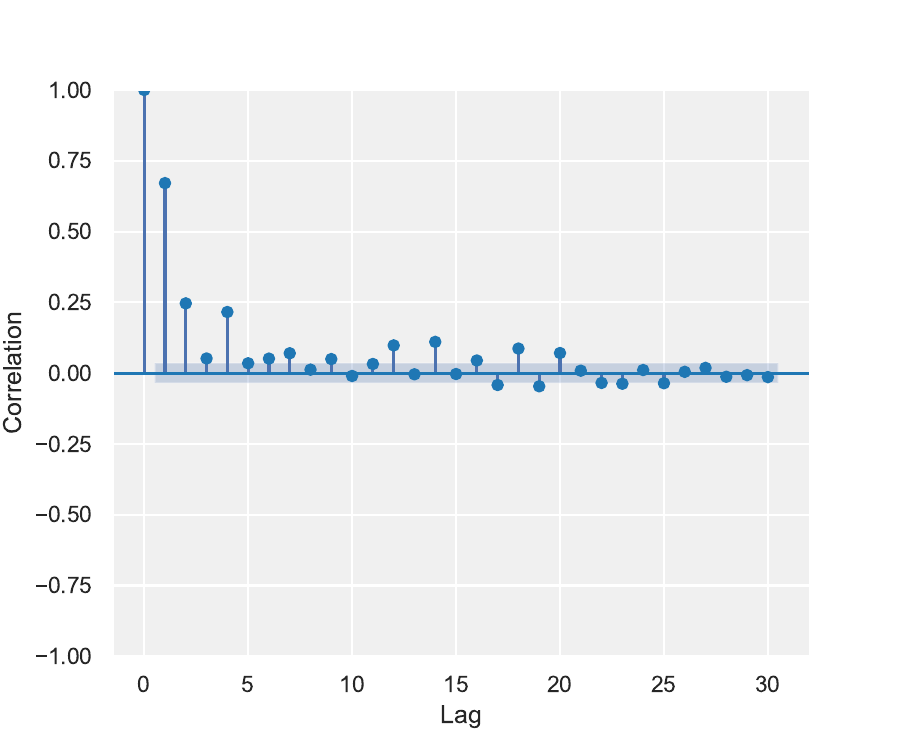}
\end{subfigure}
\caption{The following plot illustrates autocorrelation and partial autocorrelation plots for the returns and volatility of the BTCUSDT pair. The ACF plots show the extent of a linear relationship between lagged values and current values, whereas the PACF plots display the correlation between a variable and its lagged value that is not explained by all shorter lags.
a) autocorrelation plots for the returns
b) partial autocorrelation plots for the returns
c) autocorrelation plots for the Realized Volatility
d) partial autocorrelation plots for the Realized Volatility}
\label{fig:snp_acf_pcf}
\end{figure}

\begin{figure}[htp]
\centering
\begin{subfigure}{.45\textwidth}
  \centering
  \caption{}
  \label{fig:dell_dist}
  \includegraphics[width=\linewidth]{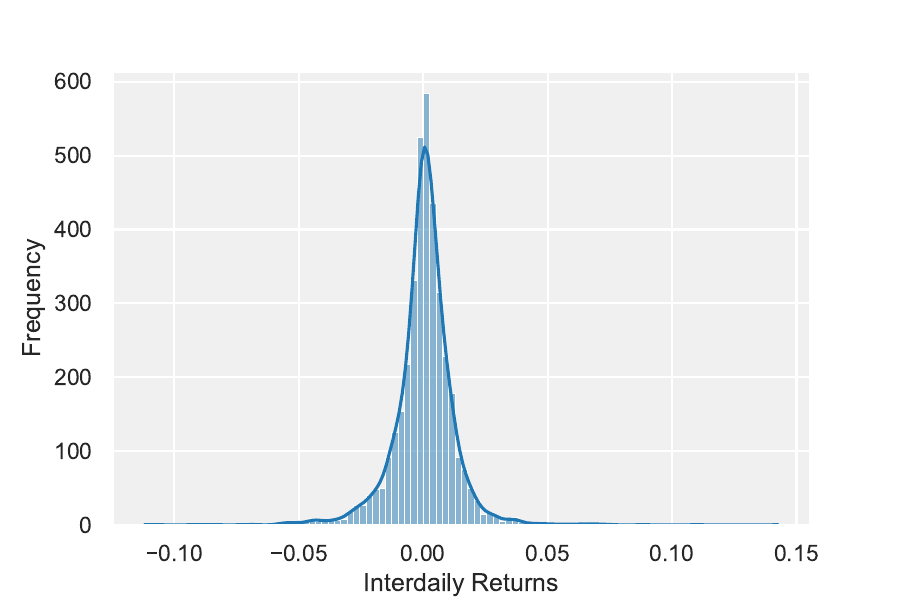}
\end{subfigure}%
\begin{subfigure}{.45\textwidth}
  \centering
  \caption{}
  \label{fig:dell_boxplot}
  \includegraphics[width=\linewidth]{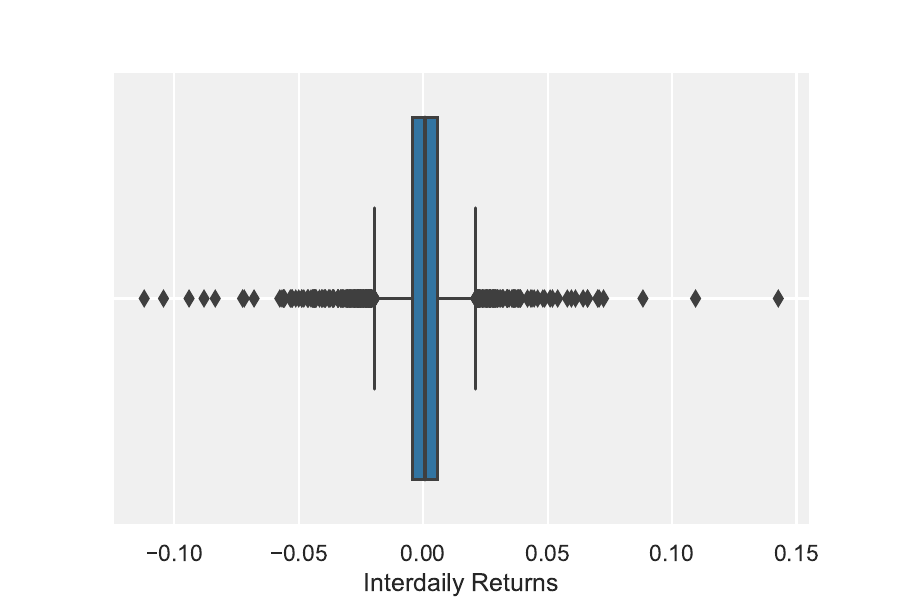}
\end{subfigure}

\begin{subfigure}{.45\textwidth}
  \centering
  \caption{}
  \label{fig:bitcoin_dist}
  \includegraphics[width=\linewidth]{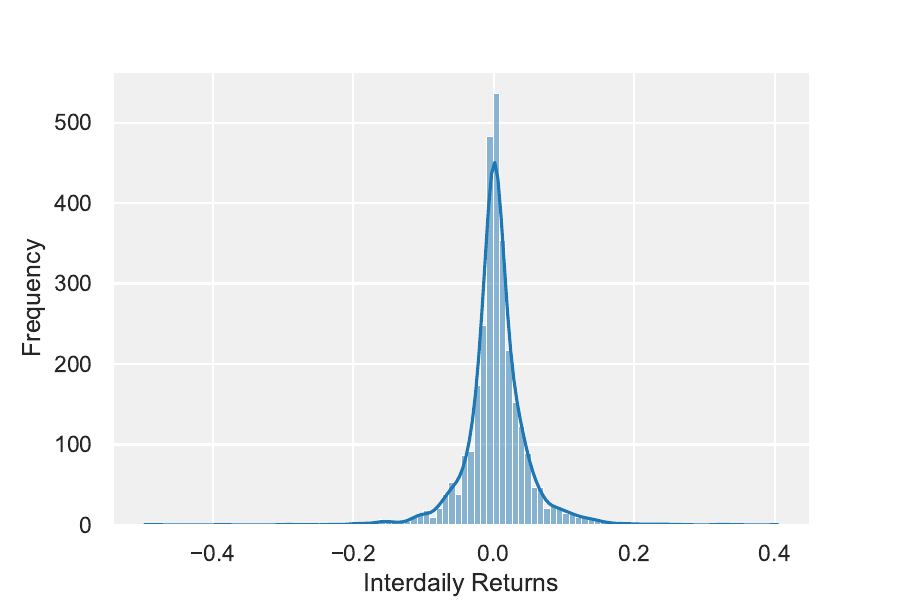}
\end{subfigure}%
\begin{subfigure}{.45\textwidth}
  \centering
  \caption{}
  \label{fig:bitcoin_boxplot}
  \includegraphics[width=\linewidth]{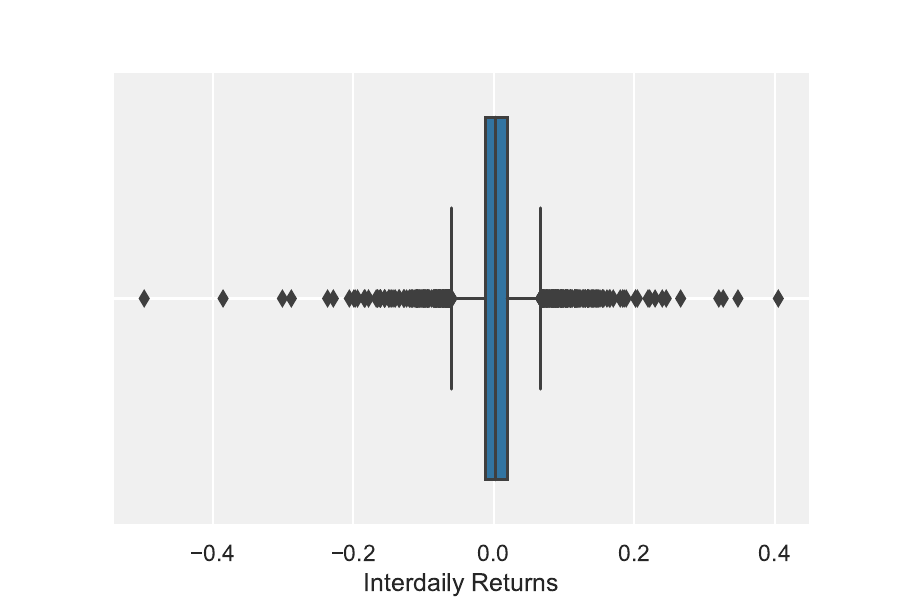}
\end{subfigure}

\caption{The following plot illustrates the distribution and Box Plot of Returns for the S\&P 500 and Bitcoin. Panel a) shows the histogram of the S\&P 500 returns, highlighting the distribution of daily returns. Panel b) provides a box plot of the S\&P 500 returns, showing the spread of data and identifying any potential outliers. Panel c) displays the histogram of Bitcoin returns, illustrating the distribution of daily returns for this cryptocurrency. Panel d) depicts a box plot of the Bitcoin returns, showcasing the dispersion of data and pointing out any outliers. These visualizations provide insights into the central tendency, dispersion, and shape of the distribution of returns for both assets.}
\label{fig:returns_info}
\end{figure}

\begin{figure}[htp]
\centering

\begin{subfigure}{.45\textwidth}
  \centering
  \caption{}
  \label{fig:}
  \includegraphics[width=\linewidth]{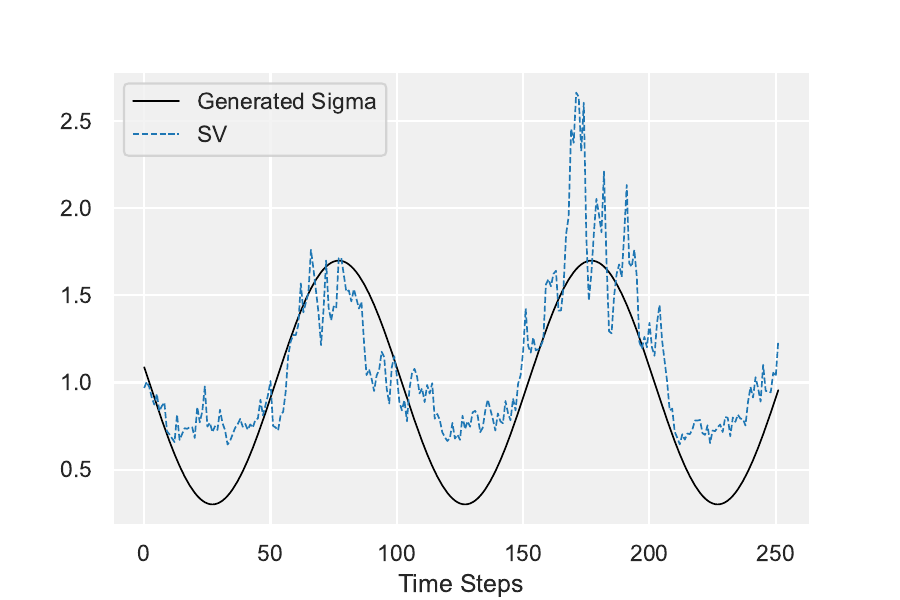}
\end{subfigure}%
\begin{subfigure}{.45\textwidth}
  \centering
  \caption{}
  \label{fig:}
  \includegraphics[width=\linewidth]{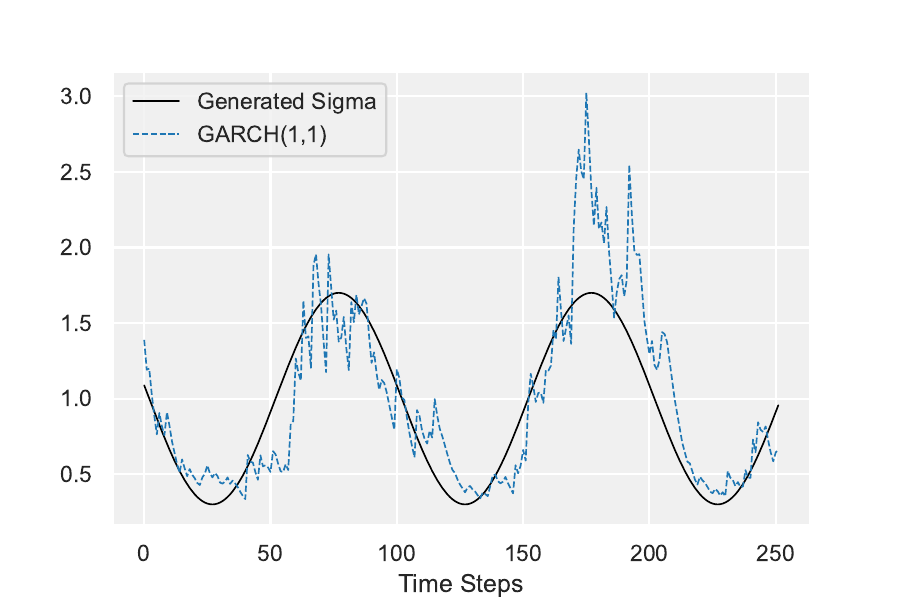}
\end{subfigure}

\begin{subfigure}{.45\textwidth}
  \centering
  \caption{}
  \label{fig:}
  \includegraphics[width=\linewidth]{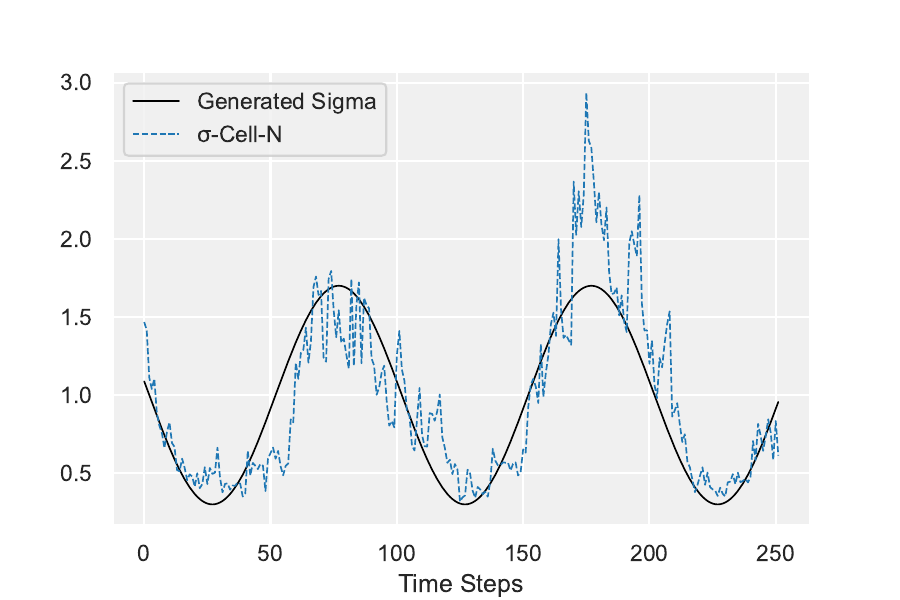}
\end{subfigure}%
\begin{subfigure}{.45\textwidth}
  \centering
  \caption{}
  \label{fig:}
  \includegraphics[width=\linewidth]{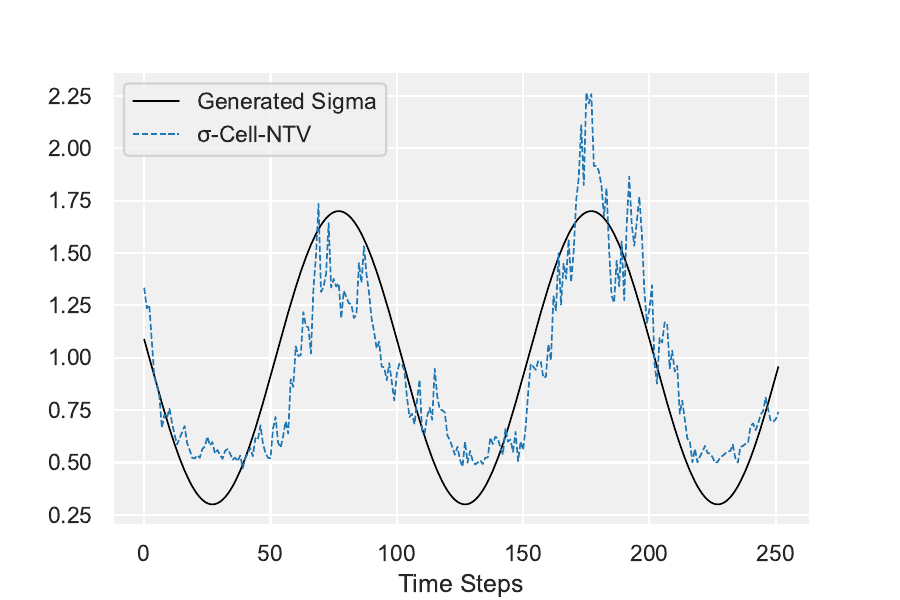}
\end{subfigure}

\begin{subfigure}{.45\textwidth}
  \centering
  \caption{}
  \label{fig:}
  \includegraphics[width=\linewidth]{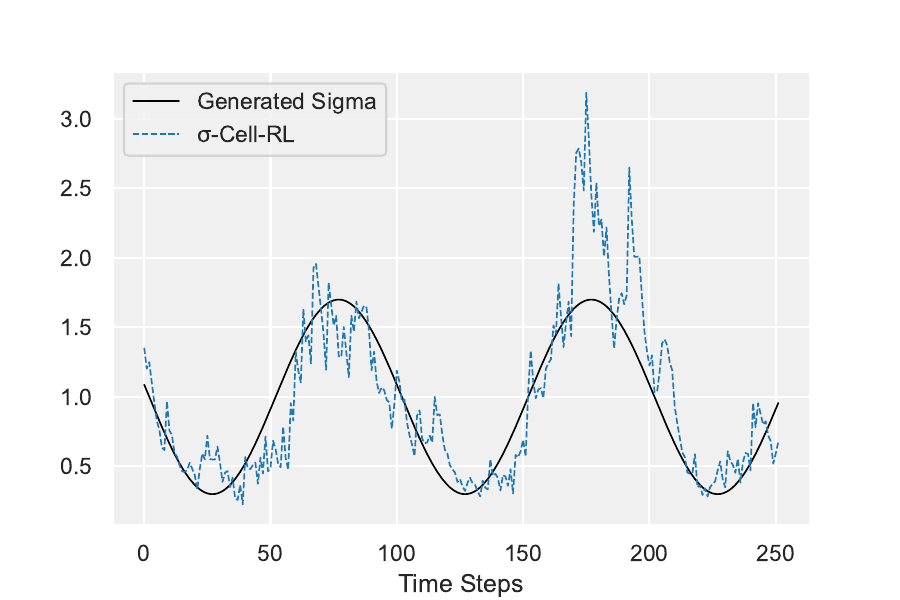}
\end{subfigure}%
\begin{subfigure}{.45\textwidth}
  \centering
  \caption{}
  \label{fig:}
  \includegraphics[width=\linewidth]{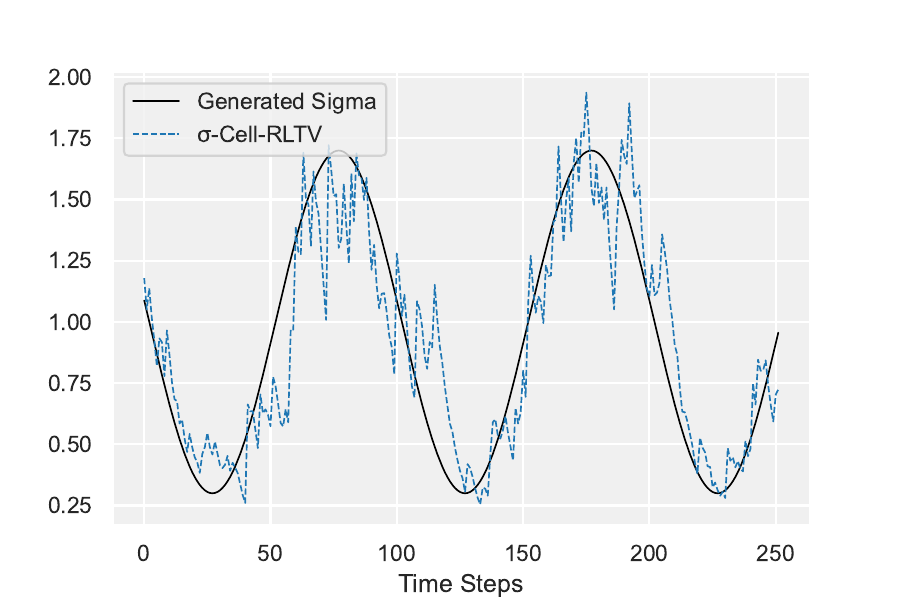}
\end{subfigure}

\caption{The following plot illustrates prediction for In-Sample Synthetic Data for Different Forecasting Models:
(a) Comparison of generated sigma values for the Stochastic Volatility (SV) model.
(a) Stochastic Volatility (SV) model's prediction of generated sigma values.
(b) GARCH(1,1) model's prediction of generated sigma values.
(c) $\sigma$-Cell-N model's prediction of generated sigma values.
(d) $\sigma$-Cell-NTV model's prediction of generated sigma values.
(e) $\sigma$-Cell-RL model's prediction of generated sigma values.
(f) $\sigma$-Cell-RLTV model's prediction of generated sigma values.
}
\label{fig:Valid_synth}
\end{figure}
\begin{figure}[htp]
\centering
\begin{subfigure}{.45\textwidth}
  \centering
  \caption{}
  \label{fig:}
  \includegraphics[width=\linewidth]{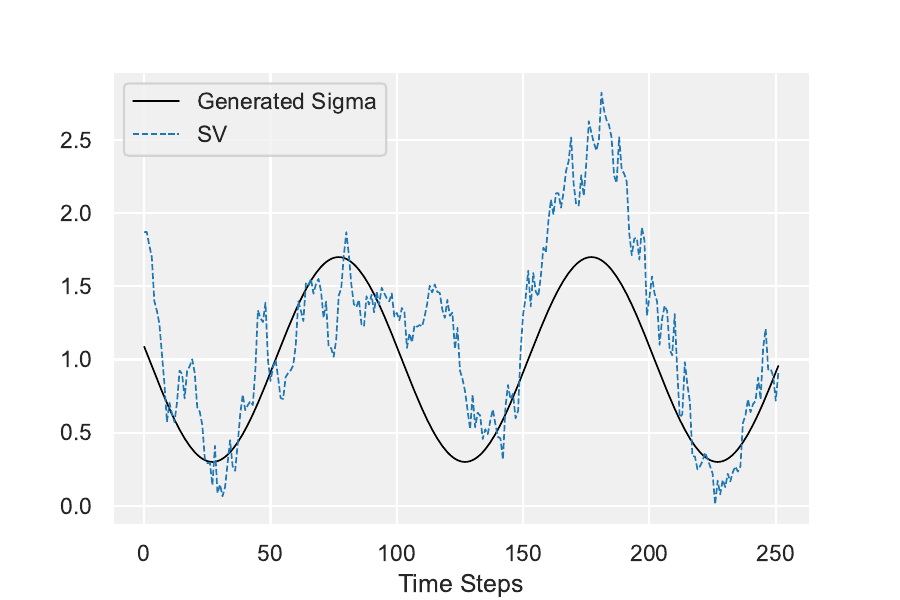}
\end{subfigure}%
\begin{subfigure}{.45\textwidth}
  \centering
  \caption{}
  \label{fig:}
  \includegraphics[width=\linewidth]{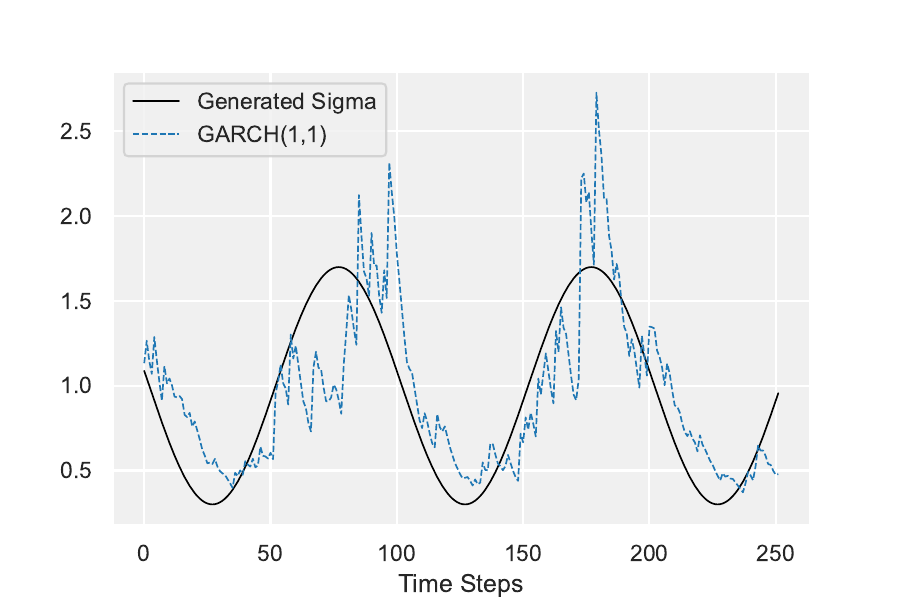}
\end{subfigure}

\begin{subfigure}{.45\textwidth}
  \centering
  \caption{}
  \label{fig:}
  \includegraphics[width=\linewidth]{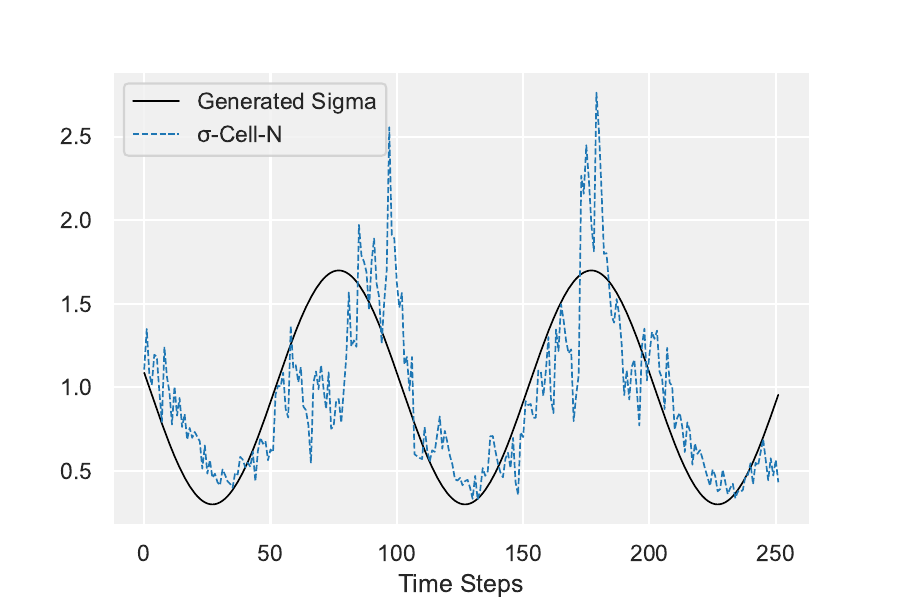}
\end{subfigure}%
\begin{subfigure}{.45\textwidth}
  \centering
  \caption{}
  \label{fig:}
  \includegraphics[width=\linewidth]{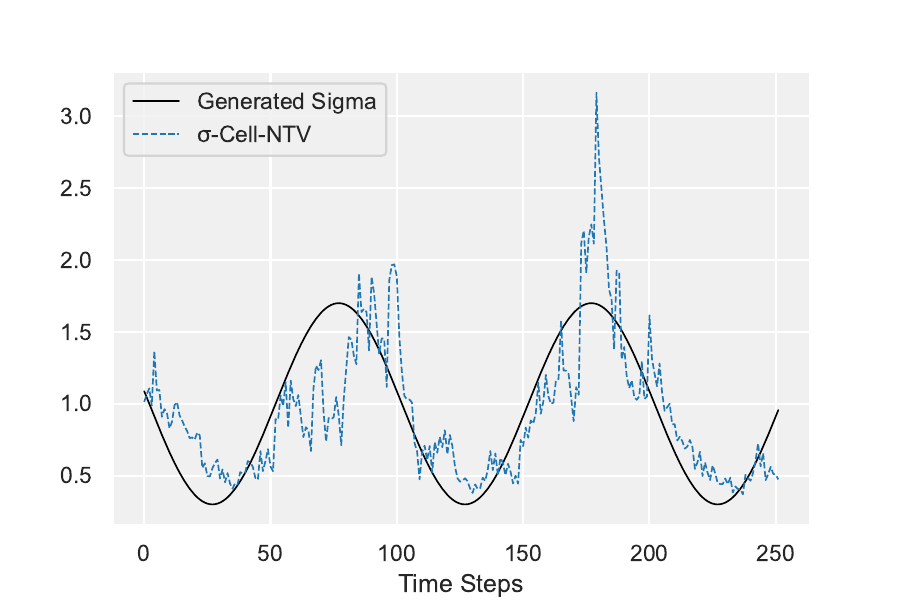}
\end{subfigure}

\begin{subfigure}{.45\textwidth}
  \centering
  \caption{}
  \label{fig:}
  \includegraphics[width=\linewidth]{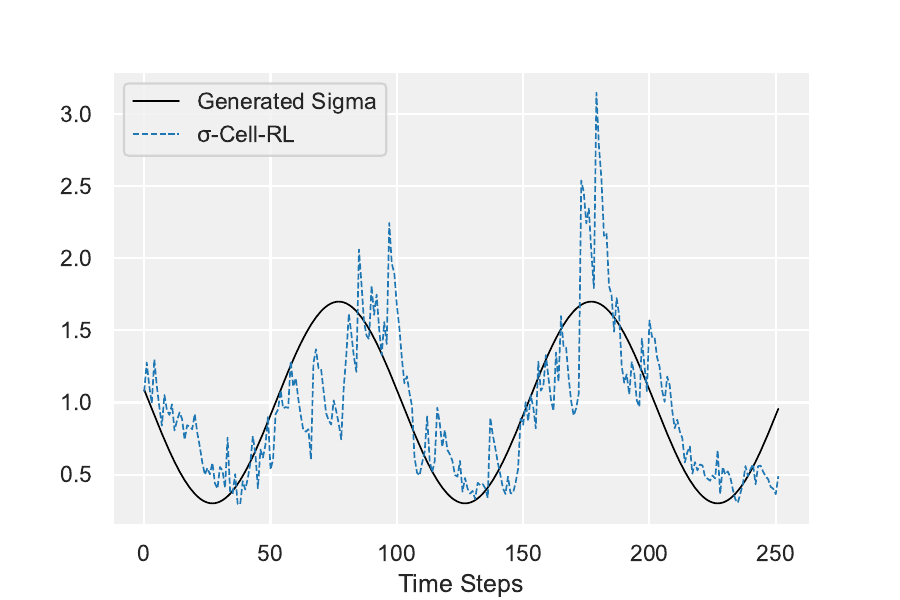}
\end{subfigure}%
\begin{subfigure}{.45\textwidth}
  \centering
  \caption{}
  \label{fig:}
  \includegraphics[width=\linewidth]{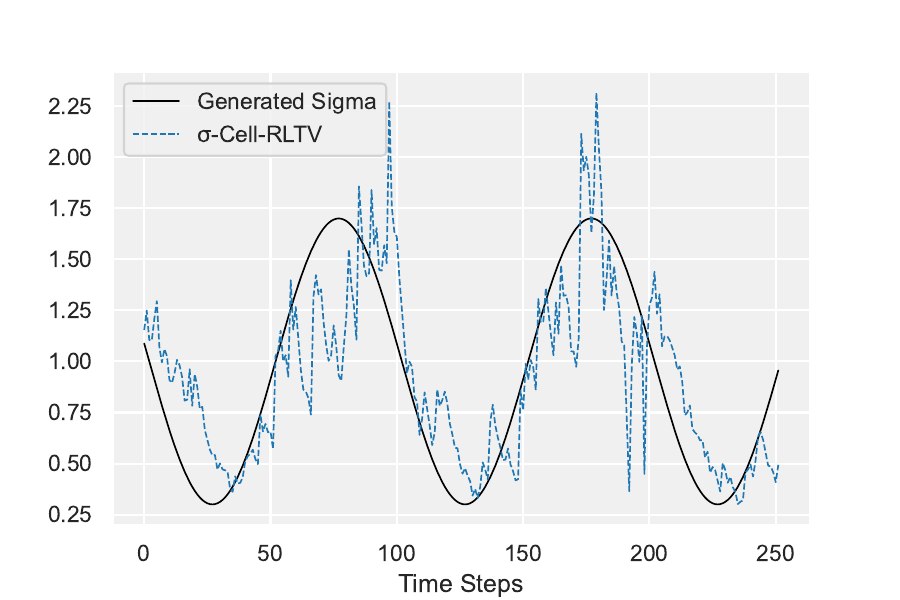}
\end{subfigure}
\caption{The following plot illustrates prediction for out-of-sample Synthetic Data for Different Forecasting Models:
(a) Stochastic Volatility (SV) model's prediction of generated sigma values.
(b) GARCH(1,1) model's prediction of generated sigma values.
(c) $\sigma$-Cell-N model's prediction of generated sigma values.
(d) $\sigma$-Cell-NTV model's prediction of generated sigma values.
(e) $\sigma$-Cell-RL model's prediction of generated sigma values.
(f) $\sigma$-Cell-RLTV model's prediction of generated sigma values.}
\label{fig:test_synth}
\end{figure}

\begin{figure}[htp]
\centering
\begin{subfigure}{.45\textwidth}
  \centering
  \caption{}
  \label{fig:a}
  \includegraphics[width=\linewidth]{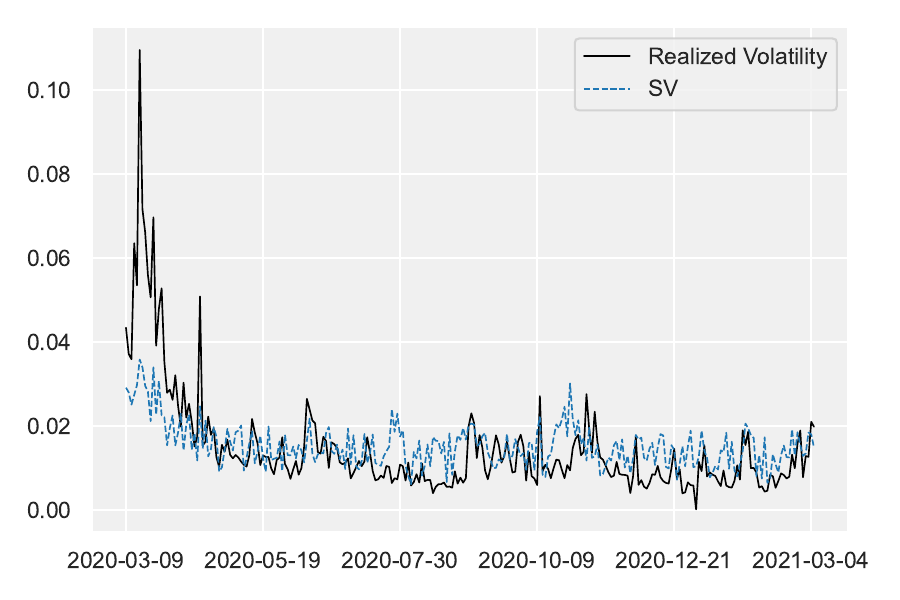}
\end{subfigure}%
\begin{subfigure}{.45\textwidth}
  \centering
  \caption{}
  \label{fig:b}
  \includegraphics[width=\linewidth]{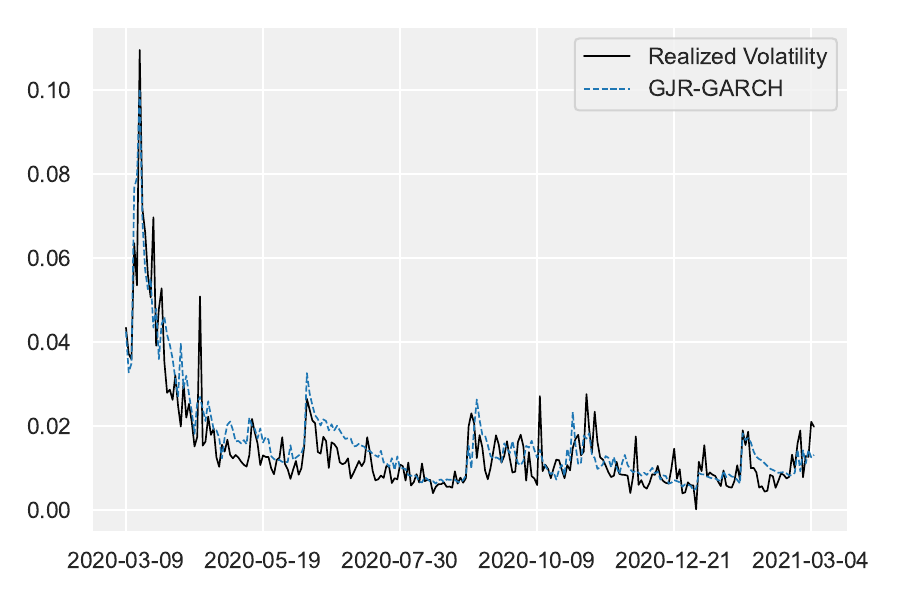}
\end{subfigure}

\begin{subfigure}{.45\textwidth}
  \centering
  \caption{}
  \label{fig:c}
  \includegraphics[width=\linewidth]{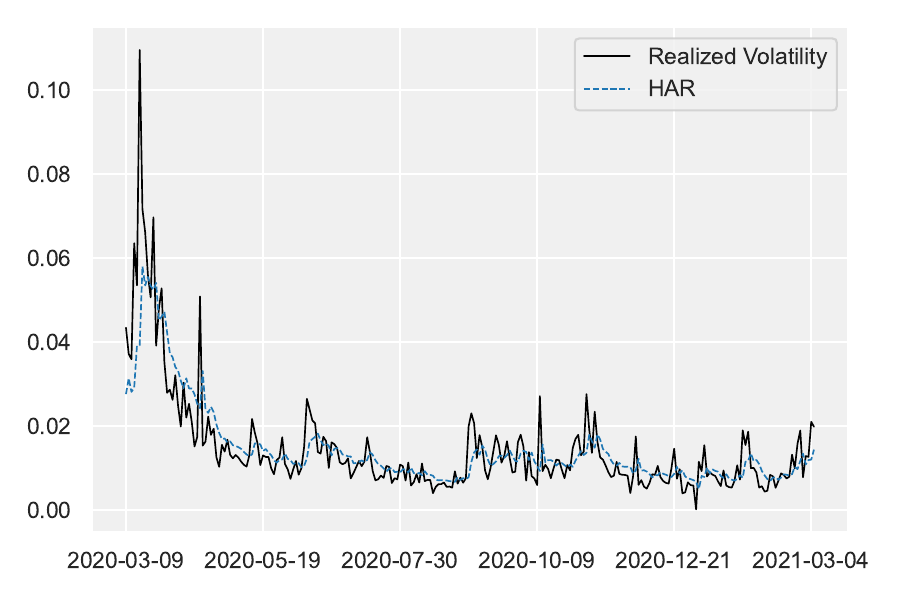}
\end{subfigure}%
\begin{subfigure}{.45\textwidth}
  \centering
  \caption{}
  \label{fig:d}
  \includegraphics[width=\linewidth]{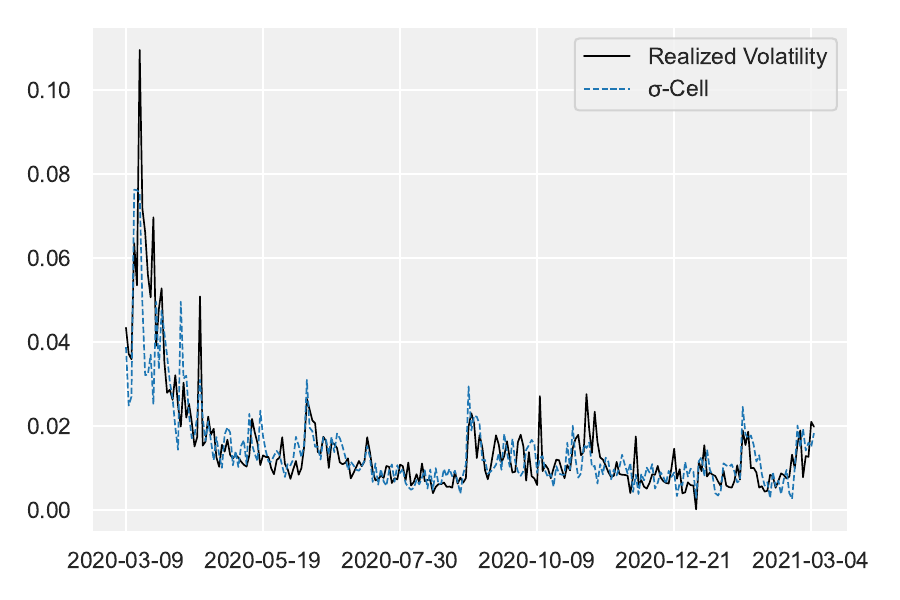}
\end{subfigure}
\caption{The following plot illustrates prediction for In-Sample Realized Volatility for the S\&P 500 index. The presented plots provide a visual assessment of the performance of various models in predicting realized volatility. Each sub-figure displays the true realized volatility along with the model's estimate. 
(a) Stochastic Volatility (SV) model
(b) GJR-GARCH model
(c) HAR model
(d) $\sigma$-Cell model
\textbf{(Continued on next page.)}}
\label{fig:valid_SnP}
\end{figure}
\begin{figure}[htp]
\ContinuedFloat
\centering
\begin{subfigure}{.45\textwidth}
  \centering
  \caption{}
  \label{fig:e}
  \includegraphics[width=\linewidth]{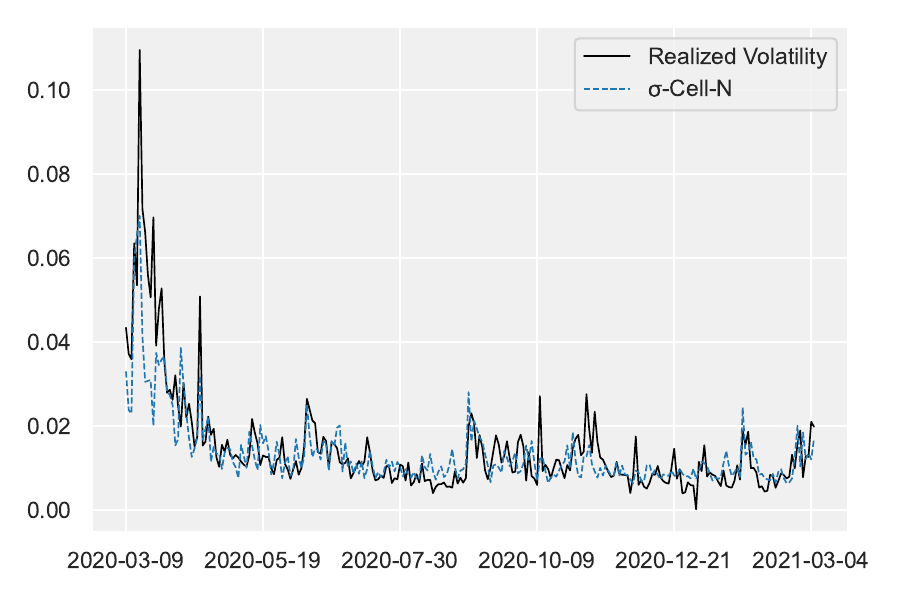}
\end{subfigure}
\begin{subfigure}{.45\textwidth}
  \centering
  \caption{}
  \label{fig:f}
  \includegraphics[width=\linewidth]{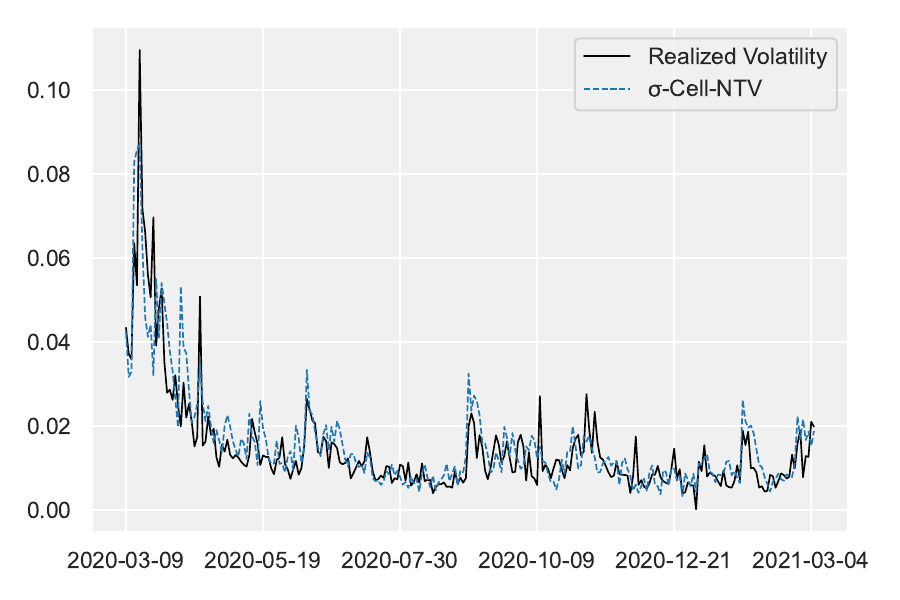}
\end{subfigure}%

\begin{subfigure}{.45\textwidth}
  \centering
  \caption{}
  \label{fig:g}
  \includegraphics[width=\linewidth]{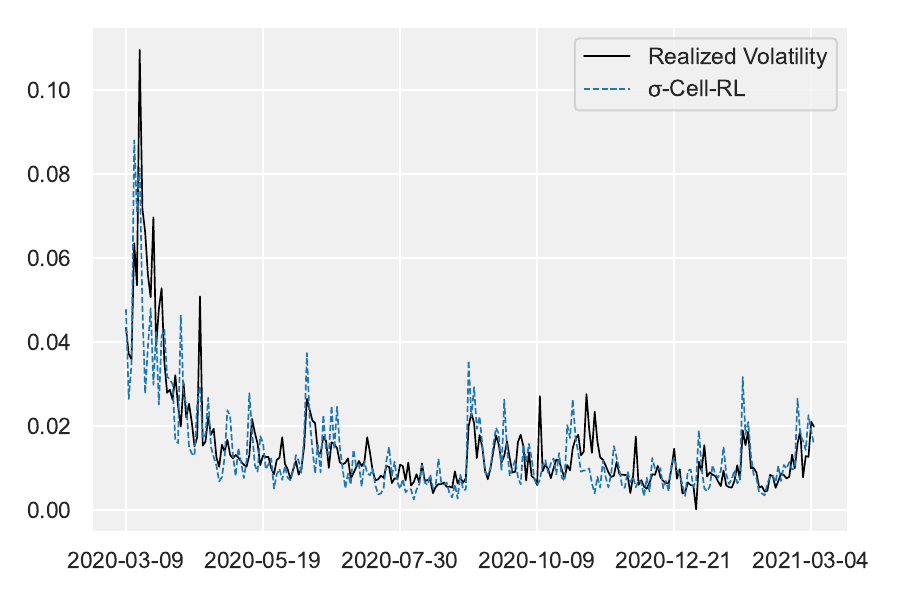}
\end{subfigure}
\begin{subfigure}{.45\textwidth}
  \centering
  \caption{}
  \label{fig:h}
  \includegraphics[width=\linewidth]{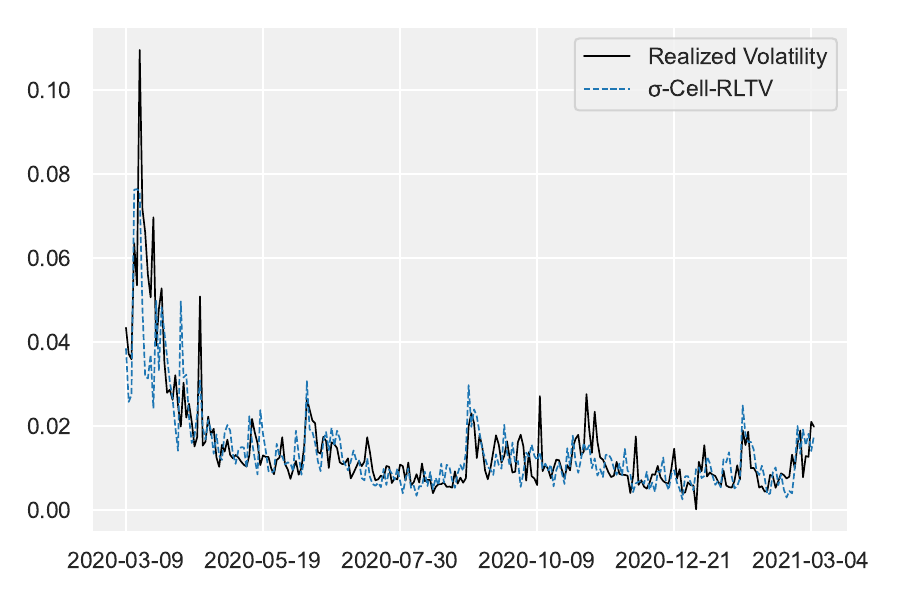}
\end{subfigure}%
\caption{(Continued from previous page.) (e) $\sigma$-Cell-N model (f) $\sigma$-Cell-NTV model (g) $\sigma$-Cell-RL model (h) $\sigma$-Cell-RLTV model}
\end{figure}

\begin{figure}[htp]
\centering
\begin{subfigure}{.45\textwidth}
  \centering
  \caption{}
  \label{fig:a}
  \includegraphics[width=\linewidth]{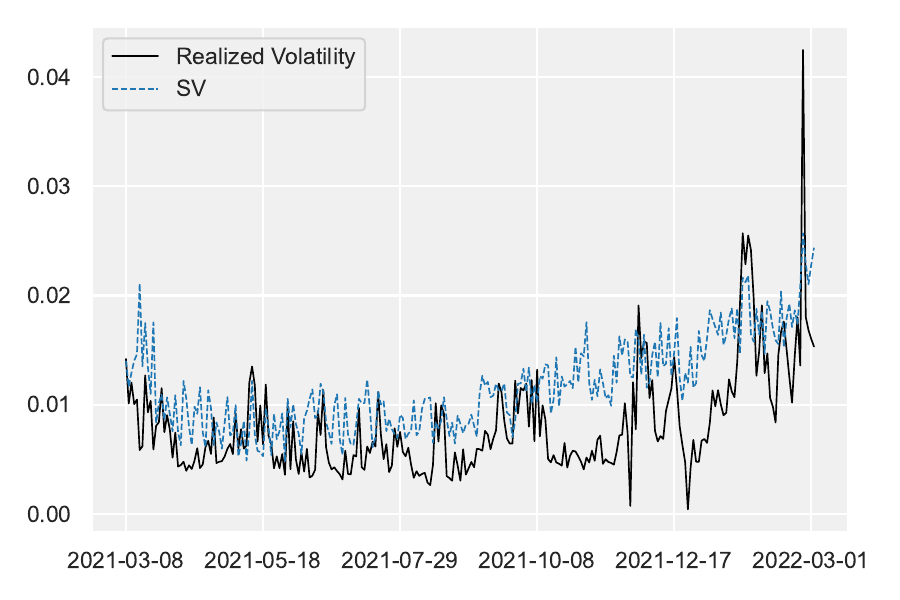}
\end{subfigure}%
\begin{subfigure}{.45\textwidth}
  \centering
  \caption{}
  \label{fig:b}
  \includegraphics[width=\linewidth]{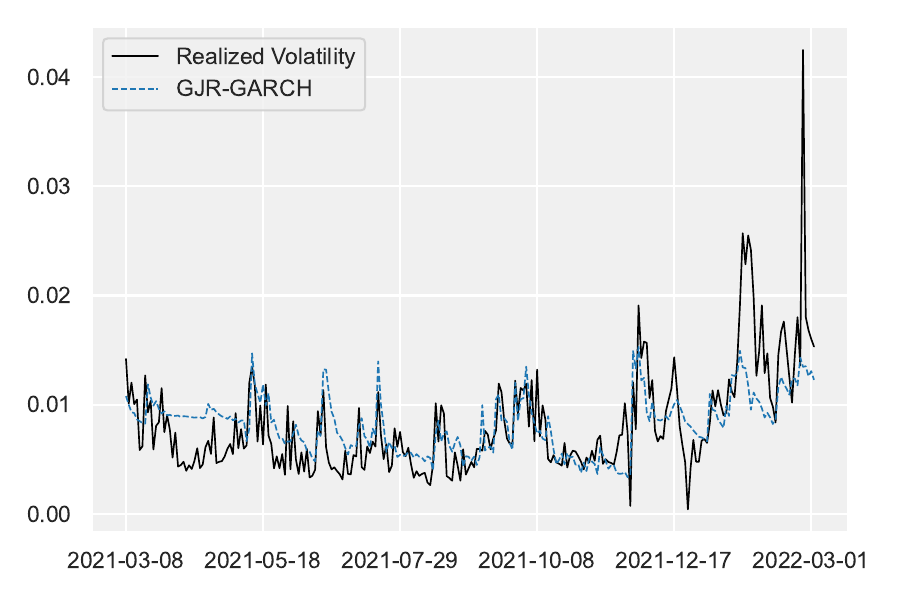}
\end{subfigure}

\begin{subfigure}{.45\textwidth}
  \centering
  \caption{}
  \label{fig:c}
  \includegraphics[width=\linewidth]{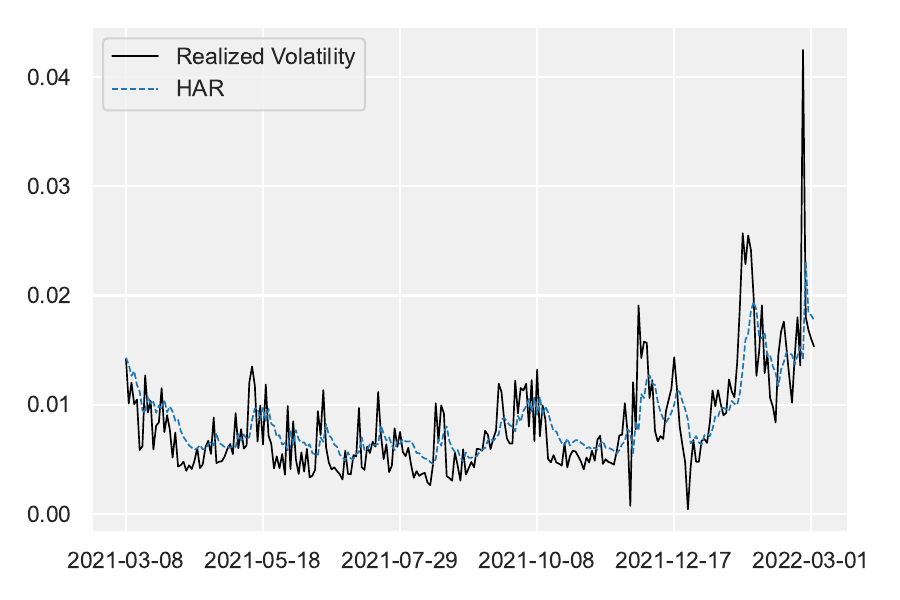}
\end{subfigure}%
\begin{subfigure}{.45\textwidth}
  \centering
  \caption{}
  \label{fig:d}
  \includegraphics[width=\linewidth]{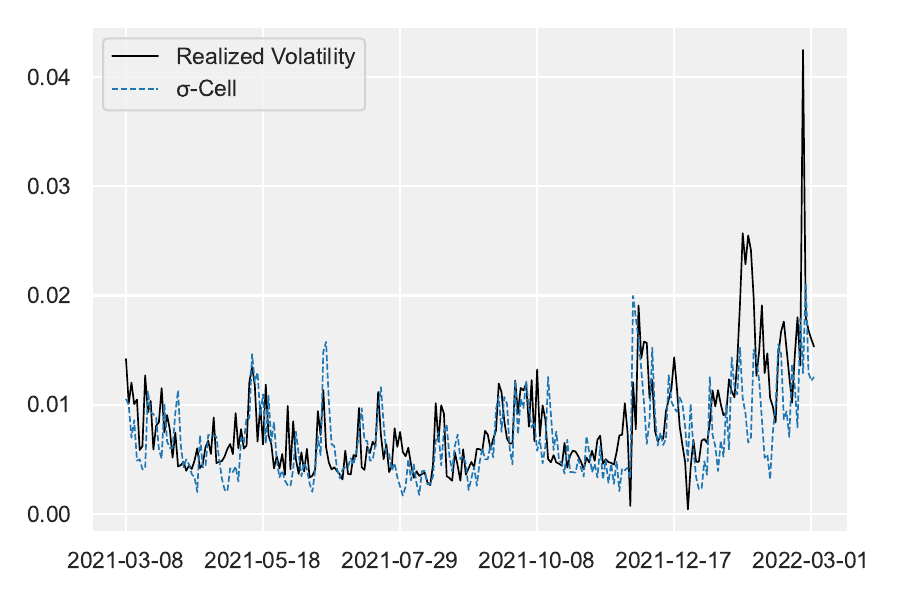}
\end{subfigure}
\caption{The following plot illustrates prediction for out-of-sample Realized Volatility for the S\&P 500 index. The presented plots provide a visual assessment of the performance of various models in predicting realized volatility. Each sub-figure displays the true realized volatility along with the model's 1-step ahead prediction. (a) Stochastic Volatility (SV) model
(b) GJR-GARCH model
(c) HAR model
(d) $\sigma$-Cell model 
\textbf{(Continued on next page.)}}
\label{fig:test_SnP}
\end{figure}
\begin{figure}[htp]
\ContinuedFloat
\centering
\begin{subfigure}{.45\textwidth}
  \centering
  \caption{}
  \label{fig:e}
  \includegraphics[width=\linewidth]{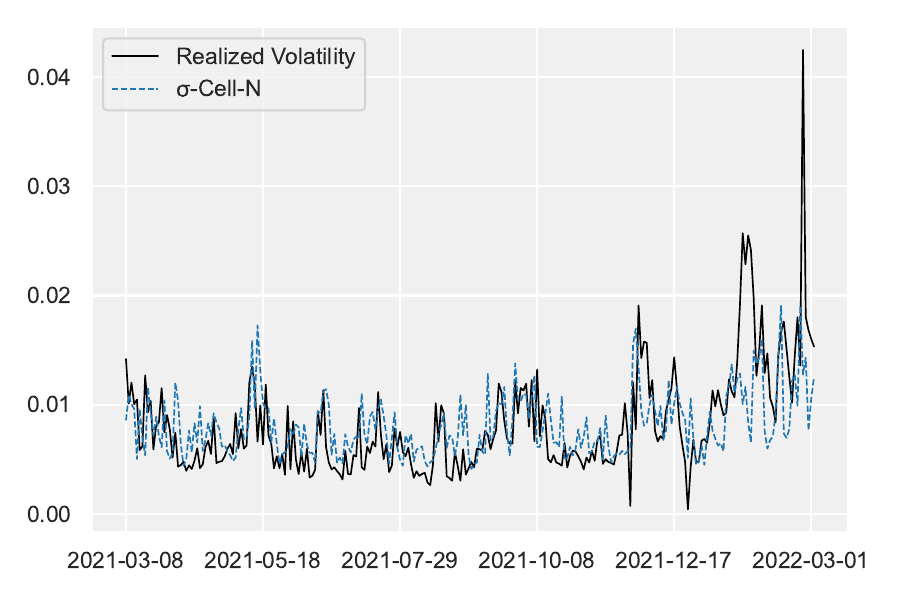}
\end{subfigure}
\begin{subfigure}{.45\textwidth}
  \centering
  \caption{}
  \label{fig:f}
  \includegraphics[width=\linewidth]{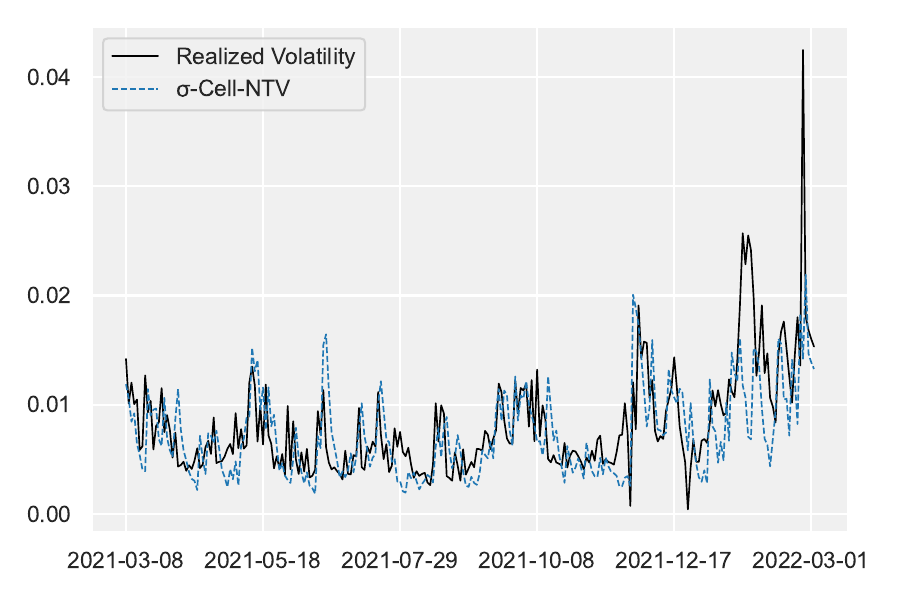}
\end{subfigure}%

\begin{subfigure}{.45\textwidth}
  \centering
  \caption{}
  \label{fig:g}
  \includegraphics[width=\linewidth]{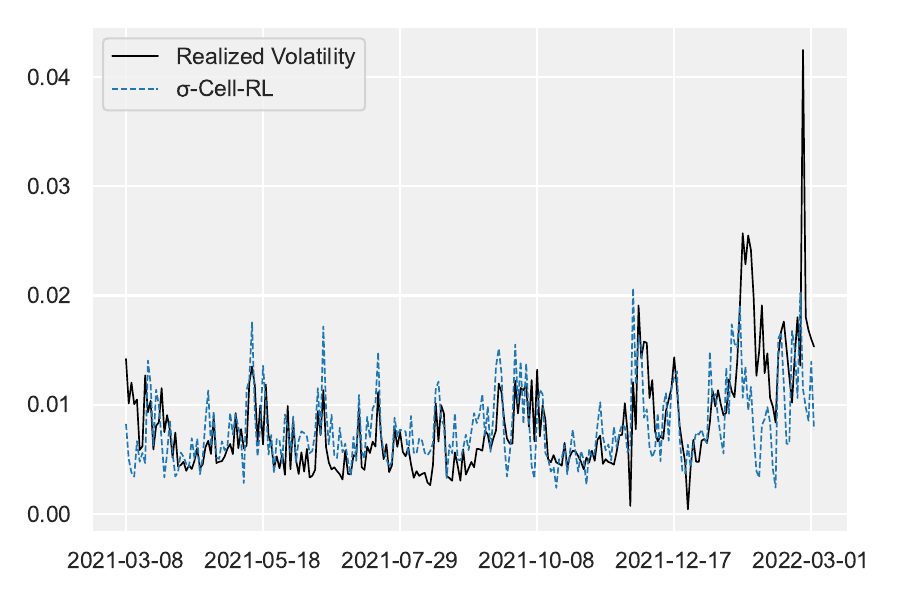}
\end{subfigure}
\begin{subfigure}{.45\textwidth}
  \centering
  \caption{}
  \label{fig:h}
  \includegraphics[width=\linewidth]{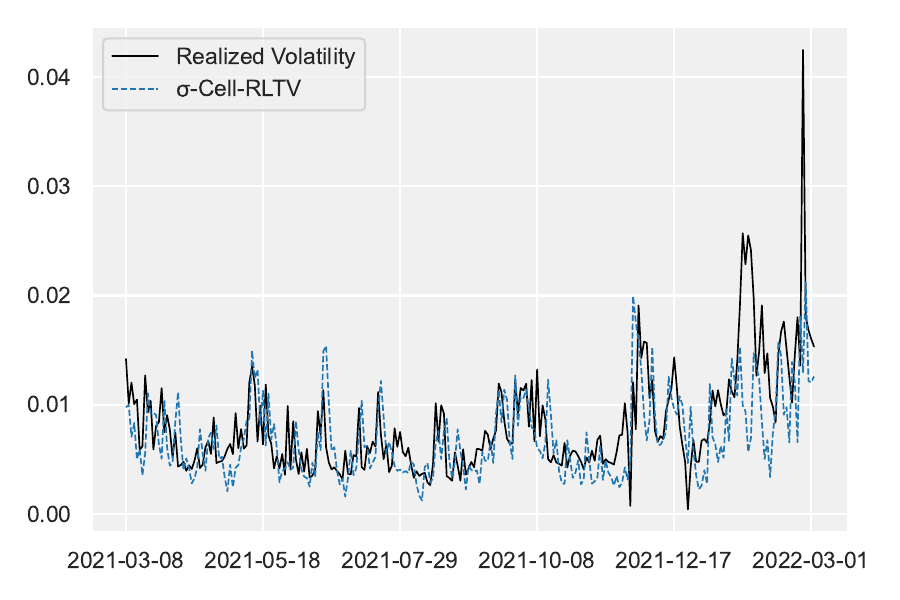}
\end{subfigure}%
\caption{(Continued from previous page.) (e) $\sigma$-Cell-N model (f) $\sigma$-Cell-NTV model (g) $\sigma$-Cell-RL model (h) $\sigma$-Cell-RLTV model}
\end{figure}

\begin{figure}[htp]
\centering
\begin{subfigure}{.45\textwidth}
  \centering
  \caption{}
  \label{fig:a}
  \includegraphics[width=\linewidth]{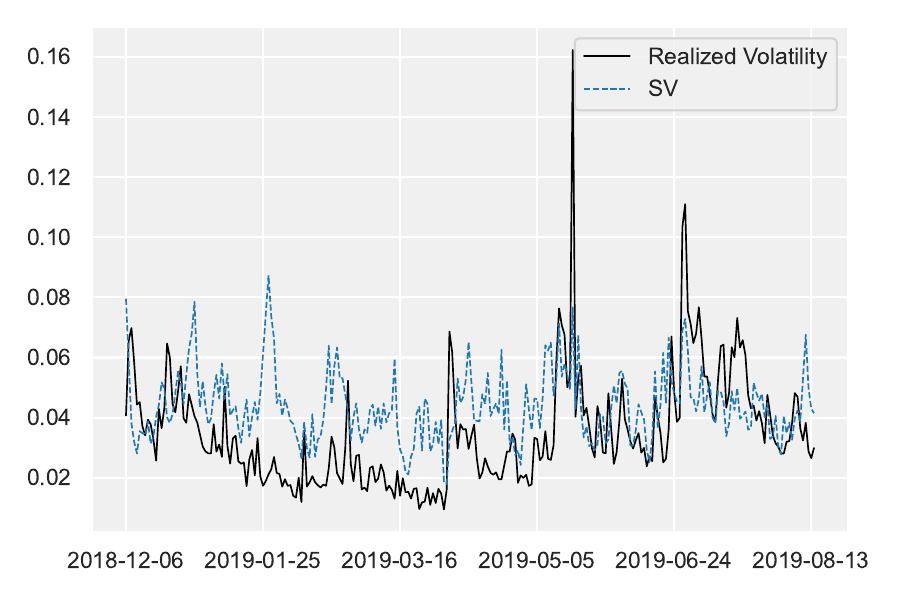}
\end{subfigure}%
\begin{subfigure}{.45\textwidth}
  \centering
  \caption{}
  \label{fig:b}
  \includegraphics[width=\linewidth]{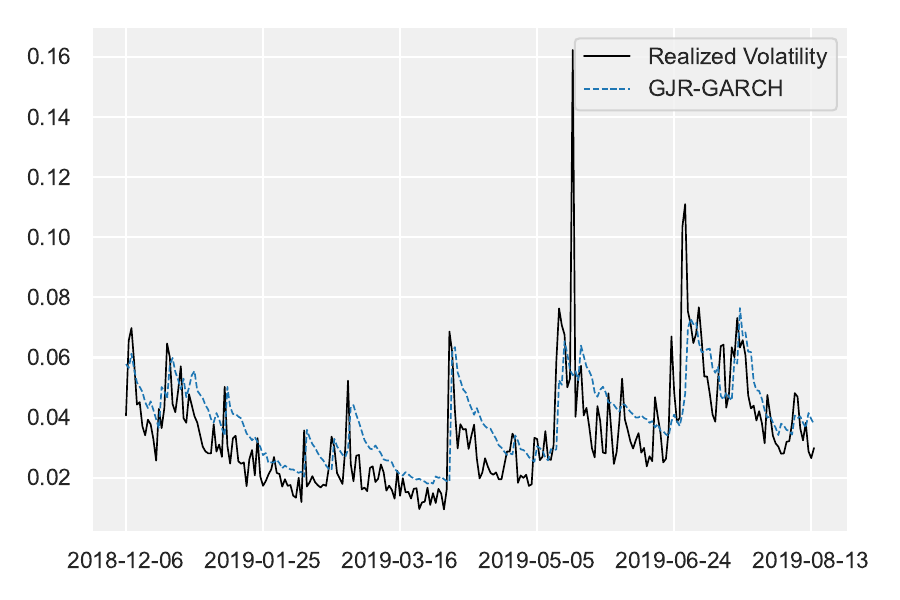}
\end{subfigure}

\begin{subfigure}{.45\textwidth}
  \centering
  \caption{}
  \label{fig:c}
  \includegraphics[width=\linewidth]{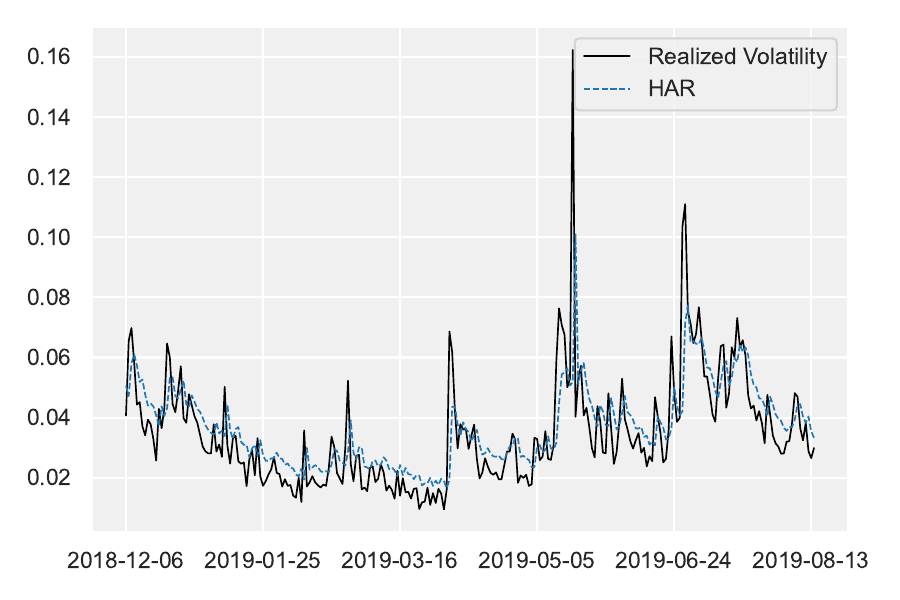}
\end{subfigure}%
\begin{subfigure}{.45\textwidth}
  \centering
  \caption{}
  \label{fig:d}
  \includegraphics[width=\linewidth]{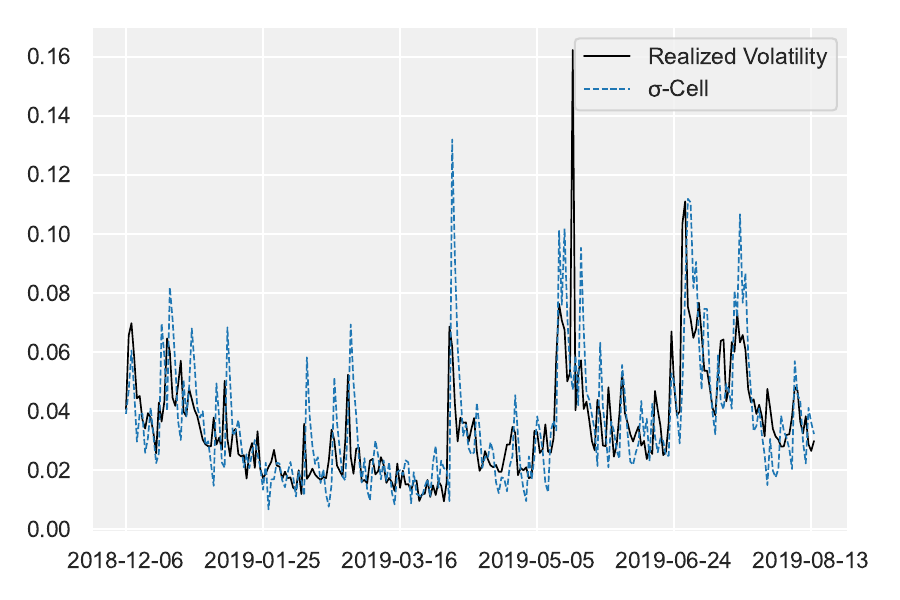}
\end{subfigure}
\caption{The following plot illustrates prediction for In-Sample Realized Volatility for the BTCUSDT. The presented plots provide a visual assessment of the performance of various models in predicting realized volatility. Each sub-figure displays the true realized volatility along with the model's estimate. 
(a) Stochastic Volatility (SV) model
(b) GJR-GARCH model
(c) HAR model
(d) $\sigma$-Cell model 
\textbf{(Continued on next page.)}}
\label{fig:valid_BTCUSDT}
\end{figure}
\begin{figure}[htp]
\ContinuedFloat
\centering
\begin{subfigure}{.45\textwidth}
  \centering
  \caption{}
  \label{fig:e}
  \includegraphics[width=\linewidth]{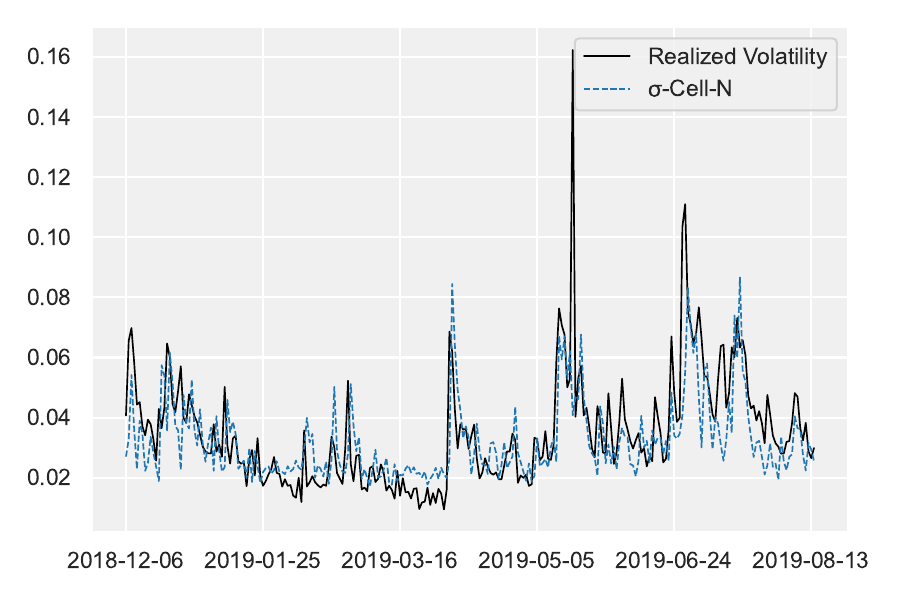}
\end{subfigure}
\begin{subfigure}{.45\textwidth}
  \centering
  \caption{}
  \label{fig:f}
  \includegraphics[width=\linewidth]{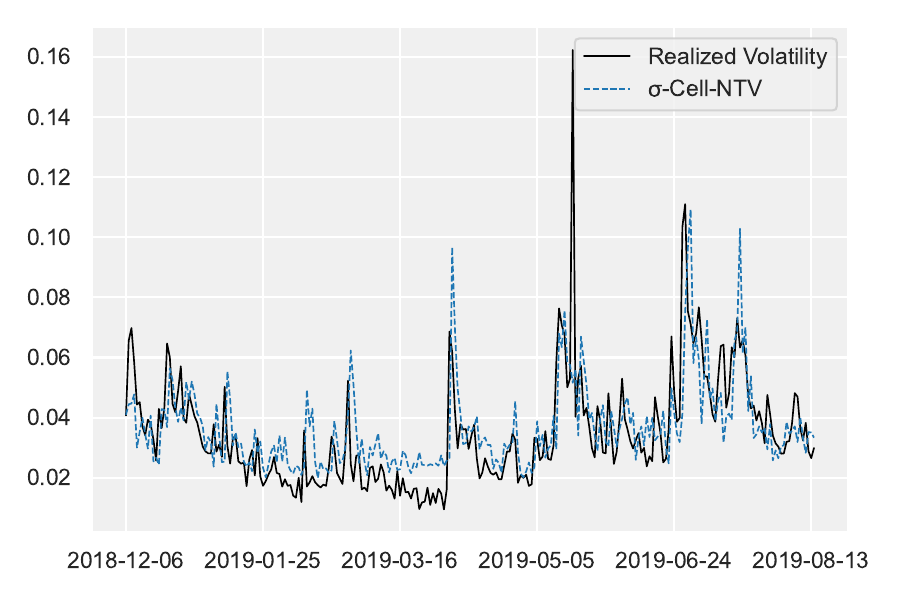}
\end{subfigure}%

\begin{subfigure}{.45\textwidth}
  \centering
  \caption{}
  \label{fig:g}
  \includegraphics[width=\linewidth]{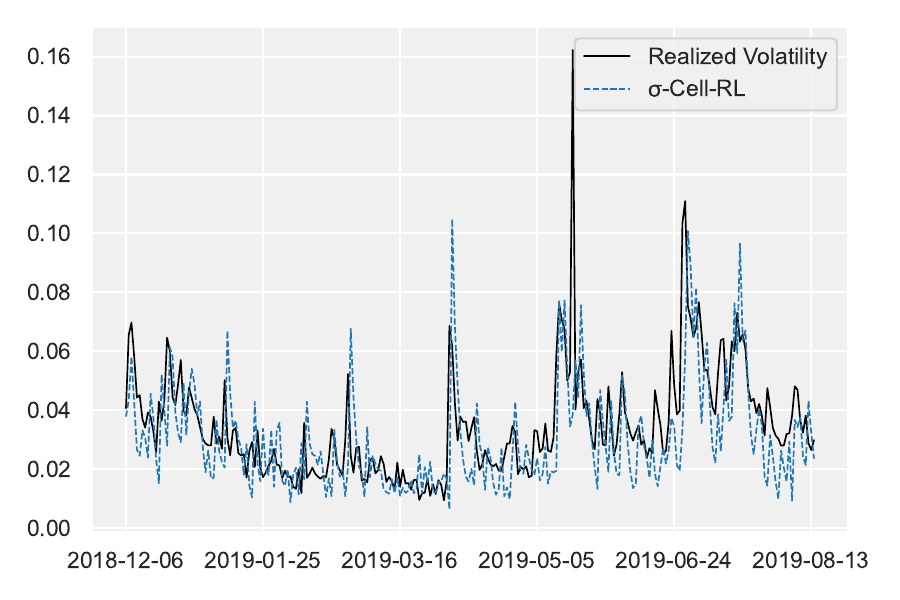}
\end{subfigure}
\begin{subfigure}{.45\textwidth}
  \centering
  \caption{}
  \label{fig:h}
  \includegraphics[width=\linewidth]{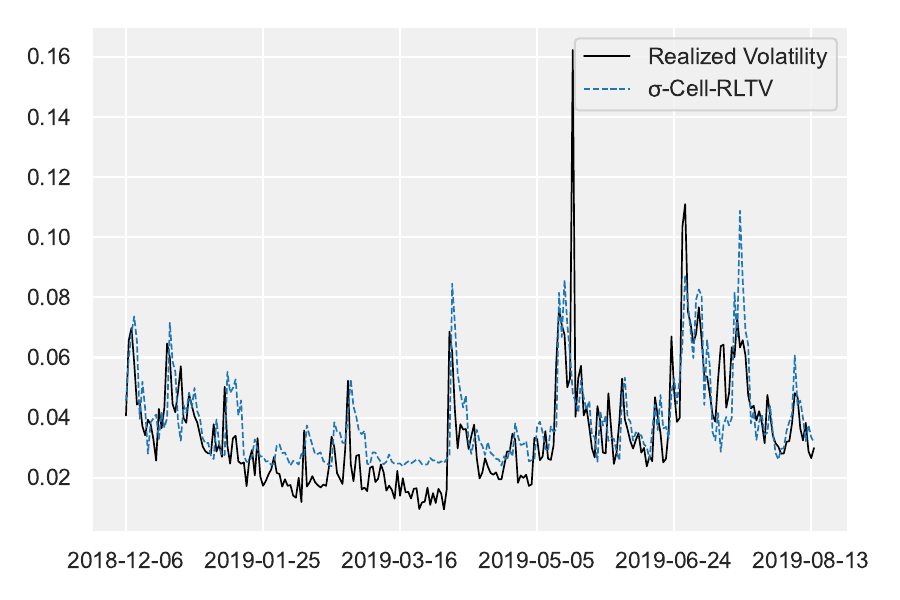}
\end{subfigure}%
\caption{(Continued from previous page.) (e) $\sigma$-Cell-N model (f) $\sigma$-Cell-NTV model (g) $\sigma$-Cell-RL model (h) $\sigma$-Cell-RLTV model}
\end{figure}

\begin{figure}[htp]
\centering
\begin{subfigure}{.45\textwidth}
  \centering
  \caption{}
  \label{fig:a}
  \includegraphics[width=\linewidth]{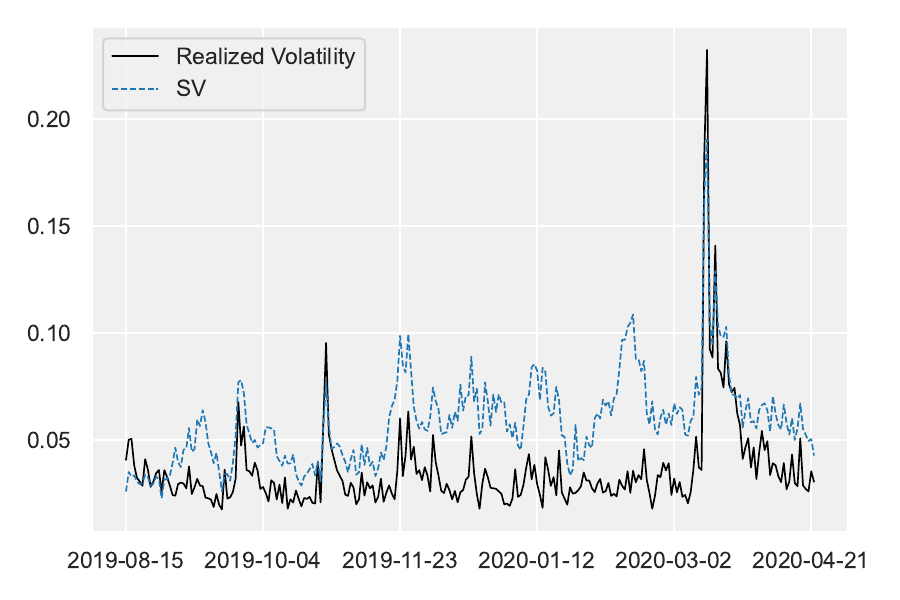}
\end{subfigure}%
\begin{subfigure}{.45\textwidth}
  \centering
  \caption{}
  \label{fig:b}
  \includegraphics[width=\linewidth]{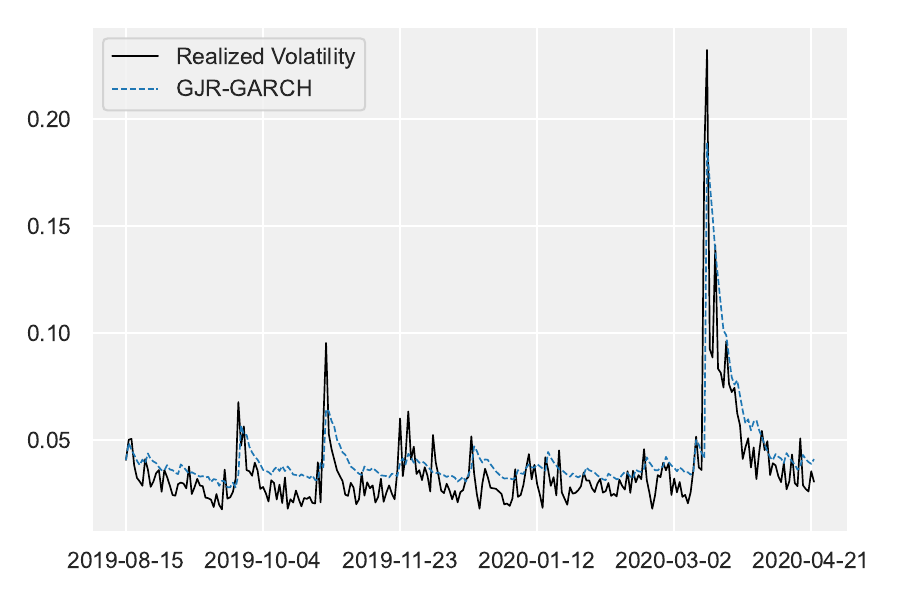}
\end{subfigure}

\begin{subfigure}{.45\textwidth}
  \centering
  \caption{}
  \label{fig:c}
  \includegraphics[width=\linewidth]{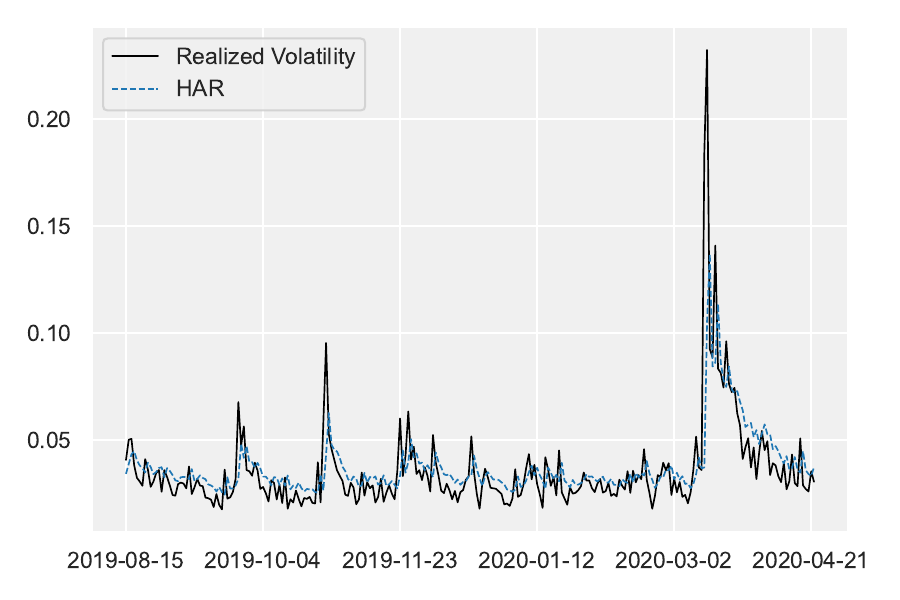}
\end{subfigure}%
\begin{subfigure}{.45\textwidth}
  \centering
  \caption{}
  \label{fig:d}
  \includegraphics[width=\linewidth]{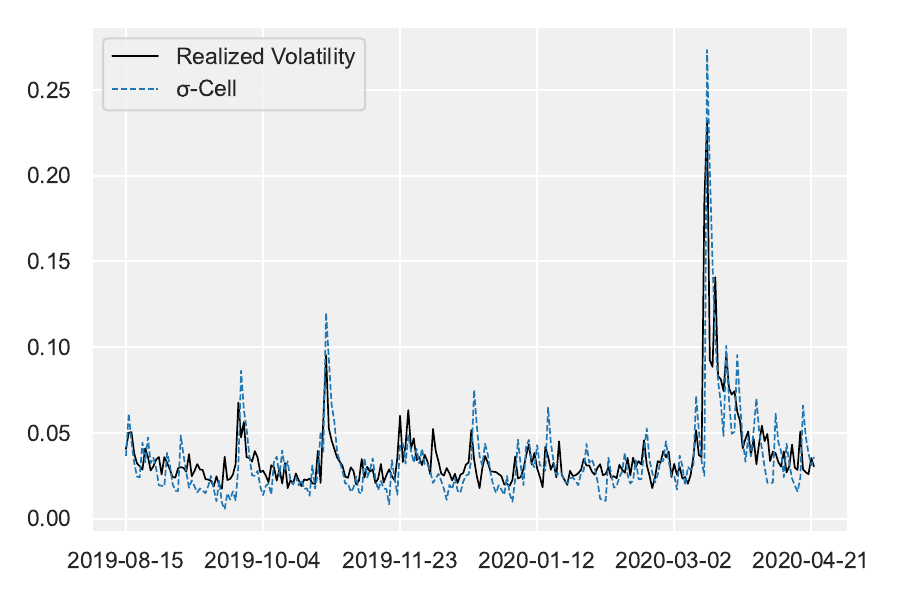}
\end{subfigure}
\caption{The following plot illustrates prediction for out-of-sample Realized Volatility for BTCUSDT pair. The presented plots provide a visual assessment of the performance of various models in predicting realized volatility. Each sub-figure displays the true realized volatility along with the model's 1-step ahead prediction.
(a) Stochastic Volatility (SV) model
(b) GJR-GARCH model
(c) HAR model
(d) $\sigma$-Cell model 
\textbf{(Continued on next page.)}}
\label{fig:test_BTCUSDT}
\end{figure}
\begin{figure}[htp]
\ContinuedFloat
\centering
\begin{subfigure}{.45\textwidth}
  \centering
  \caption{}
  \label{fig:e}
  \includegraphics[width=\linewidth]{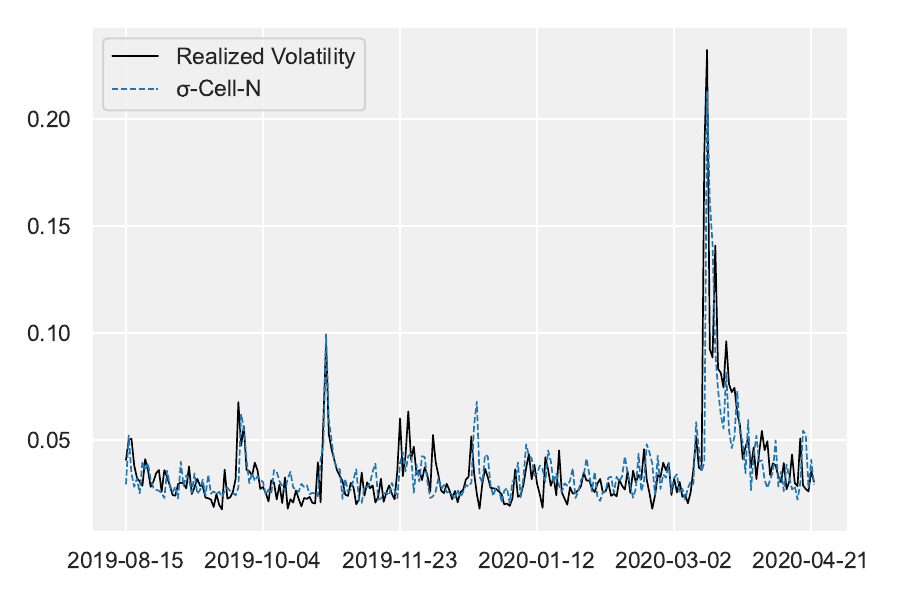}
\end{subfigure}
\begin{subfigure}{.45\textwidth}
  \centering
  \caption{}
  \label{fig:f}
  \includegraphics[width=\linewidth]{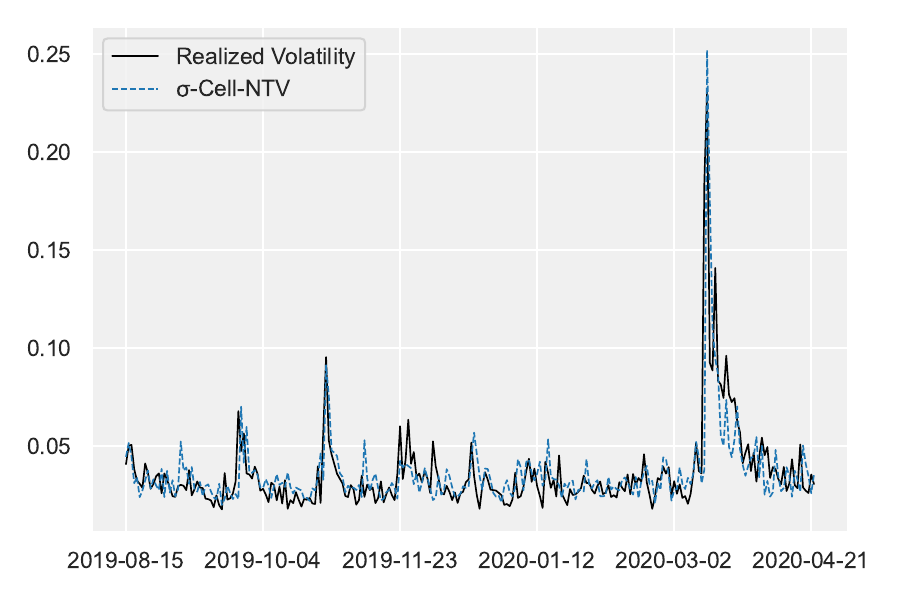}
\end{subfigure}%

\begin{subfigure}{.45\textwidth}
  \centering
  \caption{}
  \label{fig:g}
  \includegraphics[width=\linewidth]{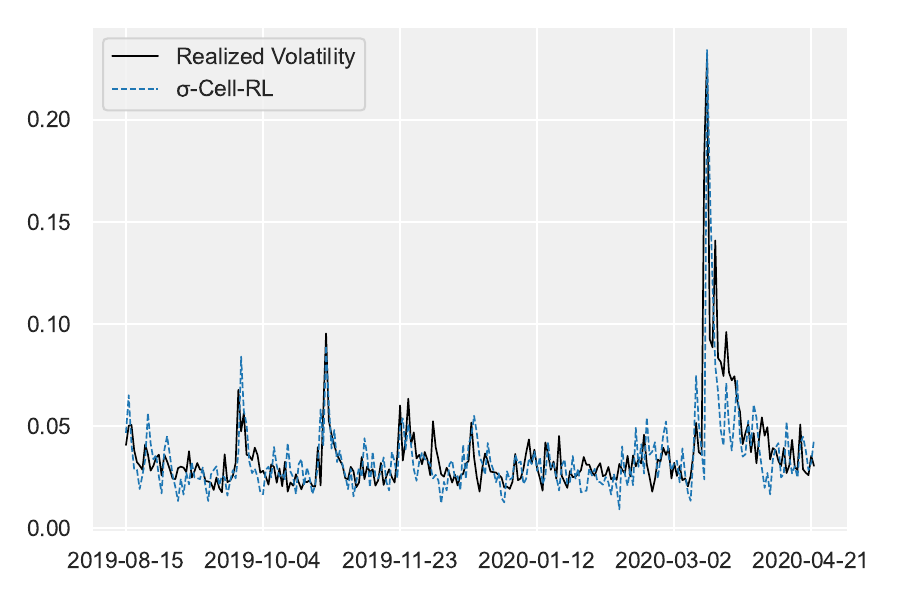}
\end{subfigure}
\begin{subfigure}{.45\textwidth}
  \centering
  \caption{}
  \label{fig:h}
  \includegraphics[width=\linewidth]{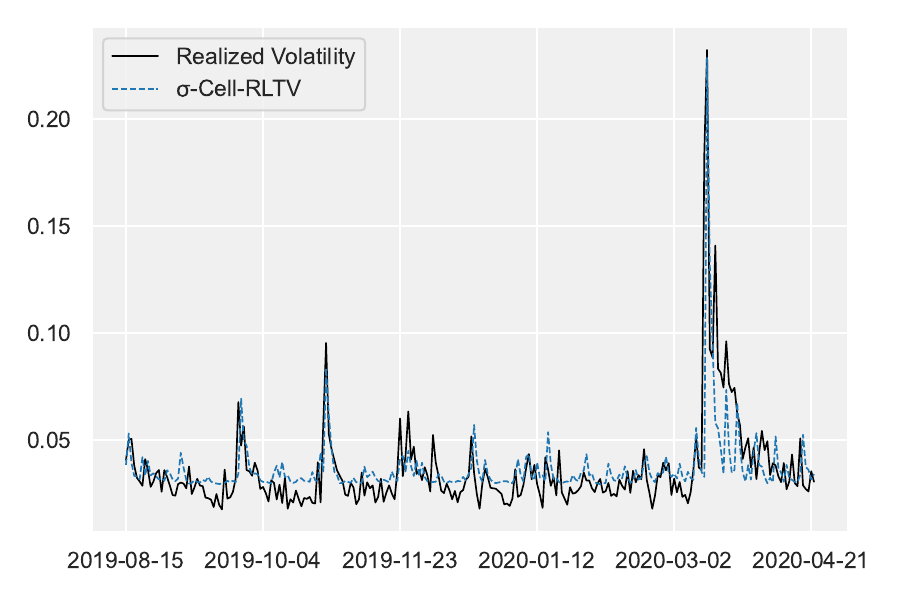}
\end{subfigure}%
\caption{(Continued from previous page.) (e) $\sigma$-Cell-N model (f) $\sigma$-Cell-NTV model (g) $\sigma$-Cell-RL model (h) $\sigma$-Cell-RLTV model}
\end{figure}

\newpage
\printbibliography

\end{document}